\documentclass{article}

\usepackage{amsmath}
\usepackage{amsfonts}
\usepackage{graphicx}
\usepackage{caption}
\usepackage{subcaption}

 \usepackage{amssymb}

\usepackage{epstopdf}
\usepackage{epsfig}
\usepackage{hyperref}

\usepackage{color}

\date{\today}

\newcommand{\insertplot}[5]{\begin{figure}
 \hfill\hbox to 0.05in{\vbox to #5in{\vfill
 \inputplot{#1}{#4}{#5}}\hfill}
 \hfill\vspace{-.1in}
 \caption{#2}\label{#3}
 \end{figure}}
 \newcommand{\inputplot}[3]{
 \special{ps: plotfile #1}
}
\newcounter{fig}

\newcommand{\ee}{\end{equation}}
\newcommand{\eea}{\end{eqnarray}}
\newcommand{\be}{\begin{equation}}
\newcommand{\bea}{\begin{eqnarray}}

\def\theequation{\arabic{equation}}

\begin{document}

\title{
Charged rotating black holes in  
Einstein--Maxwell--Chern-Simons
\\
 theory with a negative cosmological constant
}
 \vspace{1.5truecm}
\author{
{\bf Jose Luis Bl\'azquez-Salcedo$^1$},
{\bf Jutta Kunz$^1$},\\
{\bf Francisco Navarro-L\'erida$^2$},
{\bf Eugen Radu$^3$}\\
 \vspace{0.5truecm}
$^1$
 Institut f\"ur  Physik, Universit\"at Oldenburg\\ Postfach 2503,
D-26111 Oldenburg, Germany\\
$^2$
Dept.~de F\'{\i}sica At\'omica, Molecular y Nuclear, Ciencias F\'{\i}sicas\\
Universidad Complutense de Madrid, E-28040 Madrid, Spain\\
$^3$
Departamento de F\'isica da Universidade de Aveiro and CIDMA,\\
 Campus de Santiago, 3810-183 Aveiro, Portugal
}
\vspace{0.5truecm}

\vspace{0.5truecm}

\vspace{0.5truecm}
\date{
\today}

\maketitle

  \medskip

\begin{abstract}
We consider rotating black hole solutions in five-dimensional
Einstein-Maxwell-Chern-Simons theory with a negative cosmological constant 
and a generic value of the Chern-Simons coupling constant $\lambda$.
Using both analytical and numerical techniques,
we focus on cohomogeneity-1 configurations, with two equal-magnitude angular momenta,
which approach at infinity a globally AdS background.
We find that the generic solutions share a number of basic properties with the
 known  Cveti\v c, L\"u and Pope black holes which have $\lambda=1$.
New features occur as well;  for example,
when the Chern-Simons coupling constant exceeds a critical value, 
the solutions are no longer uniquely determined by their global charges.
Moreover,  the black holes possess radial excitations 
which can be
labelled by the node number of the magnetic gauge potential function.
Solutions with small values of $\lambda$ possess other distinct features.
For instance, the extremal black holes there form two disconnected branches,
while not all near-horizon solutions
are associated with global solutions.

\end{abstract}

\section{Introduction}

The study of black hole (BH) solutions with  a cosmological constant $\Lambda<0$ has
enjoyed recently a tremendous amount of interest.
The natural ground state here is the  Anti-de Sitter (AdS) spacetime,
which, from a mathematical viewpoint, can be regarded as fundamental as the Minkowski one, 
possessing the same number of Killing vectors \cite{Hawking:1973uf}.
Therefore, finding less symmetric solutions 
and contrasting the situation with the Minkowskian counterparts
is an interesting problem in itself 
which ultimately may lead to a better understanding of the real world BHs.
However, the main motivation for the study of BHs with AdS asymptotics 
comes from the proposed correspondence between physical effects
associated with gravitating fields propagating in AdS spacetime and those of a conformal
field theory (CFT) on the boundary of AdS spacetime 
\cite{Witten:1998qj,Maldacena:1997re}.
According to this conjecture,  the AdS$_D$ BHs would offer  the possibility  
of understanding the nonperturbative structure of some CFTs in $(D-1)-$dimensions.

Restricting to a globally AdS background and $D=5$ spacetime dimensions,
one remarks that the asymptotically flat Myers-Perry BH \cite{Myers:1986un}
possesses  a generalization which
has been studied by various authors,
starting with the work by Hawking ${\it et. al.}$ \cite{Hawking:1998kw}.
However, the situation is more patchy in the presence of an (Abelian) gauge field.
The only Einstein-Maxwell solution which is known in closed form is the
(electrically charged, spherically symmetric) Reissner-Nordstr\"om-AdS (RN-AdS) BH.
The basic properties of its rotating generalization  has been 
studied in \cite{Kunz:2007jq} by using numerical methods,
for the particular case of two equal angular momenta $|J_1|=|J_2|$,
while the general solutions with $J_1\neq J_2$
are still unknown. 
 
\medskip

However, in five spacetime dimensions the Maxwell
action may be supplemented by a Chern-Simons
(CS) term. 
This makes no difference to 
static configurations but it does affect the class of stationary solutions.
Rather than being merely an extension of the Einstein-Maxwell model,
the inclusion of a CS term is motivated by its presence 
(with  a particular coefficient $\lambda=\lambda_{SG}=1$) 
in the bosonic sector of $D = 5$ minimal gauged supergravity. 
Several exact solutions of this supergravity model describing charged rotating  BH solutions
have been reported in the literature. 
Of main interest here are the BH solutions found in \cite{Cvetic:2004hs} 
by Cveti\v c, L\"u and Pope (CLP).
These are the most general asymptotically AdS BHs 
which rotate in two planes with equal-magnitude angular momenta
and are free of pathologies.
As such, they possess three 
global charges: the mass $M$, the electric charge $Q$ and the angular momenta $|J_1|=|J_2|=J$.
The  parameters $M,$ $Q$, $J$ are subject to some constraints, such that
closed timelike curves 
and naked singularities are avoided.
The BHs in  \cite{Cvetic:2004hs} 
 also possess an extremal limit
which preserves some amount of supersymmetry 
\cite{Gutowski:2004ez}
(note that these solutions present non-vanishing angular momentum).
Generalizations of these configurations with  more matter fields 
and/or 
unequal angular momenta 
have been constructed in
\cite{Chong:2005hr}. 

\medskip
 
The main purpose of this paper
is to answer the question on
{\it  how general is 
the CLP solution?}
For example, when
taking a value  $\lambda \neq  \lambda_{SG}$  of the CS coupling constant
and imposing the following assumptions: 
$i)$ AdS$_5$ asymptotics;
$ii)$ two equal angular momenta; and
$iii)$ the absence of pathologies,
{\it do we recover the same qualitative features as in} \cite{Cvetic:2004hs}? 

\medskip

Although a CS term does not contribute to the Einstein equations,
its presence breaks the charge reversal invariance.
Moreover,
a value $\lambda \neq 0$ introduces a nonlinearity 
at the level of the Maxwell--Chern-Simons equations;
thus varying $\lambda$ may lead to new features of the solutions.
Indeed, as discussed in 
\cite{Blazquez-Salcedo:2013muz},
\cite{Blazquez-Salcedo:2015kja},\cite{Blazquez-Salcedo:2016hae},
the  asymptotically flat limit of these BHs possesses 
a variety of new properties
when the Chern-Simons coupling constant is large enough
[starting above the supergravity (SUGRA) value]. 
Perhaps the most unusual feature there is that the
BHs form sequences of radially excited solutions, 
that can be labeled by the node number of the magnetic gauge potential function.
Moreover, the solutions there exhibit non-uniqueness
and one finds extremal and non-extremal
BHs with the same sets of global charges
and different bulk geometries.

It is likely that some of the new features in 
\cite{Blazquez-Salcedo:2013muz},
\cite{Blazquez-Salcedo:2015kja},\cite{Blazquez-Salcedo:2016hae}
will survive in the presence of a negative cosmological constant.
Thus one can predict that the CLP solution  
(which possesses a nodeless magnetic gauge potential function)
will fail to provide an accurate qualitative description
for large enough $\lambda$.
However, the situation is less clear 
for small enough values of the CS coupling constant. 

To answer the  questions above, 
we consider the same framework as in the CLP case
(in particular the same boundary conditions at infinity)
and study Einstein-Maxwell-Chern-Simons (EMCS) BHs with $\lambda \neq \lambda_{SG}$,
a task which, to our knowledge, has not been yet undertaken in the literature. 
Our results show that new qualitative features occur both for small and
for large enough values of  the CS coupling constant.
Then one cannot safely extrapolate the features of the  CLP  solutions to the case of
a generic $\lambda$.
In particular, all unusual features found in  
\cite{Blazquez-Salcedo:2013muz},
\cite{Blazquez-Salcedo:2015kja},\cite{Blazquez-Salcedo:2016hae}
for asymptotically flat BHs in EMCS theory
survive in the presence of a negative cosmological constant.
Moreover, new features occur as well for a small $\lambda$. 

\medskip

The paper is organized as follows:
in Section 2 we introduce the general framework. 
The squashed $AdS_2\times S^3$ solutions of the EMCS model are discussed in Section 3 in conjunction with
 the near-horizon formalism.
Such configurations are of interest since 
in principle  they could emerge as near-horizon limit of the extremal global solutions.
The BH solutions are discussed in Sections 4 and 5.
Several values of $\lambda$ are considered there, the situation being
contrasted with the $\lambda=1$ CLP case. 
We end with a brief conclusion
and outlook in Section 6.
The Appendix A contains a discussion of the basic properties of the exact CLP solution.

\section{Framework}

The action for $D=5$ Einstein-Maxwell theory with negative cosmological constant 
$\Lambda=-{6}/{L^2}$
and a Chern-Simons (CS) term  is given by 
\begin{equation} 
\label{EMCSac}
I= -\frac{1}{16\pi G_5} \int_{{\cal M}} d^5x\biggl[ 
\sqrt{-g}(R +\frac{12}{L^2}
-F_{\mu \nu} F^{\mu \nu}  
)
+
\frac{2\lambda}{3\sqrt{3}}\varepsilon^{\mu\nu\alpha\beta\gamma}A_{\mu}F_{\nu\alpha}F_{\beta\gamma} 
 \biggr ],
\end{equation}
where $R$ is the curvature scalar  
and
$G_5$ is Newton's constant in five dimensions;
in the following,
to simplify the relations,
we consider units such that $G_5=1$.  
Also,
$A_\mu $ is the gauge potential with the field strength tensor 
$ F_{\mu \nu} = \partial_\mu A_\nu -\partial_\nu A_\mu $
and
${ \lambda}$ is the CS coupling constant. 
For the value $\lambda = \lambda_{\rm SG} \equiv 1$, 
the action (\ref{EMCSac}) coincides with the bosonic part of $D = 5$ minimal
gauged supergravity.  
 
The field equations of this model consist of the Einstein equations
\begin{equation}
\label{Einstein_equation}
G_{\mu\nu} + \Lambda g_{\mu\nu} =2\left(F_{\mu\rho}{F^{\rho}}_{\nu}-\frac{1}{4}F_{\rho \sigma}F^{\rho \sigma} \right),
\end{equation}
together with the Maxwell--Chern-Simons equations
\begin{equation}
\label{Maxwell_equation}
\nabla_{\nu} F^{\mu\nu} + \frac{\lambda}{2\sqrt{3}}\varepsilon^{\mu\nu\alpha\beta\gamma}F_{\nu\alpha}F_{\beta\gamma}=0.
\end{equation}

The general EMCS rotating BHs would possess two independent
angular momenta and a topology
of the event horizon which is not necessarily spherical\footnote{Vacuum 
BHs with an $S^2\times S^1$ event horizon topology  in a global AdS$_5$
have been constructed recently in \cite{Figueras:2014dta}.
One expects these solutions to possess generalizations in EMCS theory.}. 
Thus a generic Ansatz would contain metric functions and gauge potentials 
with a nontrivial dependence on more than one coordinate. 
However, this is a very hard numerical problem which
we have not yet solved.

The problem is greatly simplified 
by assuming that the solutions have two  equal-magnitude angular momenta  
and an event horizon with spherical topology.
This factorizes the angular dependence of the
problem, leading to a cohomogeneity-1 Ansatz, the resulting equations of motion
forming a set of coupled nonlinear ODEs in terms of the radial coordinate only\footnote{Note that 
a similar approach has been used 
by various authors
 to numerically construct $D=5$ spinning BHs,
for various models
where an exact solution is missing, see $e.g.$ 
\cite{Brihaye:2010wx},
\cite{Dias:2011at},
\cite{Stotyn:2011ns},
\cite{Stotyn:2013yka},
\cite{Kunz:2006eh},
\cite{Kunz:2006yp},
\cite{Brihaye:2014nba},
\cite{Brihaye:2016vkv}
(perturbative $exact$ solutions have been constructed as well within the same approach,
see 
$e.g.$ \cite{Aliev:2006yk}, \cite{Aliev:2008bh}
\cite{Allahverdizadeh:2010xx}, \cite{Allahverdizadeh:2010fn}).
}.
For such solutions the isometry group of the line element 
is enhanced from $R_t \times U(1)^{2}$
to $R_t \times U(2)$, where $R_t$ denotes the time translation.

An Ansatz with these symmetries is built  in terms of
the left-invariant
1-forms $\sigma_i$ on $S^3$, with a line element
\begin{eqnarray}
\label{metrici}
&&ds^2 = F_1(r)dr^2
  + \frac{1}{4}F_2(r)(\sigma_1^2+\sigma_2^2)+\frac{1}{4}F_3(r) \big(\sigma_3-2W(r) dt \big)^2
-F_0(r) dt^2,
\end{eqnarray}
and a gauge field
\begin{eqnarray}
\label{gauge1}
A=a_0(r)dt +a_\varphi(r) \frac{1}{2} \sigma_3,
\end{eqnarray}
where
$\sigma_1= \cos \psi d\bar \theta+\sin\psi \sin  \bar \theta d \phi$,
$\sigma_2=-\sin \psi d\bar \theta+\cos\psi \sin  \bar \theta d \phi$,
$\sigma_3=d\psi  + \cos \bar \theta d \phi$,
and $\bar \theta,\phi$ and $\psi$ are the Euler angles on $S^3$.

\section{Squashed $AdS_2\times S^3$ solutions and the attractor mechanism}
%
Some analytical expressions together with a partial
understanding of the properties 
of $extremal$ EMCS BHs can be achieved by
considering  solutions with a squashed $AdS_2\times S^3$
geometry.
In this case, the $r$-dependence of the problem 
factorizes such that
the field equations reduce to a set of algebraic relations.

Such solutions are  found by taking 
a particular expression of the general Ansatz  
(\ref{metrici}) and (\ref{gauge1}) 
with
\begin{eqnarray}
\label{sq1}
F_1=\frac{v_1}{r^2},~~F_0= {v_1}{r^2},~~F_2=v_2,~~
F_3=v_2\eta,~~W=\alpha r,~~
a_0=-q r,~~a_\varphi=p~.
 \end{eqnarray}
The resulting geometry describes a fibration of $AdS_2$ over the homogeneously squashed $S^3$ with symmetry
group $SO(2, 1)\times SU(2)\times U(1)$ \cite{Kunduri:2007qy}.
In the above relations, 
$v_1$, $v_2$, $\eta$, $\alpha$, $q$ and $p$
are six constants subject to four constraints
which result from the EMCS equations (\ref{Einstein_equation}), (\ref{Maxwell_equation}):
\begin{eqnarray}
\label{nh1}
&&
v_1=\frac{L^2v_2}{4(L^2+3v_2)},~~
\eta=\frac{L^4(2p^2-v_2)}{v_2(L^2+3v_2)^2-L^4},
\\
\nonumber
&&
\label{nh2}
q=\frac{L}{4\sqrt{3}}
\left(
2-\frac{L^2(L^2+6p^2)}{(L^2+3v_2)^2}-\frac{L^2}{L^2+3v_2}
-\frac{18 \alpha^2L^2(2p^2-v_2)}{2\alpha^2(L^2+3v_2)^2-L^4)}
\right)^{1/2},
\end{eqnarray}
together with
\begin{eqnarray}
\label{nh3}
p=\frac{\alpha \sqrt{\eta}q v_2^2}{4v_1(\sqrt{\eta}v_1+\frac{2}{\sqrt{3}}\lambda q \sqrt{v_2})}~.
\end{eqnarray}

A simple solution of the above equations exists in two limiting cases. 
First, in the absence of a gauge potential ($p=q=0$) one finds
\begin{eqnarray}
\label{MP-nh}
\eta=2(1+\frac{v_2}{L^2}),~~
v_1=\frac{v_2}{4(1+\frac{3v_2}{L^2})},~~
\alpha=\frac{1}{2(1+\frac{3v_2}{L^2})}\sqrt{\frac{1+\frac{2v_2}{L^2}}{1+\frac{ v_2}{L^2}}}~.
\end{eqnarray}
This solution describes the near-horizon geometry of the extremal Myers-Perry-AdS (MP-AdS) BHs.
 
For $q\neq 0$,
a simple analytic solution can be written for $\lambda=0$  only
($i.e.$ pure EM theory), with
\begin{eqnarray}
&&
v_1=\frac{L^2v_2}{4(L^2+3v_2)},~~
\eta=\frac{4(6v_2+4L^2)v_1^2}{L^2(4v_1^2+\alpha^2v_2^2))}
-\frac{16(L^2+3v_2)^2}{L^4 v_2}q^2, \nonumber
\\
&&
q=\frac{\sqrt{3v_2}}{4(L^2+3v_2)}
\frac{\sqrt{4\alpha^2(3v_2^2(5L^2+v_2)+L^4(L^2+7v_2))-L^4(L^2+2v_2)}}
{16\alpha^4(L^2+3v_2^2)^4-L^8}~.
\label{nh_EMAdS}
\end{eqnarray}
For $\lambda\neq 0$,
writing a solution similar to 
(\ref{nh_EMAdS})
reduces to solving a 6th-order algebraic equation,
a task which is approached numerically.
Thus
 we are left with a two-parameter family of solutions,
which we found convenient to parametrize in terms of 
 $v_2$ and $\eta$.
These constants
 measure the radius of the round $S^2$ and the squashing of the 
$S^3$ part of the metric, respectively.

\subsection{Charges}
The solutions of the algebraic system
 (\ref{nh1})-(\ref{nh3})
are expected to describe the near-horizon
limit of  asymptotically AdS$_5$ extremal BHs in EMCS theory.
Thus they are
of particular interest in conjunction with the 
 {\it entropy function formalism}
\cite{Sen:2005wa,Astefanesei:2006dd,Goldstein:2007km}.
For example, this formalism allows us to find the expression  
for some quantities of interest for the $global$ extremal solutions
without integrating the field equations in the bulk;
it also leads to some predictions for the structure of those BHs. 

The analysis is standard and 
a detailed  computation has been given in 
\cite{Blazquez-Salcedo:2015kja},
for the same framework, although in the absence of a cosmological constant.
Thus we shall present here the basic steps only.
Following  the usual approach,
we consider the action functional of the model
(with ${\cal L}$ the EMCS Lagrangian)
\begin{eqnarray}
\label{at2}
\nonumber
h(\alpha,v_1,v_2,\eta,p,q)
=\int d \bar \theta d \varphi  d\psi \sqrt{-g} {\cal L}
=
4\pi^2 
\left (
\sqrt{\eta v_2} v_1
(
4-\eta+\frac{6v_2}{L^2}+\frac{\alpha^2\eta v_2^2}{4v_1^2}
)
+v_1\sqrt{\frac{\eta}{v_2}}(\frac{v_2^2q^2}{v_1^2}-4p^2)-\frac{16\lambda p^2q}{2\sqrt{3}}
\right)~.
\end{eqnarray}
 The entropy function is the Legendre transform of the above integral
with respect to the parameters $\alpha$, $ q$ 
\begin{equation}
S=2\pi (2 \alpha \tilde{J} + \rho \tilde{q}-h),
\label{entropy}
\end{equation} 
where $\tilde{J}$ and $\tilde{q}$
are related to the angular momentum and the electric charge of the solutions, 
respectively.

Within this approach, the Einstein equations   
 correspond to
\begin{eqnarray}
\label{derv}
 \frac{\partial h}{\partial v_1}=0,~~
 \frac{\partial h}{\partial v_2}=0,~~
\frac{\partial h}{\partial \eta}=0.
\end{eqnarray}
Note that in the presence of the CS term in the action,
the corresponding analysis for the gauge potentials is more intricate \cite{Suryanarayana:2007rk},
and recovering the relations for $(p,q)$ in (\ref{nh1})-(\ref{nh3}) requires some care. 

Within the entropy function formalism,
the considered configurations
are characterized by two independent parameters,
for which we would like to choose the 
angular momentum and electric charge. 
In order to calculate them, 
one can employ the Noether charges approach
\cite{Suryanarayana:2007rk,Wald:1993nt,Lee:1990nz,Rogatko:2007pv}.  
Then the expression  of the total angular momentum is \cite{Blazquez-Salcedo:2015kja}
\begin{eqnarray}
\label{j1}
J 
=4\pi\frac{v_2^{3/2}}{v_1}
\sqrt \eta p(\rho+p\alpha) + \pi\frac{v_2^{5/2}}{v_1}\eta^{3/2}\alpha 
- \frac{16}{9}\sqrt 3 \pi p^3\lambda,
\end{eqnarray}
a result  which is also obtained from the equation
\begin{eqnarray}
\frac{\partial h}{\partial \alpha}=\tilde{J}\equiv 16\pi J.
\end{eqnarray} 
The corresponding expression of the electric charge is \cite{Blazquez-Salcedo:2015kja}
\begin{eqnarray}
\label{qe1}
Q 
= -4\pi\frac{v_2^{3/2}}{v_1}\sqrt \eta (\rho+p\alpha) 
+ \frac{8\pi\sqrt 3}{3}\lambda p^2~,
\end{eqnarray}
which is equivalent to
\begin{eqnarray}
\label{ps1}
\frac{\partial h}{\partial \rho}
=\hat q = -Q - \frac{8\pi\sqrt 3}{9}\lambda p^2.
\end{eqnarray} 
Then the solutions have an entropy
\begin{eqnarray}
S=\frac{A_H}{4}~,
\end{eqnarray}
where
\begin{eqnarray}
A_H = \frac{1}{4}\int d {\bar \theta} d \phi d\psi \sqrt{|det(g^{(3)})|} 
= 16\pi^2v_2^{3/2}\sqrt \eta ~,
\end{eqnarray}
will be identified with the  horizon area of the  global extremal solutions.
For the purposes of this work
it is also of interest
 to consider the horizon angular momentum $J_H$,
which can be calculated 
using the standard Komar formula
(with the Killing vector $\tau\equiv \partial_\psi - \partial_\phi$)
\begin{eqnarray}
J_H = \frac{1}{64 \pi}\int d {\bar \theta} d \phi d\psi 
\sqrt{-g}(\nabla^{r}\tau^{t} - \nabla^{t}\tau^{r}) 
= \pi\frac{v_2^{5/2}}{v_1}\eta^{3/2}\alpha ~.
\end{eqnarray}
Note  that although the cosmological constant is not explicitly found 
in the above expressions, 
it enters via the relations  (\ref{nh1})-(\ref{nh3}).

\subsection{Branch structure 
and predictions for the global solutions}

If we fix the CS constant $\lambda$ and the AdS length scale $L$, 
the near-horizon solutions only depend on the angular momentum $J$ 
and the electric charge $Q$. 
Unfortunately, the relations
(\ref{j1}), 
(\ref{qe1}), 
together with 
(\ref{nh1}),
(\ref{nh3})
are very complicated and thus
 it is not possible 
to give an explicit expression for the entropy
$S$  
as a function of $Q$ and $J$
(the same holds for the horizon angular momentum $J_H$). 

However, it is 
 a straightforward problem to compute those expressions numerically.
This reveals a rather complicated picture,
with the possible existence of several branches of solutions.  

\subsubsection{$\lambda=0$ case}

Let us start by addressing first the $\lambda=0$ limit of the solutions,
[$i.e.$ Einstein-Maxwell (EM)-AdS theory], in which case partial
analytic results are at hand 
(see the Eqs. (\ref{nh_EMAdS})).
In agreement with our physical intuition,
one finds two different branches of solutions, 
corresponding to two different possible ways to generate extremal EM-AdS global solutions.
One of the branches contains the near-horizon geometry of the extremal RN-AdS BH as the $J=0$ limit.
Then a whole branch is generated when adding spin to that limiting solution.
Therefore, we call this set  the  {\it RN branch}. 
The second branch contains the near-horizon geometry of the extremal MP-AdS BH as the uncharged limit. 
In this case, the whole branch is generated when electric charge is introduced into that
near-horizon geometry. 
As such, this set is called the {\it  MP branch}. 

These two branches are shown in Figure 1, with a $(J_H,A_H)$-diagram (left) and a $(J,J_H)$-diagram (right)
(note that the electric charge is fixed there;
also all results in this section are found for an AdS length scale $L=1$).
Interestingly, these two branches never intersect, $i.e.$ there are no attractor solutions connecting them. 
This leads us to predict  that the
global solutions will also be on disconnected branches.
Thus,
for a fixed value of the electric charge, we shall find two branches of extremal BHs,
which however, cannot be connected by global solutions whose near-horizon geometry is described by a
squashed $AdS_2\times S^3$ metric. 

\begin{figure}
    \centering
		
    \begin{subfigure}[b]{0.45\textwidth}
        \includegraphics[width=55mm,scale=0.5,angle=-90]{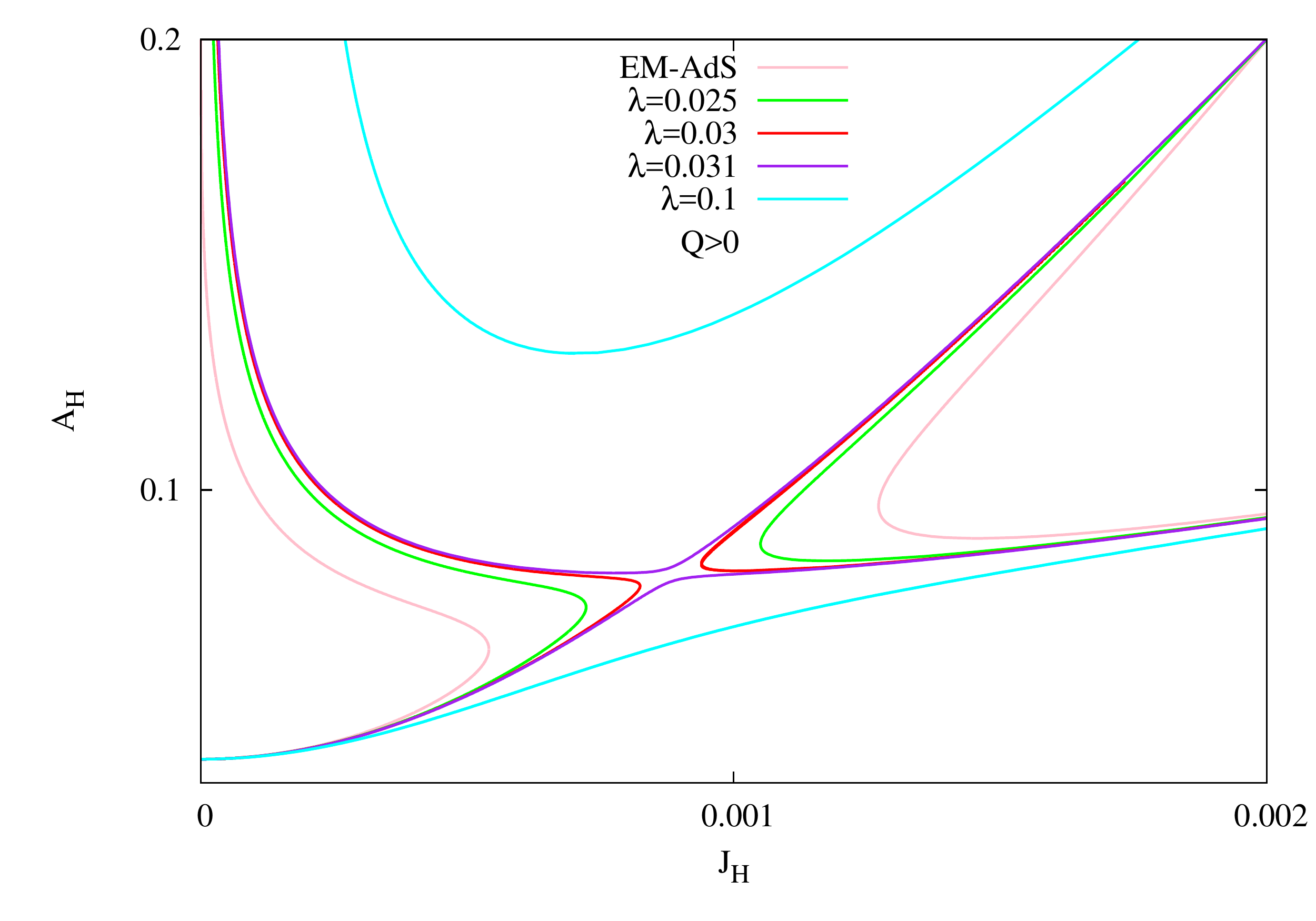}
        \caption{$Q>0$}
        \label{fig:Jh_small_lambda_p_NH}
    \end{subfigure}
    \begin{subfigure}[b]{0.45\textwidth}
        \includegraphics[width=55mm,scale=0.5,angle=-90]{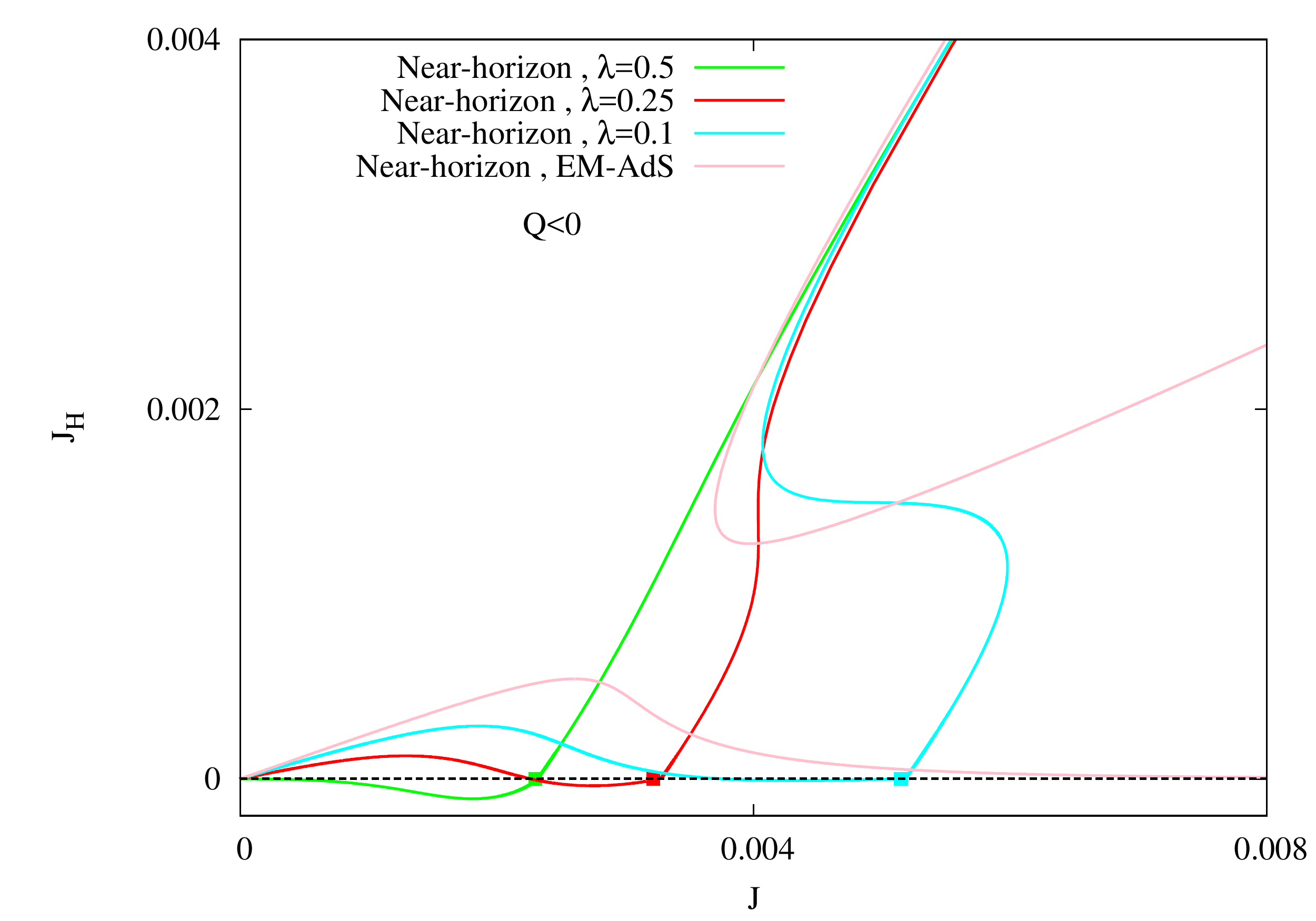}
        \caption{$Q<0$}
        \label{fig:Jh_small_lambda_n_NH}
    \end{subfigure}
		
    \caption{(a) Area $vs.$ horizon angular momentum $J_H$ for near-horizon solutions with fixed positive electric charge 
		$Q=0.044$
	and several values of $\lambda$. 
	(b) Horizon angular momentum $J_H$ $vs.$ angular momentum $J$ 
	for near-horizon solutions with negative electric charge $Q=-0.044$. The squares mark the \textit{critical} solutions.
	In both figures we include pure EM near-horizon solutions for reference.
	Also, the AdS length scale	is $L=1$. }
		
		\label{fig:Jh_small_lambda_NH}
		
\end{figure}

\subsubsection{Generic picture}

The picture for $\lambda \neq 0$ ($i.e.$ an EMCS-AdS theory)
is more complicated.
First, in this case, if we fix the CS coupling $\lambda$,
 the sign of the electric charge becomes relevant. 
In the following we will assume $\lambda>0$ without any loss of generality
and show results for several small values of $\lambda$.
For the discussion we shall fix $|Q|=0.044$ and the AdS length to $L=1$, 
although we have explored other values of these parameters and the features described below
are generic.

Let us start with the positive $Q$ case. 
Then, for the $Q=0.044$ and
 $0\leq\lambda<0.0305$, the branch structure is similar to that of the pure EM-AdS case
and one finds again two separated branches of near-horizon solutions
(see  Figure \ref{fig:Jh_small_lambda_p_NH}).
One of them, the RN branch, contains the near-horizon limit of the extremal RN-AdS BH.
The other one, the MP branch, contains the near-horizon geometry of the extremal MP-AdS BH. 
Moreover, both branches are disconnected. 
Then the prediction above for the global solutions still holds for these EMCS solutions. 

However, this structure changes drastically for $\lambda \ge 0.0305$, where a bifurcation happens.
For this specific value,
both branches connect at some particular configuration
which does not seem to possess special properties. 
When $\lambda > 0.0305$ one finds again two disconnected branches after the bifurcation. 
These features are shown in Figure \ref{fig:Jh_small_lambda_p_NH}, for $\lambda=0.031$ (purple curve) and $\lambda=0.1$ (cyan curve).  

Now let us continue with the negative $Q$ case, which presents very different properties. 
There it is more convenient to show the branch structure by plotting the horizon angular momentum 
$J_H$
$vs.$ the total angular momentum $J$, as we do in Figure  \ref{fig:Jh_small_lambda_n_NH}.
The main feature always present in $Q<0$ solutions is the appearance of a 
\textit{critical}
near-horizon solution with vanishing
horizon area (marked with a square in the Figure). 
This near-horizon solution is always separating the near-horizon geometry of the extremal MP-AdS  BH  from the near-horizon geometry of the extremal RN-AdS BH. 
Hence a prediction of the near-horizon formalism is that in this case we always find 
(at least) two branches.
For fixed $Q$, this \textit{critical} solution has a certain value of angular momentum $J_c$, 
which increases as $\lambda$ goes to zero (in fact, one finds $J_c \to \infty$ as $\lambda \to 0$). 
The horizon angular momentum of the \textit{critical} solutions also vanishes. 

For small enough values of $\lambda$, we observe also other interesting features. 
For instance, consider $0<\lambda<0.25$. 
Then the MP branch can have more than one solution with the same angular momentum. 
In Figure  \ref{fig:Jh_small_lambda_n_NH},
 this can be seen for $\lambda=0.1$ (cyan curve), 
where a vertical line of constant $J$ can intersect up to three times with the curve.
 However, as $\lambda$ increases this behavior is lost, see the  $\lambda=0.25$ curves in Figure \ref{fig:Jh_small_lambda_n_NH}. 
Yet another interesting behavior can be observed on the RN branch. 
In Figure  \ref{fig:Jh_small_lambda_n_NH} one can see  that the near-horizon formalism 
predicts the existence of counter-rotating solutions. 
For instance, consider the curves with $\lambda=0.1$ (cyan) and $\lambda=0.25$ (red). 
It can be seen that for large enough $J$, 
and for solutions satisfying $J<J_c$, 
the horizon angular momentum goes from positive to negative. 
In fact for $\lambda>0.5$, the RN branch always has negative horizon angular momentum, $J_H<0$.

We have studied as well solutions
with $0.5<\lambda<2$  and
it turns out the (qualitative) picture described above for $\lambda=0.5$
holds in that case,
in particular the existence of a $critical$ configuration with $A_H=0$.
A discussion of this property for the $\lambda=\lambda_{SG}$
case is given in Appendix A, based on the analytical CLP solution.

However, new features occur for 
 $\lambda>2$,
in which case
 the branch structure 
becomes qualitatively similar to that of the asymptotically flat case \cite{Blazquez-Salcedo:2015kja}. 
Essentially, for $Q<0$ 
the near-horizon geometry of the extremal RN-AdS BH is no longer the only solution with $J=0$, 
but there is a $J=0$ near-horizon solution, 
which is not static. 
We will comment more on that later, when discussing the global aspects of these solutions. 
The discussion of the branch structure for $\lambda>2$ can be found in \cite{Blazquez-Salcedo:2015kja} (flat case), 
and, since it is recovered here, we refer the reader to that paper.

\section{Black holes in EMCS-AdS theory. General properties}

\subsection{Parametrization and equations of motion}

To make contact with the previous numerical
work on the $D=5$ EM(CS) system
\cite{Kunz:2007jq},
 \cite{Blazquez-Salcedo:2015kja}
we introduce the new angular coordinates
\begin{eqnarray}
 \bar \theta= 2\theta,~~
\phi = \varphi_2-\varphi_1,~~
\psi= \varphi_1+\varphi_2,
\end{eqnarray}
where $\theta \in [0,\pi/2]$, $\varphi_1 \in [0,2\pi]$ 
and $\varphi_2 \in [0,2\pi]$.  
Also, we fix the metric gauge and reparametrize
the functions in (\ref{metrici}) by taking
\begin{eqnarray}
\nonumber
 F_0(r)=f(r)N(r),~~ 
 F_1(r)=\frac{m(r)}{f(r)}\frac{1}{N(r)},~~
 F_2(r)=\frac{m(r)}{f(r)}r^2 ,~~ 
 F_3(r)=\frac{n(r)}{f(r)}r^2 ,~~
W(r)=\frac{\omega(r)}{r},
\end{eqnarray}
where
\begin{eqnarray}
\nonumber
N(r)= 1+\frac{r^2}{L^2} , 
\end{eqnarray}
is a 'background' function employed to enforce AdS asymptotics 
(note that the AdS$_5$ spacetime is recovered for
$f=m=n=1$, $\omega=0$).
For completeness, we give the corresponding expression 
of the line element
\begin{eqnarray}
\label{metric}
&&ds^2 = -f(r)N(r)dt^2 + \frac{m(r) }{f(r)}\left( \frac{dr^2}{N(r)} + r^2d\theta^2 \right)  
+ \frac{n(r) }{f(r)}  r^2  \sin^2\theta \left(  d \varphi_1 -\frac{\omega(r)}{r}dt
\right)^2 \nonumber \\  
&& 
+  \frac{n(r) }{f(r)} r^2 \cos^2\theta \left(  d \varphi_2
  -\frac{\omega(r)}{r}dt \right)^2 
+ \left( \frac{m(r)-n(r) }{f(r)} \right)  r^2 \sin^2\theta \cos^2\theta
   \left( d \varphi_1  - d \varphi_2 \right)^2,
\end{eqnarray}
while the corresponding expression of the gauge potential is
\begin{equation}
A_\mu dx^\mu  = a_0(r) dt 
+ a_{\varphi}(r) (\sin^2 \theta   d\varphi_1
  +\cos^2 \theta  d\varphi_2).
\end{equation}

With this Ansatz, the Einstein equations reduce to a set of four second-order
ordinary differential equations
(ODEs) 
for the metric functions $f$, $m$, $n$ and $\omega$
\begin{eqnarray}
\nonumber
&&
f''-\frac{2f}{r^2}(1-\frac{4}{3N})+f'\left(\frac{5m'}{m}+\frac{2n'}{3n}-\frac{3f'}{2f}+\frac{N'}{N} +\frac{4}{r}\right)
+f\left(\frac{N'}{3N}(\frac{m'}{2m}+\frac{n'}{n})-\frac{m'}{3m}(\frac{n'}{n}+\frac{m'}{2m}+\frac{4}{r})\right)
\\
\nonumber
&&
-\frac{2}{3r}(\frac{n'}{n}+\frac{n}{rNm})
-\frac{7n(\omega-r\omega')^2}{6r^2 f N}
+\frac{2f}{L^2 N}\left(1-\frac{2m}{f}+\frac{L^2N'}{r} \right)
-\frac{2f^2}{r^2} \left(\frac{4 a_\varphi^2}{r^2 Nm}+\frac{5(ra_0'+wa_\varphi')^2}{3f^2N}+\frac{a_\varphi'^2}{3n} \right)
=0,
\\
\label{eqs-fin}
&&
m''+\frac{mf'}{rf}\left(1+\frac{rN'}{2N}+\frac{rn'}{3n} \right)
+\frac{m'}{3}\left(\frac{4}{r}+\frac{f'}{2f}+\frac{5N'}{2N}-\frac{5m'}{2m} \right) 
-\frac{m n'}{3r n}(\frac{1}{N}+\frac{r m'}{2m})
\\
\nonumber
&&
{~~~}
-\frac{4mn(\omega-r\omega')^2}{3r^2f^2 N}
+\frac{8m}{r^2 N}\left(\frac{r^2}{L^2}(1-\frac{m}{f})-\frac{1}{3}(1-\frac{n}{m}) \right)
-\frac{8m}{3r^2fN}\left( (ra_0+\omega a_\varphi')^2+\frac{Nf^2 a_\varphi'^2}{2n} \right)
=0,
\\
&&
\nonumber
n''+\frac{nf'}{rf}(1+\frac{mN'}{2N})
-\frac{m'n}{3m}\left(\frac{N'}{2N}+\frac{m'}{m}+\frac{5}{r} \right)
+\frac{n}{3} \left(\frac{2f'm'}{fm}-\frac{f'n'}{2fn}-\frac{m'n'}{2mn} \right)
+n'\left(\frac{8}{3r}+\frac{7N'}{6N}-\frac{n'}{2n} \right)
\\
&&
\nonumber
{~~~}
+\frac{8n}{r^2N}\left(\frac{2}{3}(1-\frac{n}{m})+\frac{r^2}{L^2}(1-\frac{m}{f}) \right)
-\frac{n^2(\omega-r\omega')^2)}{3r^2f^2N}
+\frac{8f}{r^2N}\left(\frac{1}{3}Na_\varphi'^2-\frac{2n}{r^2n}a_\varphi^2-\frac{n}{3f^2}(ra_0'+\omega a_\varphi')^2 \right)
=0,
\\
&&
\nonumber
w''+\frac{w}{r}\left(-\frac{3}{r}+\frac{5f'}{2f}-\frac{m'}{2m}-\frac{3n'}{2n} \right)
+w' \left(\frac{3}{r}-\frac{5f'}{2f}+\frac{m'}{2m}+\frac{3n'}{2n} \right)
-\frac{4fa\varphi'}{rn}(a_0'-\frac{w'a_\varphi}{r})
=0~,
\end{eqnarray}
together with the 1st-order constraint equation
\begin{eqnarray}
&&
\frac{m'n'}{mn}-\frac{f'n'}{fn}-\frac{f'm'}{fm}
-\frac{3f'N'}{4fN}
+\left(\frac{4m'}{rm}+\frac{2n'}{rn}-\frac{3f'}{rf}\right)(1+\frac{rN'}{4N})
+\frac{m'^2}{2m^2}
+\frac{n(\omega-r\omega')^2}{2r^2f^2N}
\\
&&
\nonumber
+\frac{2}{r^2N}\left(\frac{n}{m}-\frac{6r^2n}{L^2f} \right)
-\frac{2(L^2-6r^2)}{r^2L^2N}
+\frac{2}{r^2N}\left(\frac{(ra_0'+\omega a_\varphi')^2}{f}-\frac{fN a_\varphi'^2}{n}+\frac{4f a_\varphi^2}{r^2n} \right)
=0~.
\end{eqnarray}
The gauge potentials $a_0$, $a_\varphi$
satisfy the 2nd-order ODEs  
\begin{eqnarray}
&&
\label{eqa0}
a_0''+\frac{wa_\varphi'}{r^2}\left(1-\frac{rN'}{N}+\frac{rn'}{n}-\frac{2rf'}{f} \right)
+\frac{n\omega}{rN}\left(\frac{a_0'(r\omega'-\omega)}{f^2}+\frac{4a_\varphi}{r^2m}-\frac{\omega^2a_\varphi'}{rf^2} \right)
\\
&&
\nonumber
{~~~}
+a_0'\left(\frac{3}{r}-\frac{3f'}{2f}+\frac{m'}{2m}+\frac{n'}{2n} \right)
+\frac{a_\varphi'\omega'}{r}(1+\frac{n\omega^2}{f^2N} )
-\lambda \frac{8a_\varphi}{r^3\sqrt{3fmn}}\left(\frac{n\omega}{N}(r a_0'+\omega a_\varphi')-f^2 a_\varphi \right)=0,
\\
\nonumber
\label{eqaphi}
&&
a_\varphi''-\frac{4a_\varphi n}{r^2N m}
+a_\varphi'\left( \frac{1}{r}+ \frac{N'}{N}+ \frac{f'}{2f}+ \frac{m'}{2m}- \frac{n'}{2n}+ \frac{n\omega}{rf^2N}(r\omega'-\omega) \right)
\\
\nonumber
&&
{~~~}
+ \frac{na_0'}{f^2N}(\omega-r\omega')
+ \lambda \frac{8a_\varphi(ra_0'+\omega a_\varphi')}{r^2 N\sqrt{ \frac{3fm}{n}}}=0~.
\end{eqnarray}
The equations for $a_0$ and $w$ have a total derivative structure,
which implies the existence of the first integrals  
\begin{eqnarray}
&&
a_0'+\frac{\omega}{r}a_\varphi'- \frac{4\lambda}{\sqrt{3}}\frac{f^{3/2}a_{\varphi}^2}{r^3\sqrt{mn}}=
{\frac {2\,{f}^{3/2}}{\sqrt {m n} {r}^{3}\pi }}Q ,
\label{conQ}
\\
&&
\nonumber
 {\frac {8\,Q}{\pi }}\,a_{\varphi}
+\frac{16\lambda}{3\sqrt{3}}a_{\varphi}^3-{\frac {{n}^{3/2}\sqrt {m}r^3}{{f}^{5/2}}} (r \omega'-\omega)
=\frac{16}{\pi}J,
\end{eqnarray}
with the constants of integration 
$Q$, $J$, corresponding to the angular momentum and electric charge  of the solutions, respectively 
(see the relations  (\ref{Kang2}) and  (\ref{charge})). 

\subsection{Far field asymptotics and global charges}
%
 The far field expression of the solutions can be constructed
in a systematic way,  
the first terms of the expansion at infinity being
\begin{eqnarray}
\nonumber
f(r) &=& 1 +  \frac{\alpha}{r^4} - \frac{2}{21}\frac{9 L^2 \pi^2 \alpha-12L^2Q^2-2\pi^2 {\hat \mu}^2}{\pi^2 r^6} + \dots, 
\\
\nonumber
m(r) &=& 1 + \frac{\beta}{r^4} - \frac{1}{21}\frac{15 L^2 \pi^2 \alpha-20 L^2 Q^2-8\pi^2 {\hat \mu}^2}{\pi^2 r^6} +\dots, 
\\
\nonumber
n(r) &=&1 +  \frac{3(\alpha-\beta)}{r^4} - \frac{5}{21}\frac{3 L^2 \pi^2 \alpha-4 L^2 Q^2+4\pi^2 {\hat \mu}^2}{\pi^2 r^6} +\dots , 
\\
\label{far-field}
\omega(r) &=& \frac{\hat{J}}{r^3}-\frac{4q}{3}\frac{\hat \mu}{r^5}+(2\beta-\alpha)\frac{\hat{J}}{r^7}+\dots,
\\
\nonumber
a_{\varphi}(r) &=& \frac{\hat \mu}{r^2} + \frac{1}{9}\frac{2L^2\sqrt{3}Q\lambda\hat\mu - 3QL^2\hat{J}+3\pi\alpha\hat\mu-6\pi\beta\hat\mu}{\pi r^6} + \dots,
\\
\nonumber
a_{0}(r) &=& -\frac{Q}{\pi r^2} - \frac{1}{9}\frac{2\sqrt{3}\pi\lambda{\hat \mu}^2+3\hat\mu \hat{J}\pi + 3Q\beta}{\pi r^6} + \dots . 
\end{eqnarray}
The expression of the higher-order terms is rather complicated, 
and it has not been possible to identify
a general pattern for the coefficients.
Here we mention only that
this asymptotic expansion contains five undetermined parameters 
$\{\alpha, \beta, \hat J, \hat \mu, Q \}$,
which encode the global charges of the solutions.  

The total angular momenta of the BH can be calculated using the standard Komar integral
\begin{equation}
J_{(k)} =   \int_{S_{\infty}^{3}} \hat \beta_{(k)} 
\ , \label{Kang} \end{equation}
where $\hat \beta_{ (k) \mu_1 \mu_2 \mu_3} \equiv \epsilon_{\mu_1 \mu_2 \mu_3
  \rho \sigma} \nabla^\rho \eta_{(k)}^\sigma$ (with $\eta_{(k)}\equiv\partial/ \partial_{\varphi_{(k)}}$).
These configurations have equal-magnitude angular momenta, 
$|J_{(k)}|=|J|$, $k=1, 2$. 
Then one finds the following expression
\begin{equation}
J = \frac{\pi}{4}\hat{J} .
\label{Kang2}
\end{equation}
The computation of the electric charge is also standard, 
 $Q$ being obtained from
\begin{equation}
Q= - \frac{1}{2} \int_{S_{\infty}^{3}} \left( \tilde F 
+\frac{\lambda}{\sqrt{3}} A \wedge F \right)~,
  \label{charge}
\end{equation}
with 
${\tilde F}_{\mu_1 \mu_2 \mu_3} \equiv  
  \epsilon_{\mu_1 \mu_2 \mu_3 \rho \sigma} F^{\rho \sigma}$.

The computation of the total mass $M$, however,
requires special care, since the result from a naive application of the Komar integral diverges already 
in the vacuum case without the gauge field.
However, $M$ can be computed $e.g.$
by using the Ashtekar-Magnon-Das conformal mass
definition
\cite{Ashtekar:1984zz}, 
which results in
\begin{eqnarray}
\label{mass-ct}
M = -\frac{\pi}{8}\frac{(3\alpha+\beta)}{L^2}~.
\end{eqnarray}
 $M$ can also be computed by employing the boundary
counterterm approach in \cite{Balasubramanian:1999re}, 
wherein it is the conserved charge associated with Killing
symmetry $\partial/\partial t$ of the induced boundary metric, found for a large constant value of $r$.
This results in the same expression (\ref{mass-ct}),
plus a constant Casimir term $M_0=\frac{3\pi}{32}L^2$ \cite{Balasubramanian:1999re},
which we shall ignore in what follows.
We mention that $J$ can also be computed by using the approach
in \cite{Ashtekar:1984zz}  
or the one in \cite{Balasubramanian:1999re},
the results coinciding with
(\ref{Kang2}). 
Let us also note that 
$M$ and $J$ are evaluated relative to a frame which is $nonrotating$ at infinity.

The solutions possess also a
 magnetic moment $\mu_{\rm mag}$ which is 
fixed by the constant $\hat{\mu}$
which enters the 
asymptotic expansion of the gauge potential $a_{\varphi}$,
\begin{equation}
\mu_{\rm mag} = \pi \hat{\mu}  .
\end{equation}
Thus, one can define a
  gyromagnetic ratio $g$ 
\begin{equation}
{\mu_{\rm mag}}=g \frac{Q J}{2M}
\ . \label{gyro} 
\end{equation}

\subsection{Properties of the event horizon}

In the quasi-isotropic coordinates we are employing,
the BH horizon ${\cal H}$ resides at $r=r_H \geq 0$
(where the function $f$ vanishes),
and rotates with angular velocity $\Omega_H$. 
This is a Killing horizon, since the Killing vector 
\begin{eqnarray}
\nonumber
\zeta = \partial_t + \Omega_H ( \partial_{\varphi_1} +
 \partial_{\varphi_2})
\end{eqnarray}
becomes null  and  
orthogonal to the other Killing vectors on it,
$
(\zeta^2)|_{\cal H}=0,
$
$
(\zeta\cdot\partial_t)|_{\cal H}=0,
$
$
(\zeta\cdot\partial_{\varphi_{(k)}})|_{\cal H}=0.
$

For nonextremal solutions, the following expansion
holds near the event horizon:
\begin{eqnarray}
\nonumber 
f(r) &=& f_2 (r-r_H)^2 -f_2(\frac{1}{r_H}+\frac{3r_H}{L^2+r_H^2}) (r-r_H)^3+ O\left(r-r_H\right)^4,
\nonumber
\\
 m(r) &=&  m_2 (r-r_H)^2 -3m_2(\frac{1}{r_H}+\frac{r_H}{L^2+r_H^2}) (r-r_H)^3+ O\left(r-r_H\right)^4,
\\
\label{eh-expansion}
n(r)  &=&  n_2 (r-r_H)^2  -3n_2(\frac{1}{r_H}+\frac{r_H}{L^2+r_H^2}) (r-r_H)^3+ O\left(r-r_H\right)^4,
\\ 
\nonumber
\omega(r)  &=&  \omega_0 +\frac{\omega_0}{r_H}  (r-r_H)   + O\left(r-r_H\right)^2,
\\ 
\nonumber
a_{0}(r)  &=&  a_{0 }^{(0)} + a_{0 }^{(2)} (r-r_H)^2+ O\left(r-r_H\right)^3, 
\\ 
\nonumber
a_{\varphi}(r)  &=&   a_{\varphi }^{(0)} +   a_{\varphi }^{(2)} (r-r_H)^2+ O\left(r-r_H\right)^3, 
\end{eqnarray}
where $\{f_2,m_2,n_2,\omega_0;a_{0}^{(0)},a_{0, 2},a_{\varphi }^{(0)},a_{\varphi }^{(2)}\}$ are numerical coefficients
subject to the constraint 
\begin{eqnarray}
 &&
\big(
54r_H^4+\frac{71}{2}L^2r_H^2+\frac{29}{4}L^4
\big)f_2 m_2
+10L^2(L^2+2r_H^2)f_2 n_2
+12 r_H^2(L^2+2r_H^2)m_2^2
\\
\nonumber
&&
-L^2r_H^2 (5L^2+13 r_H^2)w_2\frac{m_2 n_2}{f_2}
-8 L^2 r_H^4(a_0^{(2)}+\Omega_H a_\varphi^{(2)})m_2
+\frac{4L^2(5L^2+9r_H^2)}{r_H^2}a_\varphi^{(0)2}f_2^2=0.
\end{eqnarray}
Note that for extremal BHs 
the event horizon is located at $r_H=0$,
in which case the near-horizon expansion is more complicated 
\cite{Blazquez-Salcedo:2015kja}.
Also, this results in a different expression of the horizon quantities
as compared to the one above. 

Restricting to the non-extremal case,
the area of the horizon $A_{H}$ and the Hawking temperature are given by
\begin{equation}
A_{H}=\int_{{\cal H}} \sqrt{|g^{(3)}|}=
2\pi^2r_H^3 \frac{m_2}{f_2}\sqrt{\frac{n_2}{f_2}},~~~~
T_H= \frac{1}{2\pi}\left(1+\frac{r_H^2}{L^2} \right)\frac{f_2}{\sqrt{m_2}}~.
\end{equation} 
The horizon angular velocity is obtained in terms of the inertial dragging
\begin{eqnarray}
\Omega_H = \frac{\omega_0}{r_H} .
\end{eqnarray}
Further, the horizon electrostatic potential $\Phi_{H}$ is defined by
\begin{equation}
\Phi_{H} = \left. \zeta^\mu A_\mu \right|_{r=r_{H}} 
\ = a_{0 }^{(0)}+\Omega_H a_{\varphi }^{(0)}, \label{Phi} 
\end{equation}
being constant at the horizon.

It is also of interest to compute the 
 horizon mass 
$M_{H}$ and the horizon angular momenta
$J_{{H} (k)}$,
which are given by the standard Komar integrals
(with  $\hat \alpha_{\mu_1 \mu_2 \mu_3} \equiv \epsilon_{\mu_1 \mu_2 \mu_3
  \rho \sigma} \nabla^\rho \xi^\sigma$):
\begin{eqnarray}
&&
M_{H}  =- \frac{3}{2} \int_{{\cal H}}\hat \alpha = 
\frac{3}{16} \pi r_H^3 \sqrt{\frac{m_2 n_2}{f_2^3}} 
\left (
 2f_2(1+\frac{r_H^2}{L^2})-\frac{2r_H n_2 \Omega_H w_2}{f_2}
\right),
\label{Hmass} 
\\ 
&&
J_{{H} (k)}=  \int_{{\cal H}} \hat \beta_{(k)} 
=
-\frac{1}{8}\pi \sqrt{\frac{m_2 n_2^3}{f_2^5}}r_H^4 w_2~.
\label{Hang} 
\end{eqnarray}
In the case we are interested here both horizon angular momenta have the same magnitude so we can refer to them simultaneously as the horizon angular momentum $J_H$, with $|J_H|=|J_{H (1)}|=|J_{H (2)}|$.

Note that the quantities above satisfy the horizon Smarr formula
\begin{equation}
\frac{2}{3}M_H = \frac{\kappa A_H}{8\pi} + 2\Omega_H J_H.
\end{equation}
However, different from the asymptotically flat case, no simple Smarr-type
relation can be written for asymptotically AdS configurations, 
in particular for the solutions in this work.
A proposed generalized Smarr-type relation could include the cosmological constant as
a negative pressure term \cite{Kastor:2009wy}. 
This possibility has been explored
for MP-AdS, RN-AdS and CPL BHs in \cite{Kastor:2009wy}, \cite{Cvetic:2010jb}.
 
Finally, we mention that the EMCS charged spinning BHs satisfy the 1st law
of thermodynamics
\begin{equation}
\label{1st}
dM=\frac{1}{4}T_H dA_H+2\Omega_H dJ+\Phi_{H} dQ.
\end{equation}
An extra term involving variations of the cosmological constant 
($\Theta d\Lambda \equiv -V dP$)
 can also be added to this formula \cite{Kastor:2009wy}, \cite{Cvetic:2010jb}. The
 conjugate variable to the pressure $P$ can be identified with 
 the volume $V$ of the space-time outside the event horizon. 
 However in our
calculations we will always consider families of configurations with 
a fixed value of the cosmological
constant.  
The extension of the results in \cite{Kastor:2009wy}, \cite{Cvetic:2010jb}
for numerical solutions (in particular for those in this work)
remains an interesting open problem.

\section{Black holes in EMCS-AdS theory. Numerical results}

\subsection{General remarks}

\subsubsection{Method}
Unfortunately, no exact EMCS-AdS closed-form solution is known apart from
the CLP one (with $\lambda=1$)
with its RN-AdS ($J=0$) and  MP-AdS (Q=0) limits.
The basic features of this special solution are discussed in Appendix A.

The BHs with $\lambda \neq 1$ are found numerically.
The numerical methods
we have used are similar to those used in the literature to find numerically other
$D=5$ BH solutions with equal-magnitude angular momenta,
 $e.g.$ those in 
\cite{Blazquez-Salcedo:2015kja},
\cite{Brihaye:2010wx},
\cite{Kunz:2006eh},
\cite{Kunz:2006yp}. 
In our scheme, we choose to solve a 
 system of four second-order  differential equations (ODEs) for the functions 
$(f,~m,~n,~a_\varphi)$
 (\ref{eqs-fin}), (\ref{eqa0}),
together with the two first-order ODEs for $\omega,a_0$, (\ref{conQ}).
Thus, in the generic case,
the input parameters are $\lambda, L$ together with $r_H,J,Q$.
The equations are solved by 
using the software package COLSYS 
\cite{COLSYS}, subject to appropriate boundary conditions
compatible with the asymptotics 
 (\ref{far-field}), (\ref{eh-expansion}). 
This solver employs a collocation method for boundary-value ordinary differential equations
and a  damped 
Newton method of quasi-linearization.
 A linearized problem 
is solved at each iteration step,
by using a spline collocation at Gaussian points.  
The package COLSYS possesses an
adaptive mesh selection procedure, such that the equations are solved on a sequence of
meshes until the successful stopping criterion is reached.

The solutions reported in this work  have a 
typical relative accuracy of $10^{-10}$.  
The number of mesh points used in our calculation
was  around $10^4$, distributed non-equidistant on 
$x$, where $x=1-r_H/r$ is a compactified radial coordinate
employed in the non-extremal case
(for extremal solutions we have used $x=r/(1+r)$).

\begin{figure}
    \centering
        \includegraphics[width=55mm,scale=0.5,angle=-90]{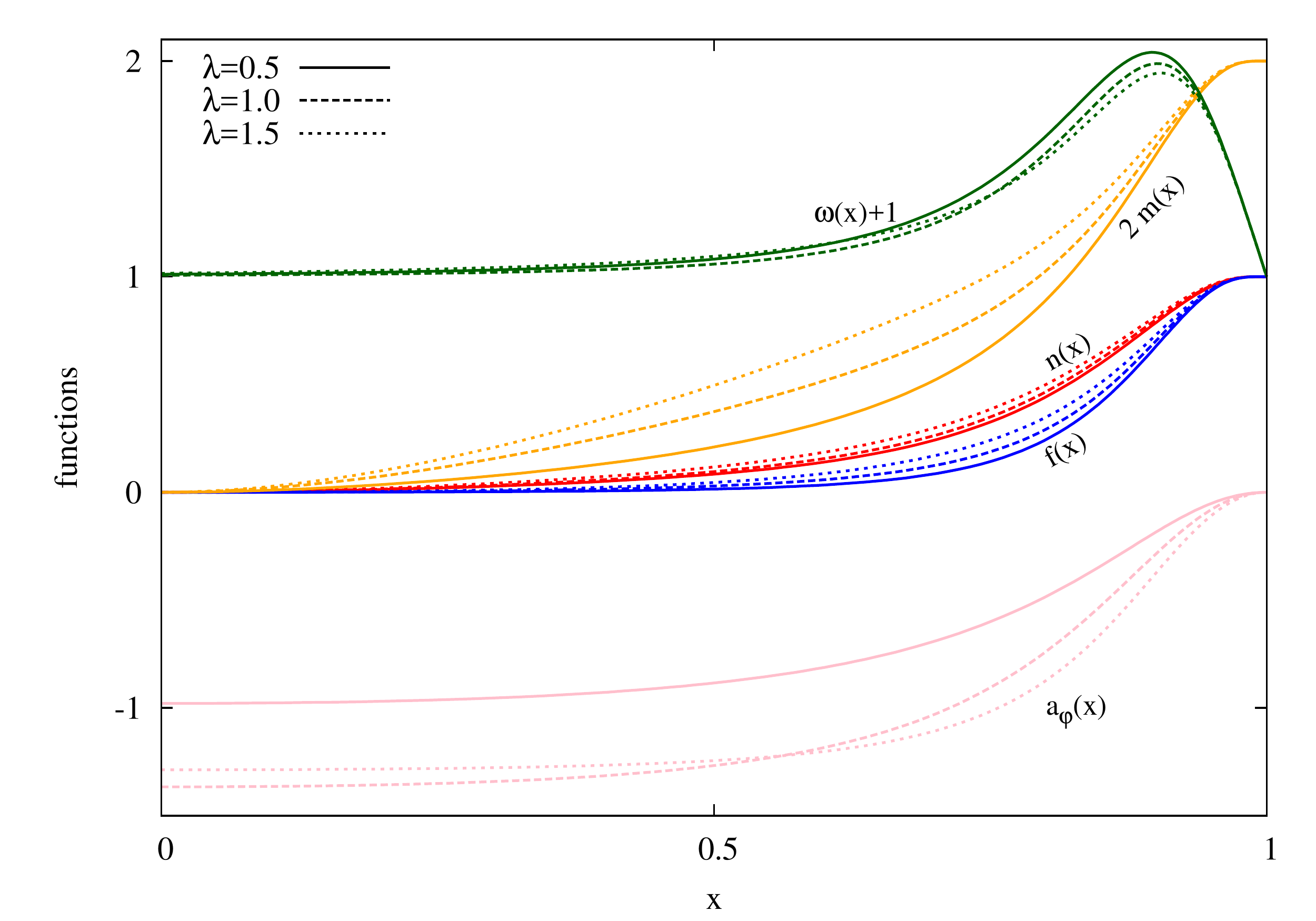}
    \caption{The profiles of typical charged rotating black holes with several values of $\lambda$
and $Q=-0.044$, $J=0.00148$, $r_H=0.4$, $L=1$
		are shown as a function of the compactified coordinate $x=1-r_H/r$.}
            \label{fig:profiles}
\end{figure}

\subsubsection{The profile of solutions and some generic features}

The profiles of three typical solutions\footnote{As discussed below, 
for large enough values of $\lambda$, new sets of $excited$ solutions occur, with a node structure for
both $a_\varphi$ and $\omega$.} with 
$\lambda=0.5$,
$\lambda=\lambda_{SG}=1$,
$\lambda=1.5$
and fixed values $Q=-0.044$, $J=0.00148$, $r_H=0.4$, $L=1$,
are shown in  Figure \ref{fig:profiles}. 
Varying $\lambda$
does not seem to lead to new qualitative features;
in particular the metric functions $f$, $m$ and $n$
always exhibit a monotonic behavior.
Also, we have noticed that the difference between solutions' profiles
for different $\lambda$ and the same $Q,J,r_H$ 
becomes more transparent when the BHs are 
close to extremality.

We mention that 
all solutions reported in this work have 
$
g^{tt}=-f<0~,
$
for any $r>r_H$, while $m$ and $n$ remain strictly positive.
Thus $t$ is a global time function
and the BHs are free of 
closed timelike (or null) curves \cite{Cvetic:2005zi}.
Also, they show no sign of a singular behavior\footnote{The exception here appears to be
 the $gap$ set of extremal solutions 
discussed  in Subsection 5.3. }
on the horizon or outside of it, that would manifest itself in 
the Ricci or Kretschmann scalars (which were monitored for most of the solutions). 
In addition, all the solutions we have analyzed present an ergoregion, 
inside of which
the observers cannot remain stationary  and will 
move in the direction of rotation. 
The ergoregion is located between the horizon and the ergosurface $r=r_c$, with $g_{tt}(r_c)=0$, $i.e.$
\begin{eqnarray} 
\label{er}  
\frac{n(r_c)}{f(r_c)}\omega^2(r_c)- f(r_c) \left(1+\frac{r_c^2}{L^2}\right)=0~,
\end{eqnarray} 
(note that, in contrast to $D=4$ Kerr-like BHs, 
the ergosurface does not touch the horizon).

The determination of the full domain of existence of the solutions
would be a huge task. 
In this work, we will only attempt to sketch its shape by analyzing the pattern of solutions on some
generic surfaces in the space of parameters.
Also, to simplify the study, 
we set the AdS 
length scale $L=1$,
such that all quantities are given in these units.
Moreover, without any loss of generality,
we consider 
values  $\lambda \geq 0$ for the CS coupling constant, only
(as such, we have to consider both signs for the electric charge).
We have considered solutions with 
a large set of $\lambda$ ranging between 0 and 50.
However,
solutions with  larger $\lambda$ are very likely to exist
and we conjecture the absence of an upper bound for the CS coupling constant.

We mention also that the numerical results exhibited
 in the  Figures in  Section 5.2 were found  
by extrapolating to the continuum
the results from discrete sets of around
one thousand 
solutions 
for each $\lambda$.
The solutions  there were found by considering first a fixed angular momentum  (Subsection 5.2.1)
and then a fixed electric charge  (Subsection 5.2.2).
Those plots are (typically) projections of 3D surfaces
which encode the dependence of the Hawking temperature $T_H$
 on two other quantities which enter the 1st law (\ref{1st}). 

As such, viewed together, 
they provide a description of the thermodynamics of the solutions,
together with the domain of existence.
For example, one can consider the thermodynamic stability
in the canonical ensemble, where the charge and angular
momentum are fixed parameters,
the response function being the heat capacity
 $C=T_H\left(\frac{\partial A_H}{\partial T_H}\right)_{J,Q}.$ 
We have found that for any value of $\lambda$,
the solutions with small values of $|J|$, $|Q|$
exhibit the pattern of the Schwarzschild-AdS BHs \cite{Hawking:1982dh}, 
only the large size BHs possessing a positive
specific heat $C>0$.
However,  the solutions become more thermally
stable as $|Q|$ and/or $|J|$ increase,
with $C>0$ for large enough values of these charges
even for small size BHs.

Apart from the quantities displayed in the Figures in  Section 5.2,
we have also considered the  gyromagnetic ratio $g$ as
resulting from  (\ref{gyro}).
A known result here is that, 
unlike in four dimensions,
the value of $g$ is not universal
in higher dimensions \cite{Aliev:2004ec},
while the AdS asymptotics further 
 introduces new features \cite{Aliev:2006tt}.
We have computed the gyromagnetic ratio 
for a large part of the solutions reported 
in this work and 
could not identify any clear pattern,
with $g$ taking a large range of values. 

Finally, let us mention that in the numerical study we have paid special attention
to extremal BHs, which have been constructed directly. 
Such configurations are important in themselves;
 they are also interesting as a
test of the predictions in Section 3 within the near-horizon formalism. 

\subsection{Three values of $\lambda$: a comparison}

To clarify the question asked in the Introduction
on {\it "how general the features of the CLP solution are"},
we shall present in what follows the results for three intermediate values of the Chern-Simons coupling constant.
Apart from  the SUGRA case $\lambda=1$, 
we shall exhibit results for a smaller value, $\lambda=0.5$,
and also for a larger one, $\lambda=1.5$.

\subsubsection{A fixed angular momentum: the  generic picture}

Starting with extremal BHs with a  $Q>0$, 
 one finds one single branch of solutions
which connects continuously  the extremal MP-AdS BH ($Q=0$) 
with the extremal RN-AdS BH (in the limit $Q\to\infty$).
This fact agrees with the prediction from the near-horizon formalism
in Section 3. 

For $Q<0$ the situation changes drastically. 
In agreement with the prediction from the near-horizon formalism, 
this set is characterized by the existence of two different branches, 
the MP one and the RN one,  
separated by a \textit{critical} solution. 
The MP branch starts with the extremal MP-AdS solution ($Q=0$), 
and extends for $Q\in(Q_0,0]$, where $Q_0<0$. 
In particular, for $J=0.0295$, one finds 
$Q_0=-0.0522$ for  $\lambda=0.5$, 
$Q_0=-0.0659$ in the SUGRA case, 
and $Q_0=-0.0755$ for $\lambda=1.5$. 
On the other hand, the RN branch exists for $Q \in(-\infty,Q_0)$.

\begin{figure}
    \centering
		
   \begin{subfigure}[b]{0.25\textwidth}
				\centering
        \includegraphics[width=30mm,scale=0.5]{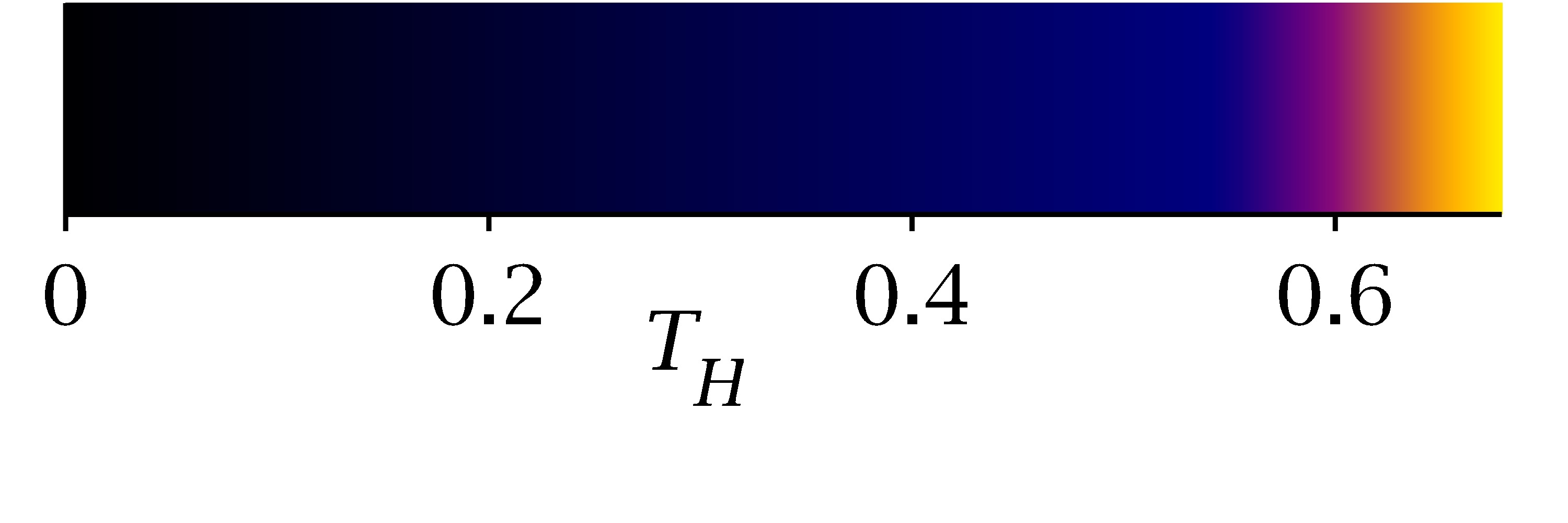}
        \label{fig:T_1.5_J0}
    \end{subfigure}		
		
    \begin{subfigure}[b]{0.3\textwidth}
        \includegraphics[width=50mm,scale=0.5]{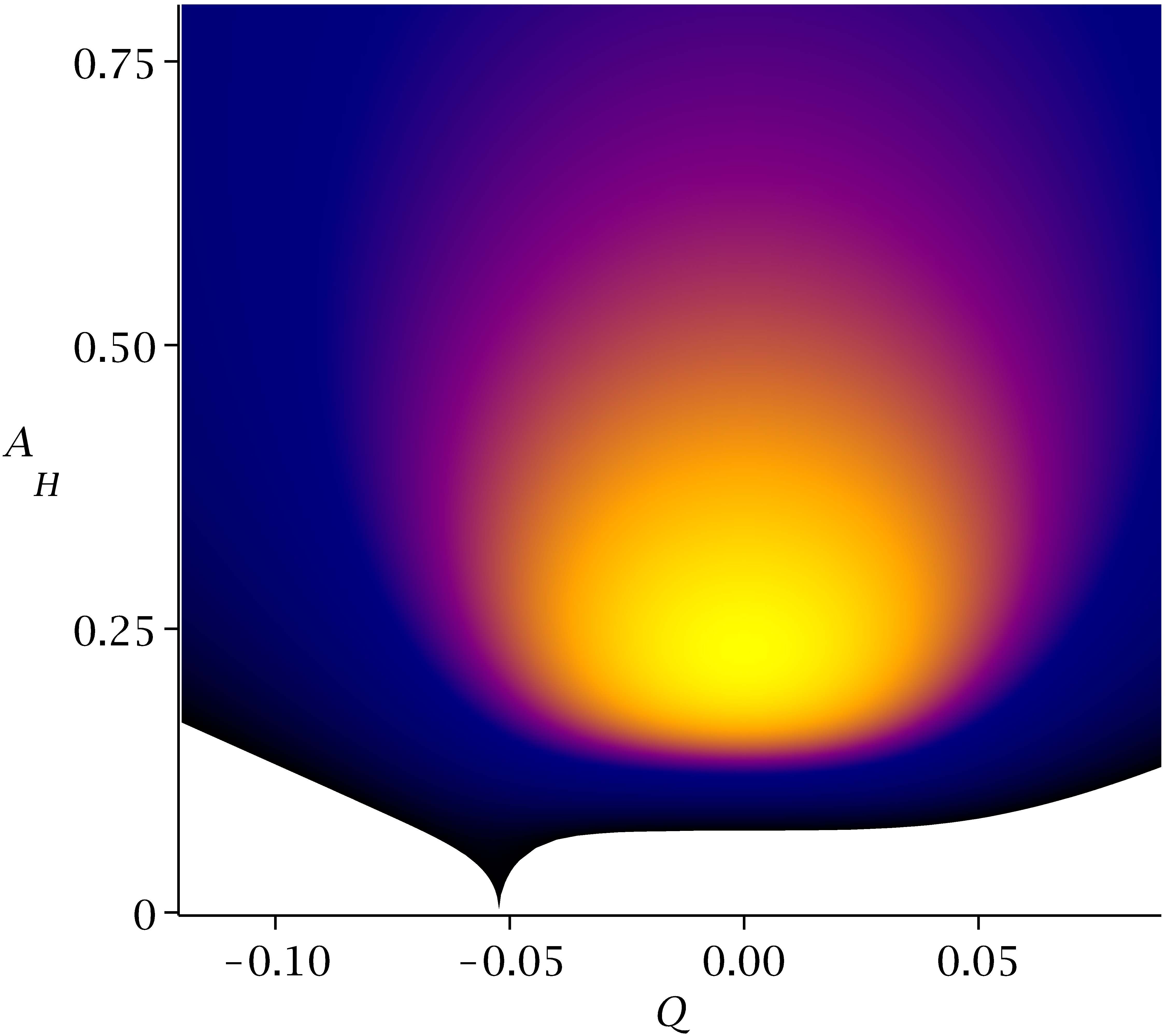}
        \caption{$\lambda = 0.5$}
        \label{fig:Ah_0.5_J0}
    \end{subfigure}
    \begin{subfigure}[b]{0.3\textwidth}
        \includegraphics[width=50mm,scale=0.5]{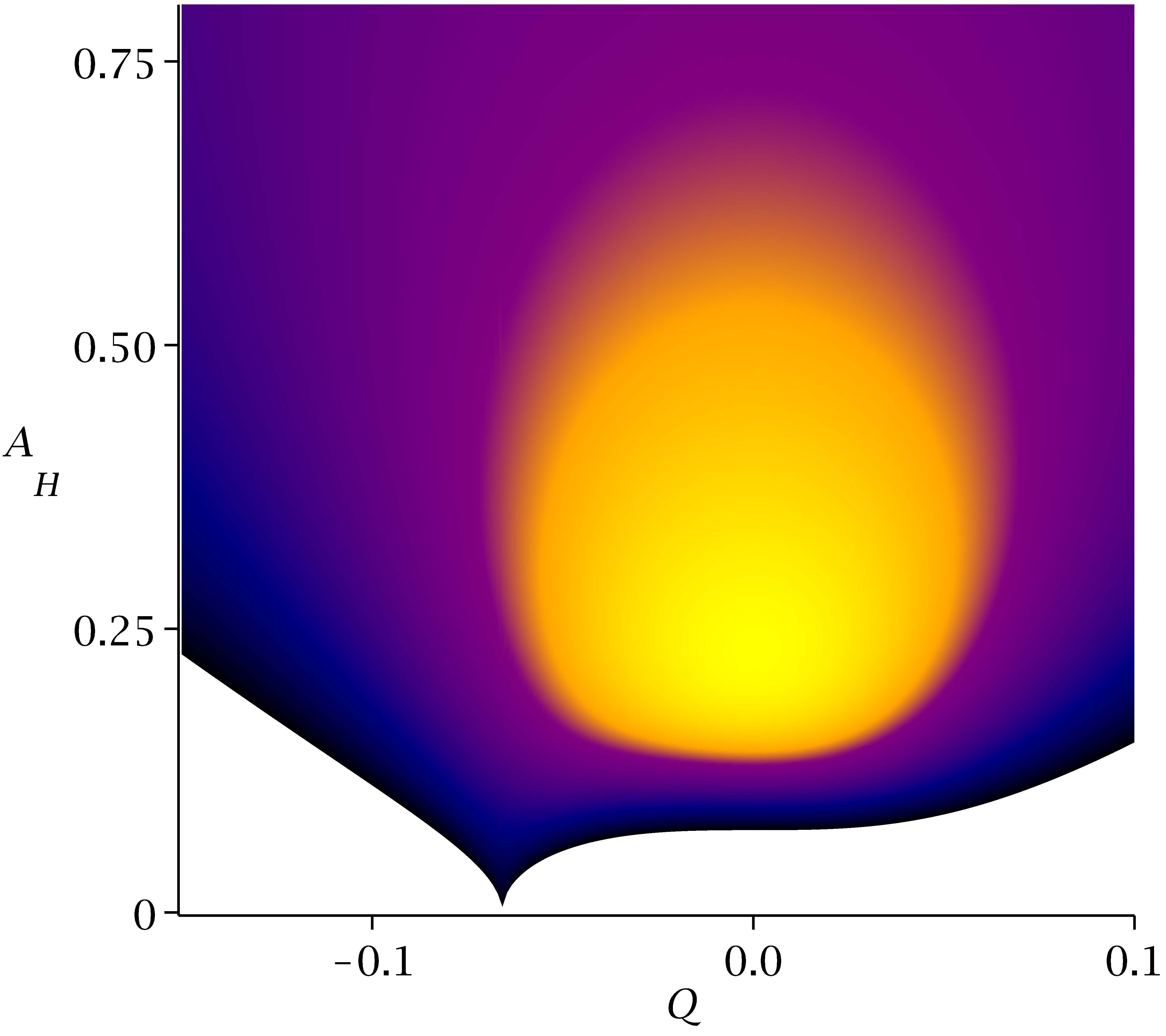}
        \caption{SUGRA}
        \label{fig:Ah_1.0_J0}
    \end{subfigure}
    \begin{subfigure}[b]{0.3\textwidth}
        \includegraphics[width=50mm,scale=0.5]{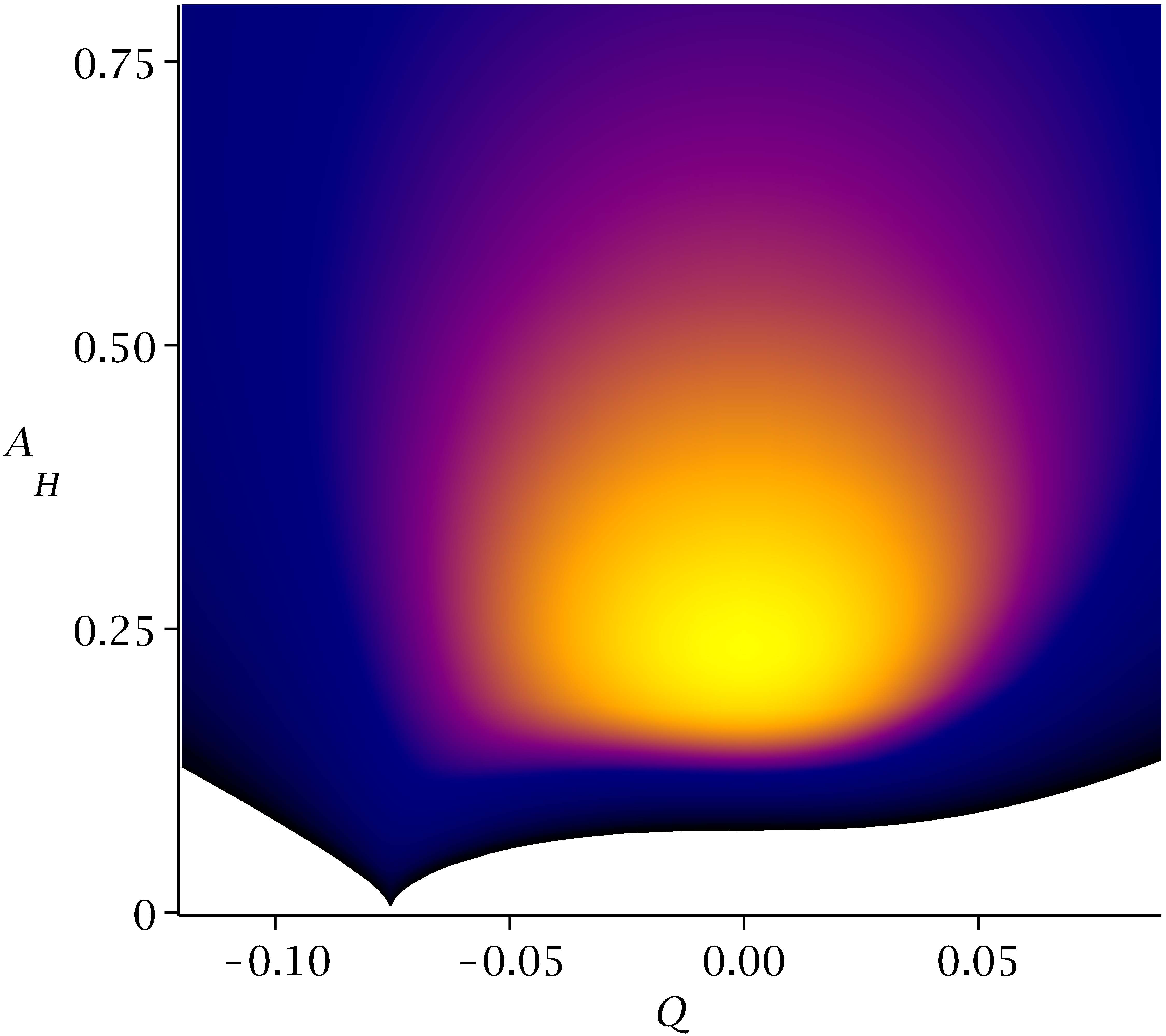}
        \caption{$\lambda = 1.5$}
        \label{fig:Ah_1.5_J0}
    \end{subfigure}
		
    \caption{Horizon area $A_H$ $vs.$ electric charge $Q$ with different temperatures $T_H$ for black holes with fixed angular momentum $J=0.00296$ and $L=1$, for $\lambda = 0.5$ (a), SUGRA $\lambda = 1$ (b), and $\lambda = 1.5$ (c). The lower bound of the area is given by the set of $T=0$ extremal solutions. Extremal solutions with negative electric charge possess a \textit{critical} solution with 
		$A_H=0$
		at $Q=Q_0$, where $Q_0=-0.0522,-0.0659,-0.0755$ for $\lambda=0.5$, SUGRA and $\lambda=1.5$, respectively. 
}
		\label{fig:Ah_J0}
		
\end{figure}
These features are shown in Figure \ref{fig:Ah_J0} where we give the $(A_H,Q;T_H)$ plot 
for BHs with $J=0.0295$. 
One can see that, in all cases, the horizon area is minimized by the set of extremal solutions. 
One can also notice 
the different behavior of the $T_H=0$ BHs with positive and negative $Q$.
For extremal solutions with $Q>0$, the area always increases with the electric charge.  
However, for $Q<0$, 
 the horizon area becomes zero at the \textit{critical} solution, which is reached for some critical electric charge $Q_0$. 

Let us discuss now the non-extremal solutions. 
In Figure \ref{fig:Ah_J0} we can see that all three cases possess a local maximum of the temperature. 
This local maximum is at $T_H=0.68$ and is always found for the $Q=0$ MP-AdS set. 
However, for higher values of the horizon area, the temperature increases again (not displayed in these Figures). 
Also, there is no upper bound for the area.
Another interesting feature that one can see in Figure \ref{fig:Ah_J0} 
is that it is possible to define closed sets of charged BHs with the same temperature (isothermal). 
These sets can only be found around the local maximum of temperature.

From these Figures we conclude that the general behavior 
of the area does not change much with respect to
the SUGRA solution. The effect of changing $\lambda$ reduces 
to a modification of the position of the \textit{critical} solution 
(essentially, increasing $\lambda$, leads to a larger magnitude of the electric charge of this configuration).

\begin{figure}
    \centering
		
   \begin{subfigure}[b]{0.25\textwidth}
				\centering
        \includegraphics[width=30mm,scale=0.5]{{T_1.5_J0}.jpg}
        
    \end{subfigure}		
		
    \begin{subfigure}[b]{0.3\textwidth}
        \includegraphics[width=50mm,scale=0.5]{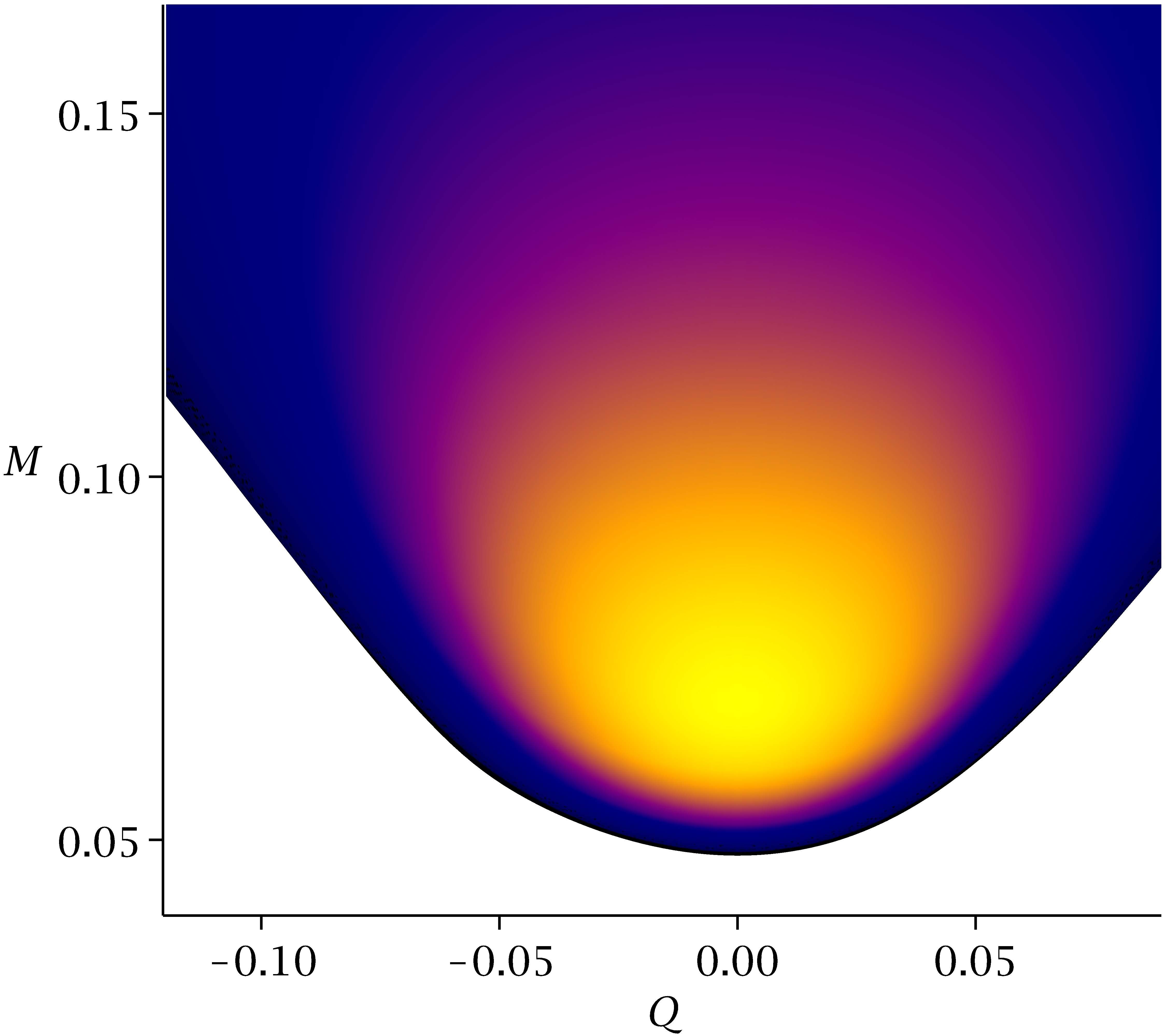}
        \caption{$\lambda = 0.5$}
        \label{fig:M_0.5_J0}
    \end{subfigure}
    \begin{subfigure}[b]{0.3\textwidth}
        \includegraphics[width=50mm,scale=0.5]{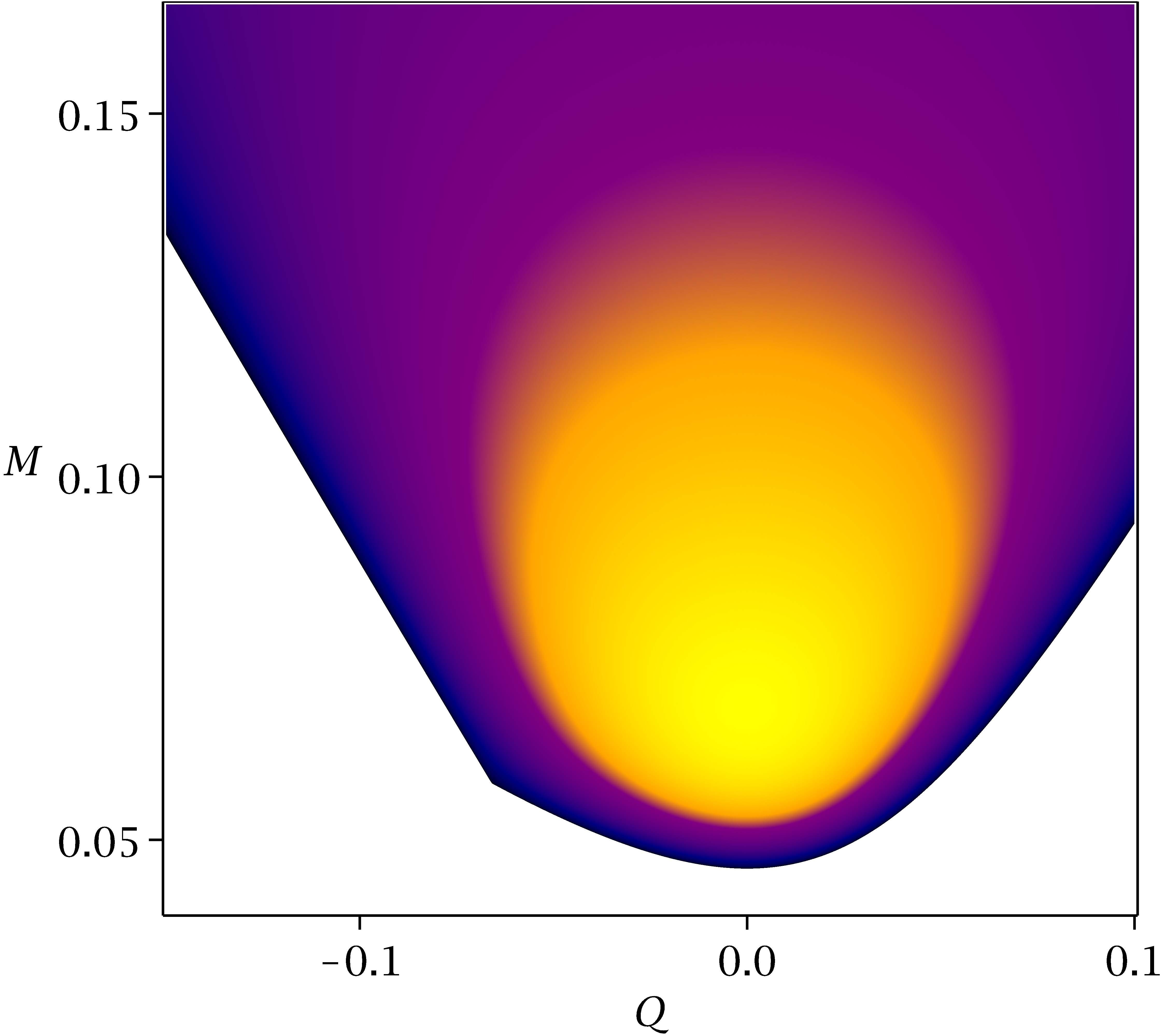}
        \caption{SUGRA}
        \label{fig:M_1.0_J0}
    \end{subfigure}
    \begin{subfigure}[b]{0.3\textwidth}
        \includegraphics[width=50mm,scale=0.5]{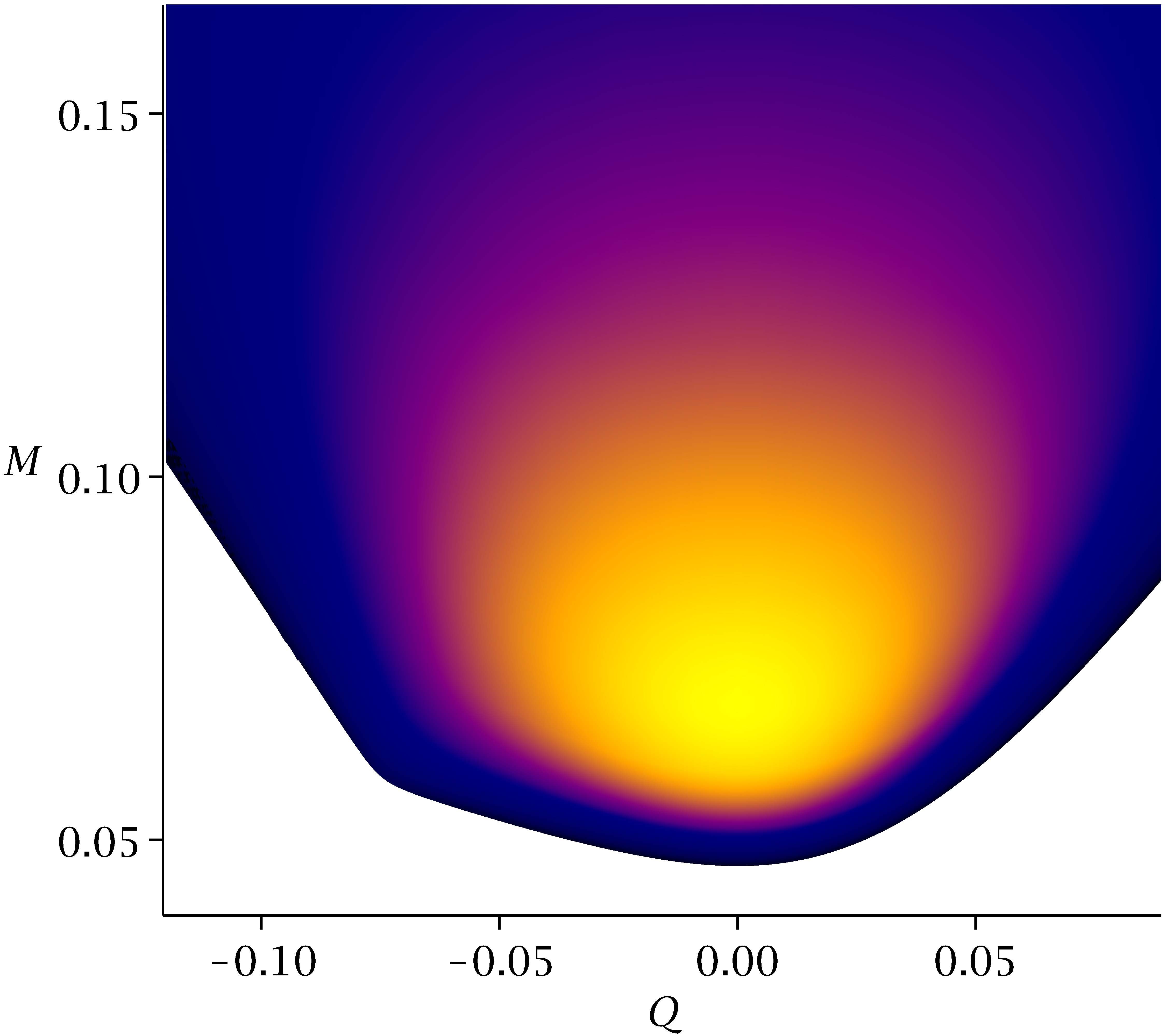}
        \caption{$\lambda = 1.5$}
        \label{fig:M_1.5_J0}
    \end{subfigure}
		
    \caption{Total mass $M$ $vs.$ electric charge $Q$ with different temperatures $T_H$ 
		for black holes with fixed angular momentum $J=0.00296$ and $L=1$, 
		for $\lambda = 0.5$ (a), SUGRA $\lambda = 1$ (b), and $\lambda = 1.5$ (c). 
		The lower bound of the mass is given by the set of extremal solutions. 
		The value of the mass of the extremal solutions always increases with the absolute value of the electric charge. 
		The \textit{critical} solution with zero area at $Q_0<0$ 
		can be identified here at the point where the lower bound exhibits a kink. 
		}
		
		\label{fig:M_J0}
		
\end{figure}

In Figure \ref{fig:M_J0} we show the $(M,Q;T_H)$ plot 
for the same BHs with $J=0.0295$. 
Similarly to the horizon area, the mass is minimized in the extremal case.
Here, however, the minimum mass is reached for the extremal MP-AdS BH ($Q=0$). 
For negative $Q$,  two different branches of extremal BHs can be identified 
at both sides of the \textit{critical} solution.  
However, the mass always increases with the absolute value of the electric charge, 
contrary to what happens for $A_H$.

Non-extremal solutions show a similar behavior for the mass as for the horizon area, 
and we can clearly identify the local maximum of temperature in the MP-AdS set. 
The mass of the non-extremal solutions with fixed $J$ can be increased to infinity 
(which also increases the temperature and the horizon area). 

The main difference between the solutions with different 
$\lambda$ 
is found close to the \textit{critical} solution. 
Here we see a jump on the slope of the $M-Q$ curve;
this jump becomes more pronounced
when increasing the value of the coupling beyond the SUGRA value. 
Below $\lambda=1$,
the Figure is softened 
(and the general behavior becomes more symmetric in $Q$).

\begin{figure}
    \centering
		
   \begin{subfigure}[b]{0.25\textwidth}
				\centering
        \includegraphics[width=30mm,scale=0.5]{{T_1.5_J0}.jpg}
        
    \end{subfigure}		
		
    \begin{subfigure}[b]{0.3\textwidth}
        \includegraphics[width=50mm,scale=0.5]{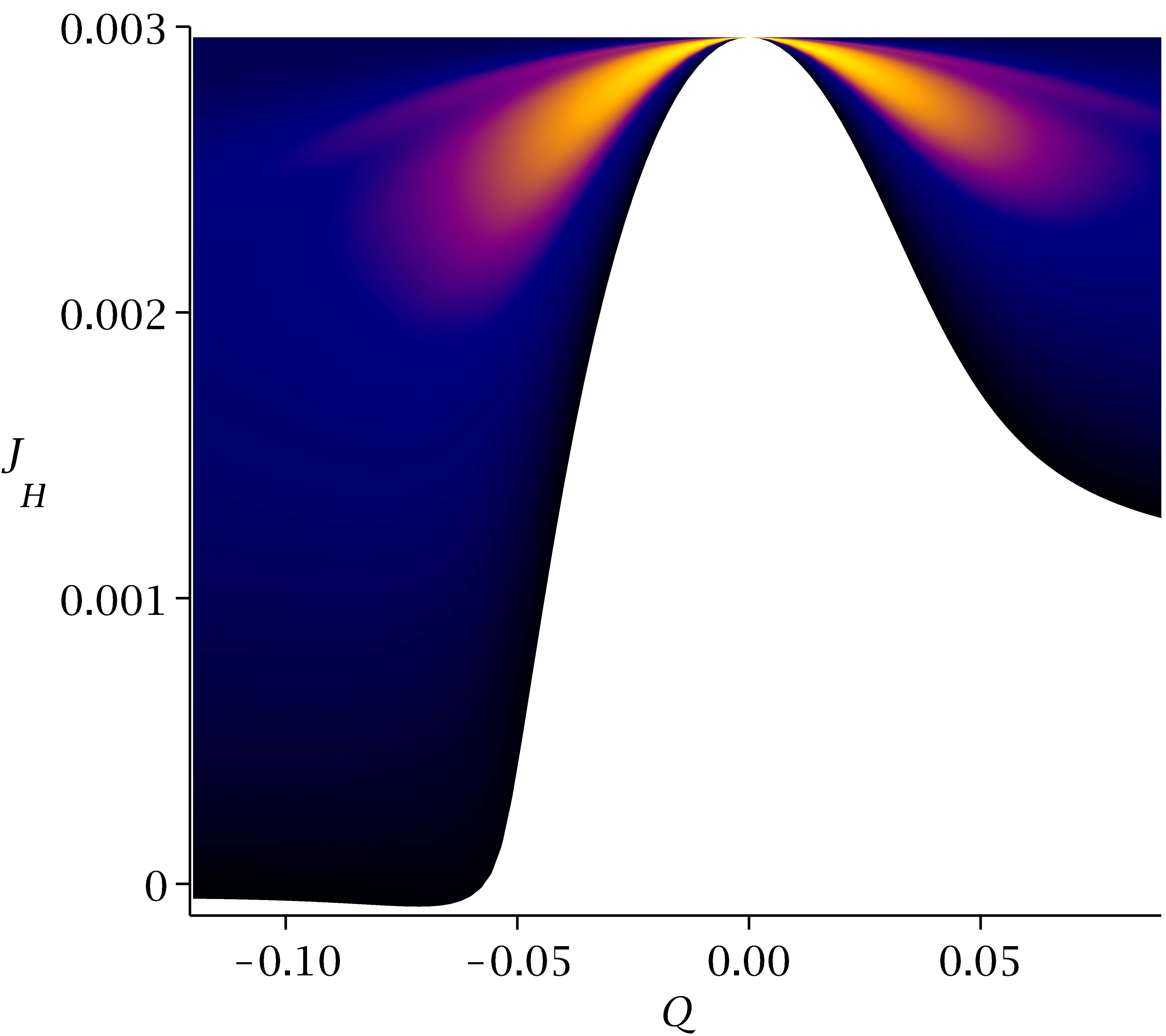}
        \caption{$\lambda = 0.5$}
        \label{fig:Jh_0.5_J0}
    \end{subfigure}
    \begin{subfigure}[b]{0.3\textwidth}
        \includegraphics[width=50mm,scale=0.5]{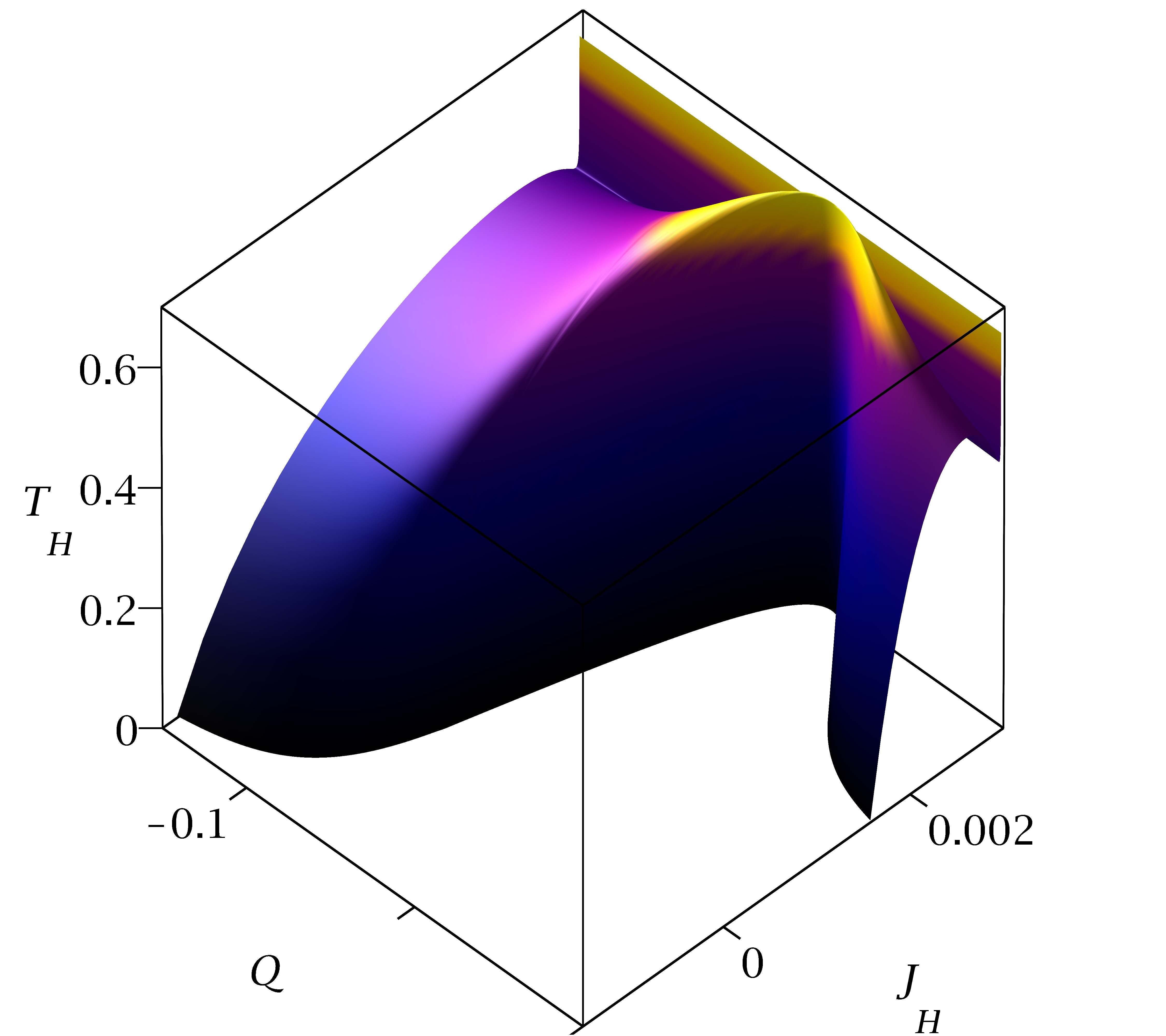}
        \caption{SUGRA}
        \label{fig:Jh_1.0_J0}
    \end{subfigure}
    \begin{subfigure}[b]{0.3\textwidth}
        \includegraphics[width=50mm,scale=0.5]{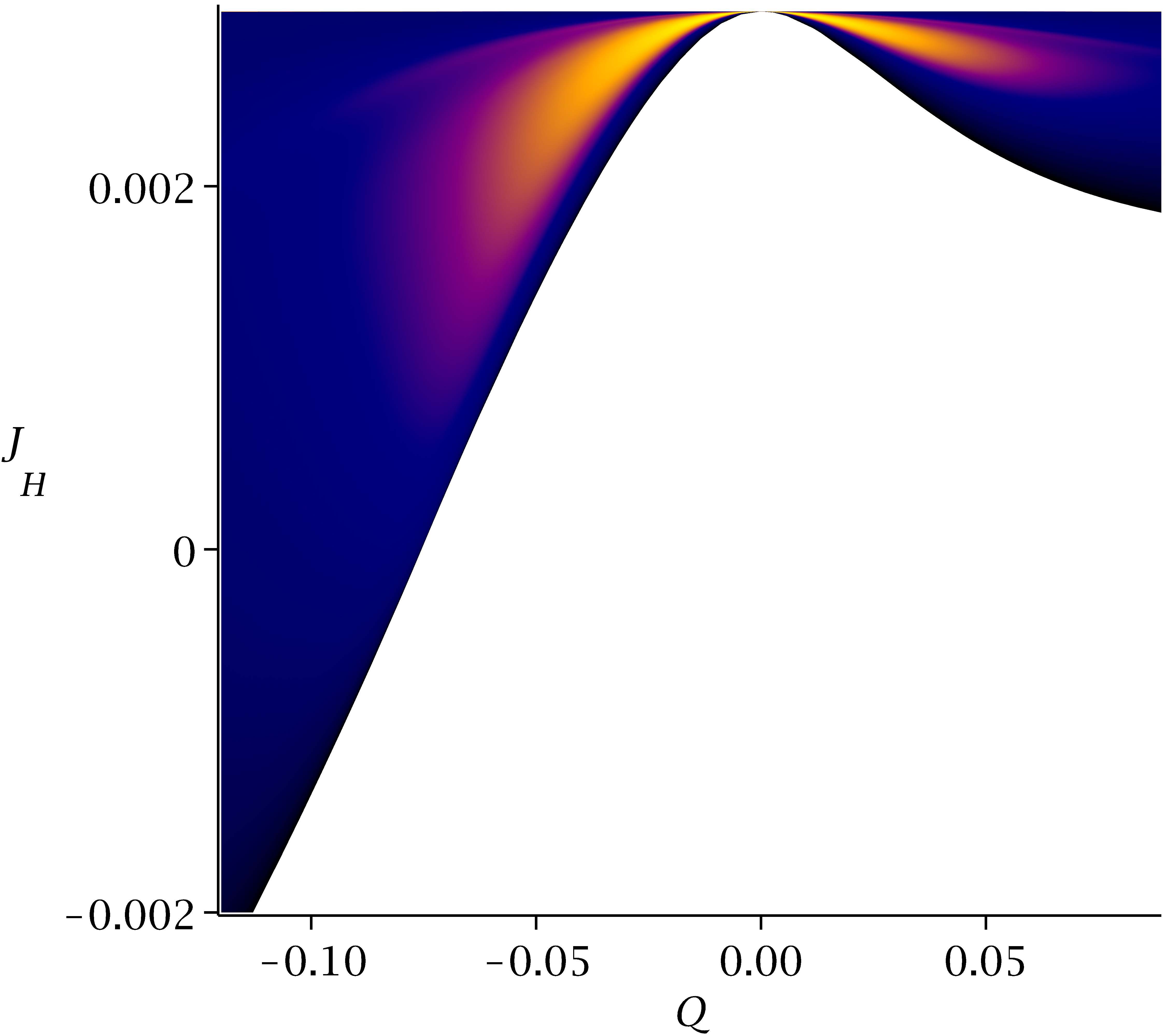}
        \caption{$\lambda = 1.5$}
        \label{fig:Jh_1.5_J0}
    \end{subfigure}
		
    \caption{Horizon angular momentum $J_H$ $vs.$ electric charge $Q$ with different temperatures $T_H$ 
		for black holes with fixed angular momentum $J=0.00296$ and $L=1$, for $\lambda = 0.5$ (a), SUGRA $\lambda = 1$ (b), and $\lambda = 1.5$ (c). 
		Note that the lower bound of the horizon angular momentum is given by the set of extremal solutions.
		But now we can see an upper bound, at the line $J_H=J=0.00296$. 
		This line is reached asymptotically as the mass and temperature of the non-extremal black holes are increased. 
		The \textit{critical} solution with zero area at $Q_0<0$ can be identified here at the point where the horizon angular momentum is zero. }
		
		\label{fig:Jh_J0}
		
\end{figure}

Turning now to the horizon quantities,
 we present in Figure \ref{fig:Jh_J0} 
the plot for $(J_H,Q;T_H)$
for the same set of configurations. 
This plot is interesting 
because it gives us an idea about the full domain of existence of the global solutions. 
For instance, consider Figures \ref{fig:Jh_0.5_J0} and \ref{fig:Jh_1.5_J0}. 
The lower bound of the horizon angular momentum is given by the set of extremal solutions. 
The upper bound is given by the line of constant $J_H=J=0.00296$. 
This line is reached asymptotically as the mass of the non-extremal BHs 
is increased to infinity. 
To better understand these aspects, 
 we present in Figure \ref{fig:Jh_1.0_J0} the 3D plot $(J_H,Q;T_H)$ for solutions in the SUGRA case.
Then one can see more clearly that, close to $J_H=J=0.00296$,
the temperature drops from the local maximum and then it increases again without limit. 
The mass and the horizon area also increase and the non-extremal BHs 
become more and more massive.
Moreover, they have almost all the angular momentum stored 
behind the horizon (hence reaching the limit $J_H=J$). 
Note the surface degenerates into a single line at 
$Q=0$, $J=J_H$, with $T\in[0,\infty)$, 
where the full set of MP-AdS BHs is recovered for the three  considered cases.

Concerning the extremal solutions, it is interesting to note the difference between the $Q>0$ and $Q<0$ cases. For positive electric charge, adding $Q$ to an extremal MP-AdS solution
decreases  the horizon angular momentum, and it goes to zero as the electric charge goes to infinity.  For negative electric charge, however, the two extremal branches present different properties.
In the MP branch, the electric charge can be increased only up to the \textit{critical} solutions with $Q=Q_0$, and $J_H$ decreases with $|Q|$. At $Q=Q_0$, the horizon angular momentum vanishes.
Combining this result with the $Q>0$ case, one can say that the horizon angular momentum of extremal solutions with $Q\in(Q_0,\infty)$ satisfies the relation $0<J_H\leq J$, saturating the relation only at $Q=0$.   

Another interesting feature one can notice in  Figure \ref{fig:Jh_J0} 
is the existence of counter-rotating configurations 
(this holds for all considered values of $\lambda$).
For example, take the extremal solutions on the RN branch ($Q<Q_0$): 
they have $J>0$, however $J_H$ is negative.
Moreover, this is not unique to extremal solutions: BHs with $T_H>0$ 
can also become counter-rotating 
for low enough temperatures and $Q<Q_0$. 

Changing the coupling $\lambda$ has a particularly relevant effect on these counter-rotating configurations. Note that in the Figure we can see how reducing the coupling below SUGRA reduces the size of the space of solutions with counter-rotation (we will comment on this again in the following sections).

\begin{figure}
    \centering
		
   \begin{subfigure}[b]{0.25\textwidth}
				\centering
        \includegraphics[width=30mm,scale=0.5]{{T_1.5_J0}.jpg}
        
    \end{subfigure}		
		
    \begin{subfigure}[b]{0.3\textwidth}
        \includegraphics[width=50mm,scale=0.5]{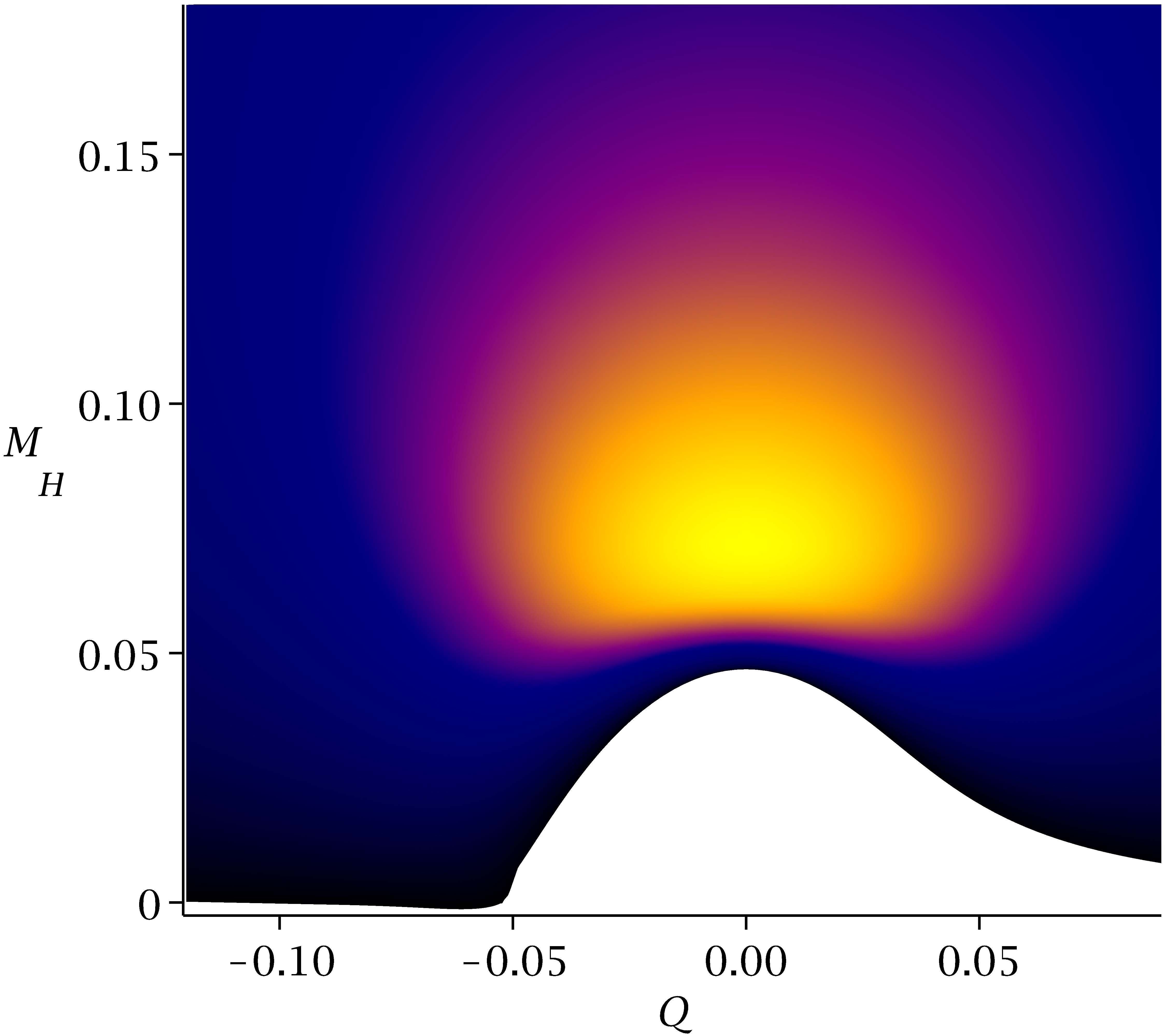}
        \caption{$\lambda = 0.5$}
        \label{fig:Mh_0.5_J0}
    \end{subfigure}
    \begin{subfigure}[b]{0.3\textwidth}
        \includegraphics[width=50mm,scale=0.5]{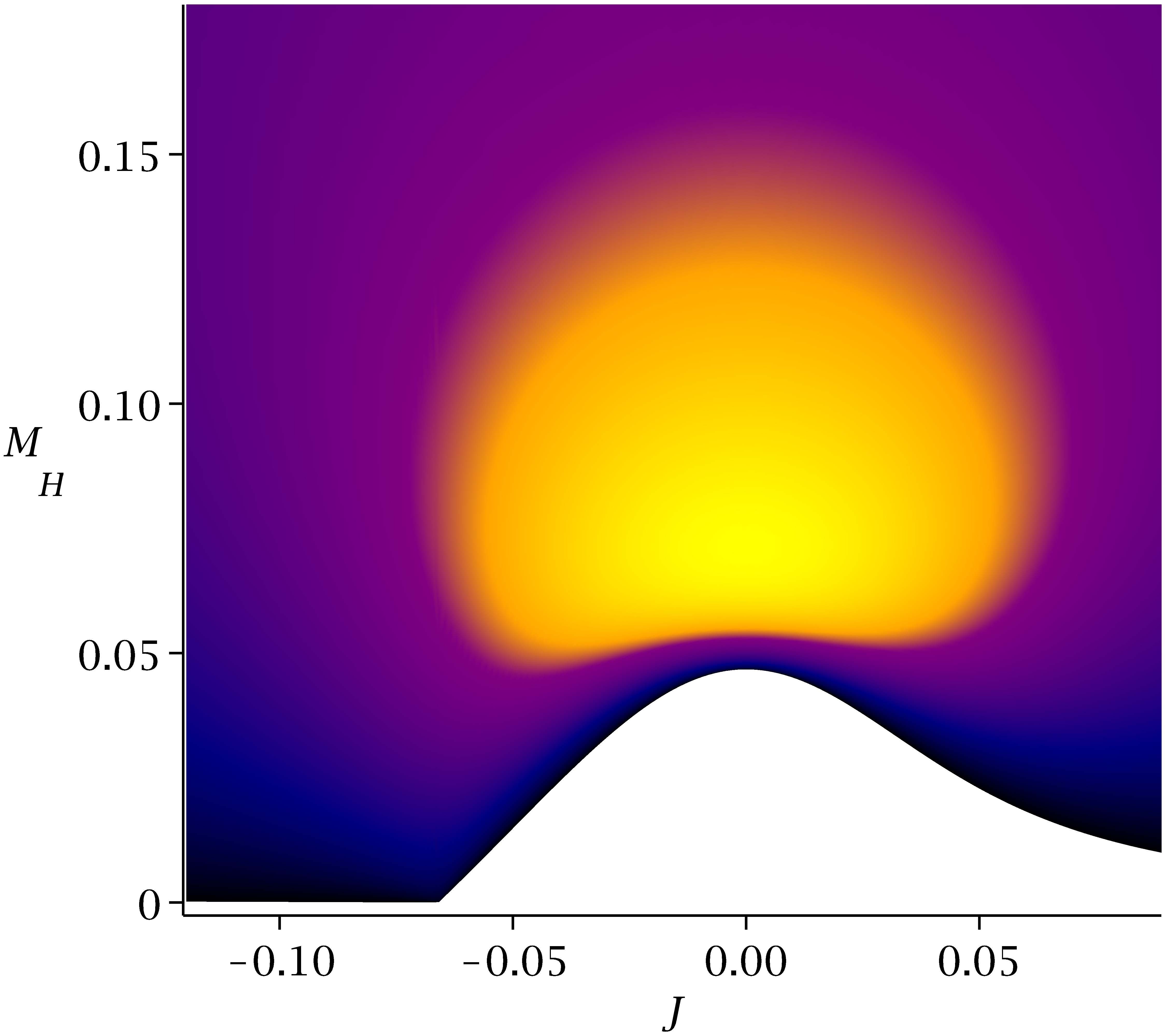}
        \caption{SUGRA}
        \label{fig:Mh_1.0_J0}
    \end{subfigure}
    \begin{subfigure}[b]{0.3\textwidth}
        \includegraphics[width=50mm,scale=0.5]{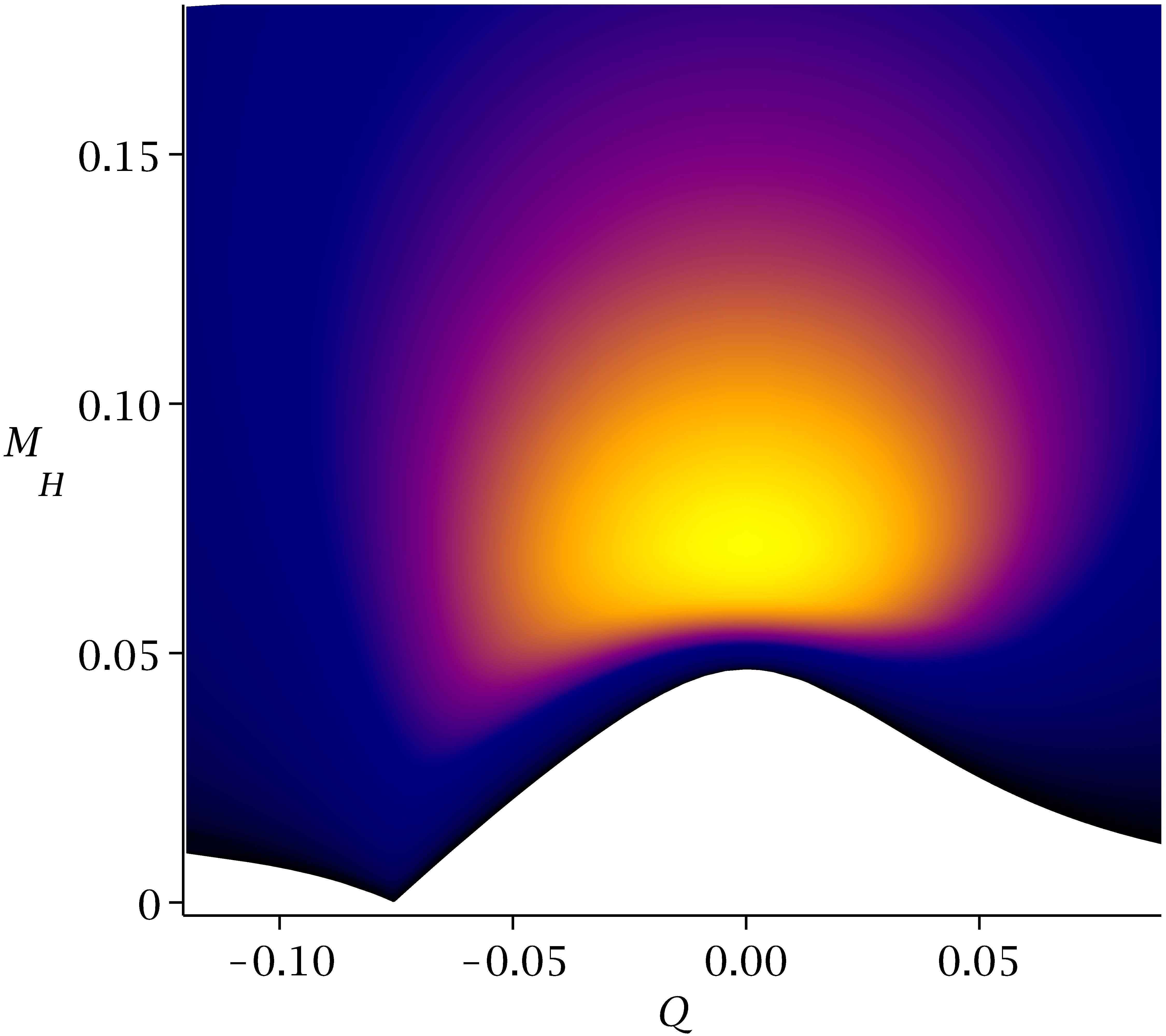}
        \caption{$\lambda = 1.5$}
        \label{fig:Mh_1.5_J0}
    \end{subfigure}
		
    \caption{Horizon mass $M_H$ $vs.$ electric charge $Q$ with different temperatures $T_H$ 
		for black holes with fixed angular momentum $J=0.00296$ and $L=1$, for 
		$\lambda = 0.5$ (a), SUGRA $\lambda = 1$ (b), and $\lambda = 1.5$ (c).
		The main difference between these values of $\lambda$ 
		is in the horizon mass of the extremal and near extremal configurations. 
		In particular the difference is found in the RN branch: note that in (a), when $Q<-0.0522$ the horizon mass can be negative; in (b) when $Q<Q_0=-0.0659$ it is always zero; in (c) when $Q<-0.0755$ it is positive and increases with the absolute value of the electric charge.}
		
		\label{fig:Mh_J0}
		
\end{figure}

We continue with 
 Figure \ref{fig:Mh_J0},
where  we show the $(M_H,Q;T_H)$ plot.
One interesting feature of the $\lambda=0.5$ set is the existence of solutions with negative horizon masses.
In the extremal case, 
this happens for solutions  on the RN branch  and  $Q<Q_0$. 
Also, some non-extremal solutions close to this set share the same property. 
Moreover,  one can notice that
the horizon mass of the \textit{critical} solution is always zero, independently of the value of $\lambda$. 
Surprisingly, in the SUGRA case,
the  extremal solutions on the RN branch ($Q<Q_0$
) always have $M_H=0$, 
as one can see in Figure \ref{fig:Mh_1.0_J0}. 

Hence the main effect of changing the coupling in this case is on the behavior of the horizon mass of the $Q<Q_0$ solutions: below SUGRA we can find $M_H<0$ configurations, and beyond SUGRA the horizon mass is positive. SUGRA is a very particular case in which $Q<Q_0$ extremal black holes have $M_H=0$.

\begin{figure}
    \centering
		
   \begin{subfigure}[b]{0.25\textwidth}
				\centering
        \includegraphics[width=30mm,scale=0.5]{{T_1.5_J0}.jpg}
        
    \end{subfigure}		
		
    \begin{subfigure}[b]{0.3\textwidth}
        \includegraphics[width=50mm,scale=0.5]{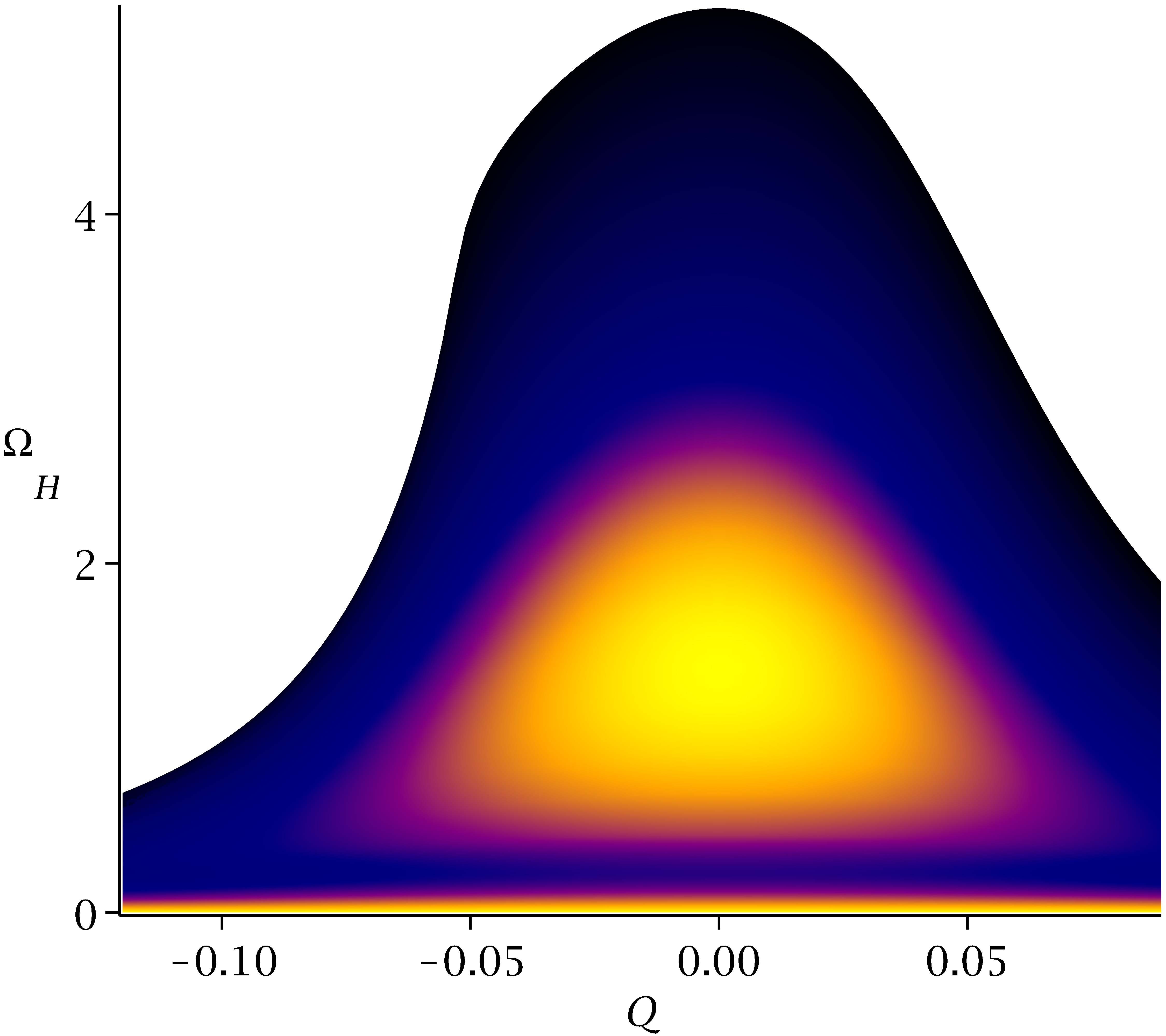}
        \caption{$\lambda = 0.5$}
        \label{fig:Omh_0.5_J0}
    \end{subfigure}
    \begin{subfigure}[b]{0.3\textwidth}
        \includegraphics[width=50mm,scale=0.5]{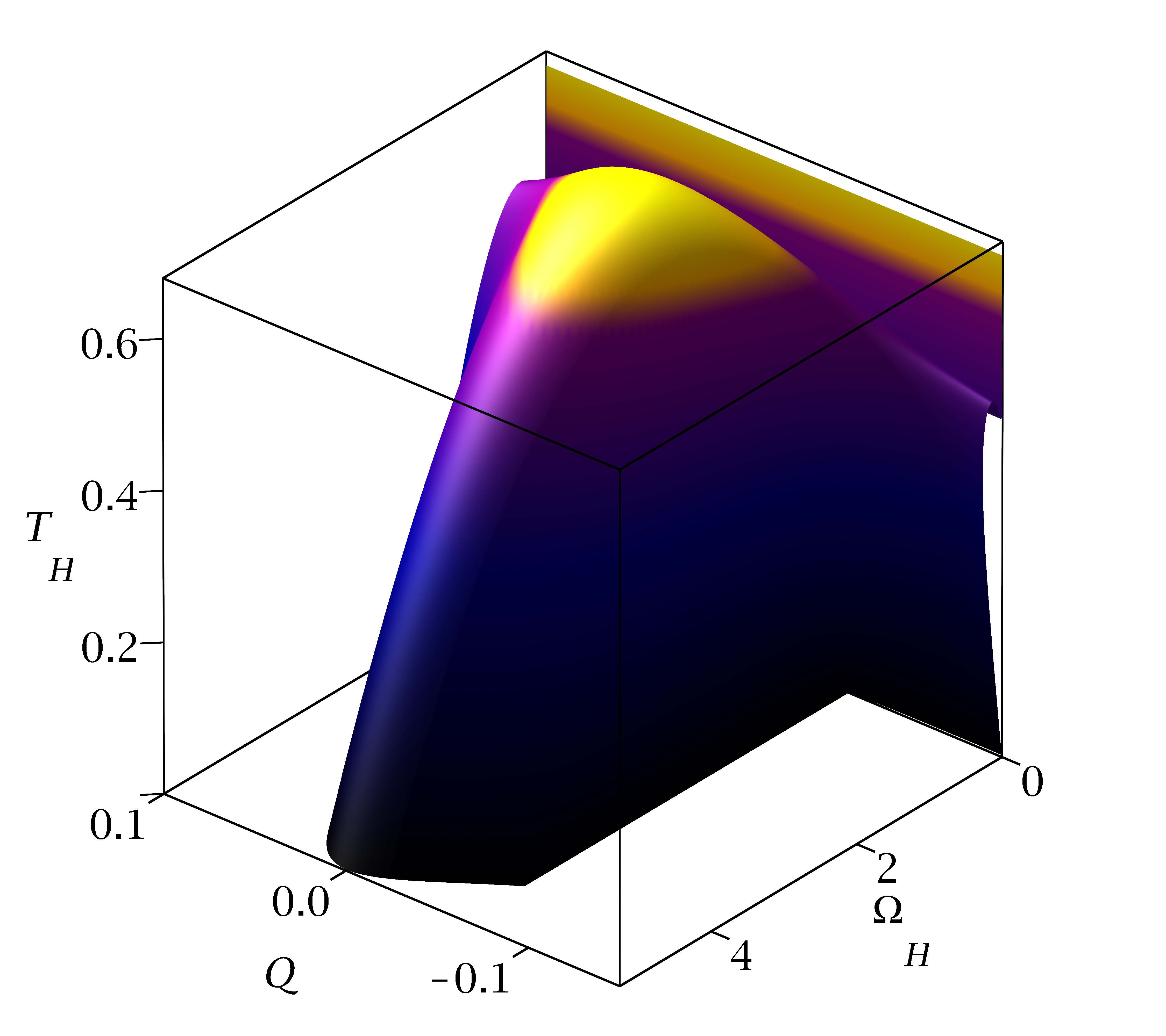}
        \caption{SUGRA}
        \label{fig:Omh_1.0_J0}
    \end{subfigure}
    \begin{subfigure}[b]{0.3\textwidth}
        \includegraphics[width=50mm,scale=0.5]{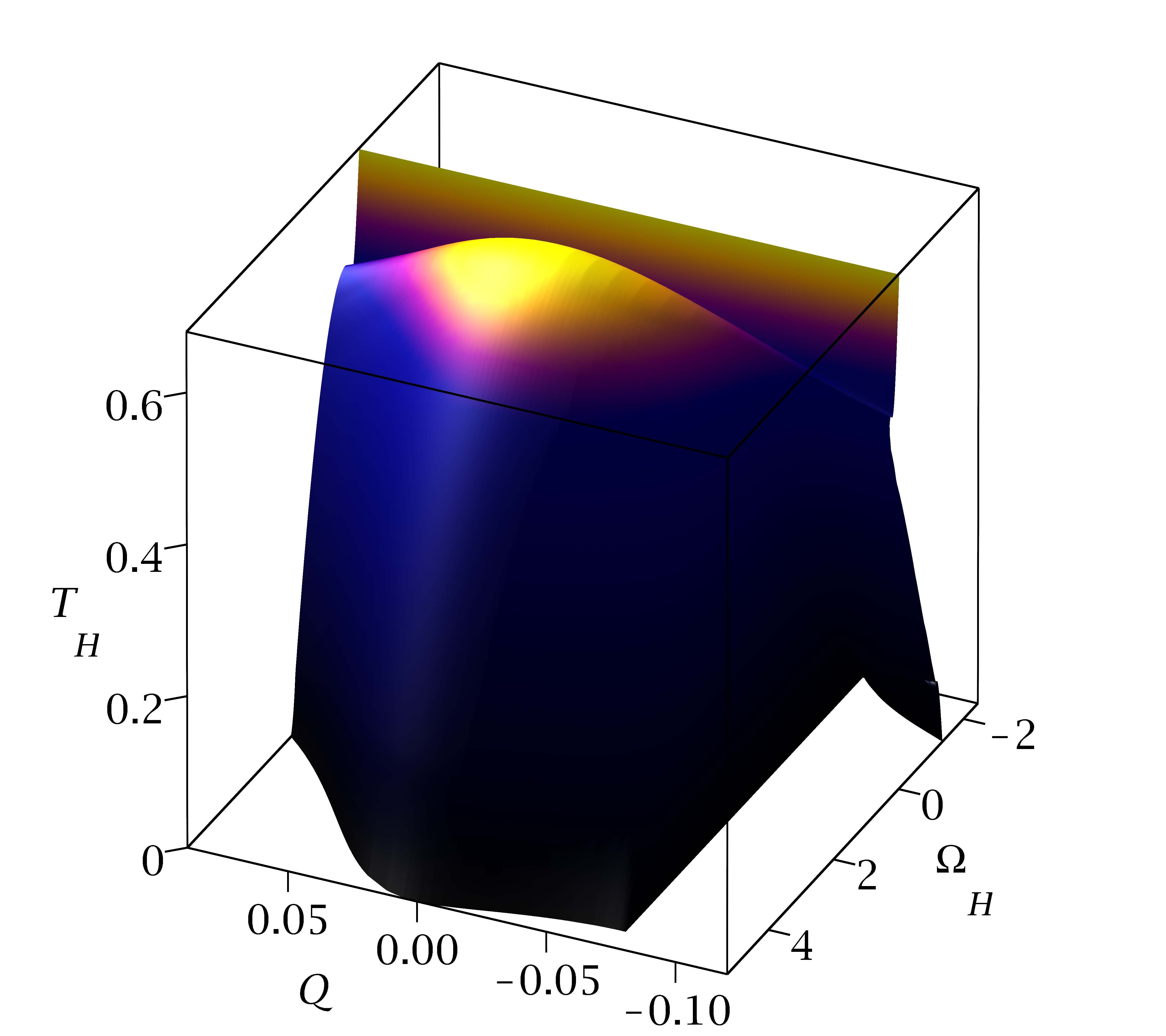}
        \caption{$\lambda = 1.5$}
        \label{fig:Omh_1.5_J0}
    \end{subfigure}
		
    \caption{Horizon angular velocity $\Omega$ $vs.$ electric charge $Q$ with different temperatures $T_H$ for black holes with fixed angular momentum $J=0.00296$ and $L=1$, for $\lambda = 0.5$ (a), SUGRA $\lambda = 1$ (b), and $\lambda = 1.5$ (c). The three cases present very different properties in their extremal limit. Note that $\Omega$ for extremal solutions in $\lambda = 0.5$ (a) is continuous. This is no longer the case in SUGRA $\lambda = 1$ (b), and $\lambda = 1.5$ (c), where the RN branch has $\Omega=0$ and $\Omega<0$ respectively. Since the MP branch has always positive $\Omega$, there is a discontinuity around the \textit{critical} solution in the angular velocity in the $T=0$ limit for these two cases.}
		
		\label{fig:Omh_J0}
		
\end{figure}

In Figure \ref{fig:Omh_J0} we present the $(\Omega_H,Q;T_H)$ diagram 
for the same configurations. 
An interesting behavior occurs here for large enough values of $\lambda$.
For example, as seen in Figure \ref{fig:Omh_1.0_J0} for the SUGRA case, 
the extremal BHs on the RN branch ($Q<Q_0$) 
always have $\Omega_H=0$. 
However, solutions on the MP branch have positive angular velocity.
This indicates the existence of a discontinuity in the angular velocity 
at the \textit{critical} solution.
This discontinuous behavior is also present for $\lambda=1.5$,
see Figure \ref{fig:Omh_1.5_J0}.
Along the RN branch ($Q<Q_0$), the BHs have a counter-rotating horizon:
the horizon angular velocity is negative, despite the total angular momentum being positive. 
However, extremal BHs on the MP branch have positive angular velocity.
 Hence, one again sees a 
discontinuous behavior in the angular velocity.
Note that non-extremal solutions are not discontinuous:
 both extremal branches can be joined by almost extremal solutions, 
for which very small variations of the electric charge cause very large (but continuous) modifications of the angular velocity,
 and can even change the sense of rotation for big values of $\lambda$.

In these Figures for the angular velocity we can see that once again changing the CS coupling has a very important effect on the properties of the solutions, in particular for $Q<Q_0$. SUGRA is a special case where these solutions have null angular velocity, but increasing the value $\lambda$ beyond SUGRA makes this set counter-rotate in the angular velocity.

\begin{figure}
    \centering
		
   \begin{subfigure}[b]{0.25\textwidth}
				\centering
        \includegraphics[width=30mm,scale=0.5]{{T_1.5_J0}.jpg}        
    \end{subfigure}		
		
    \begin{subfigure}[b]{0.3\textwidth}
        \includegraphics[width=50mm,scale=0.5]{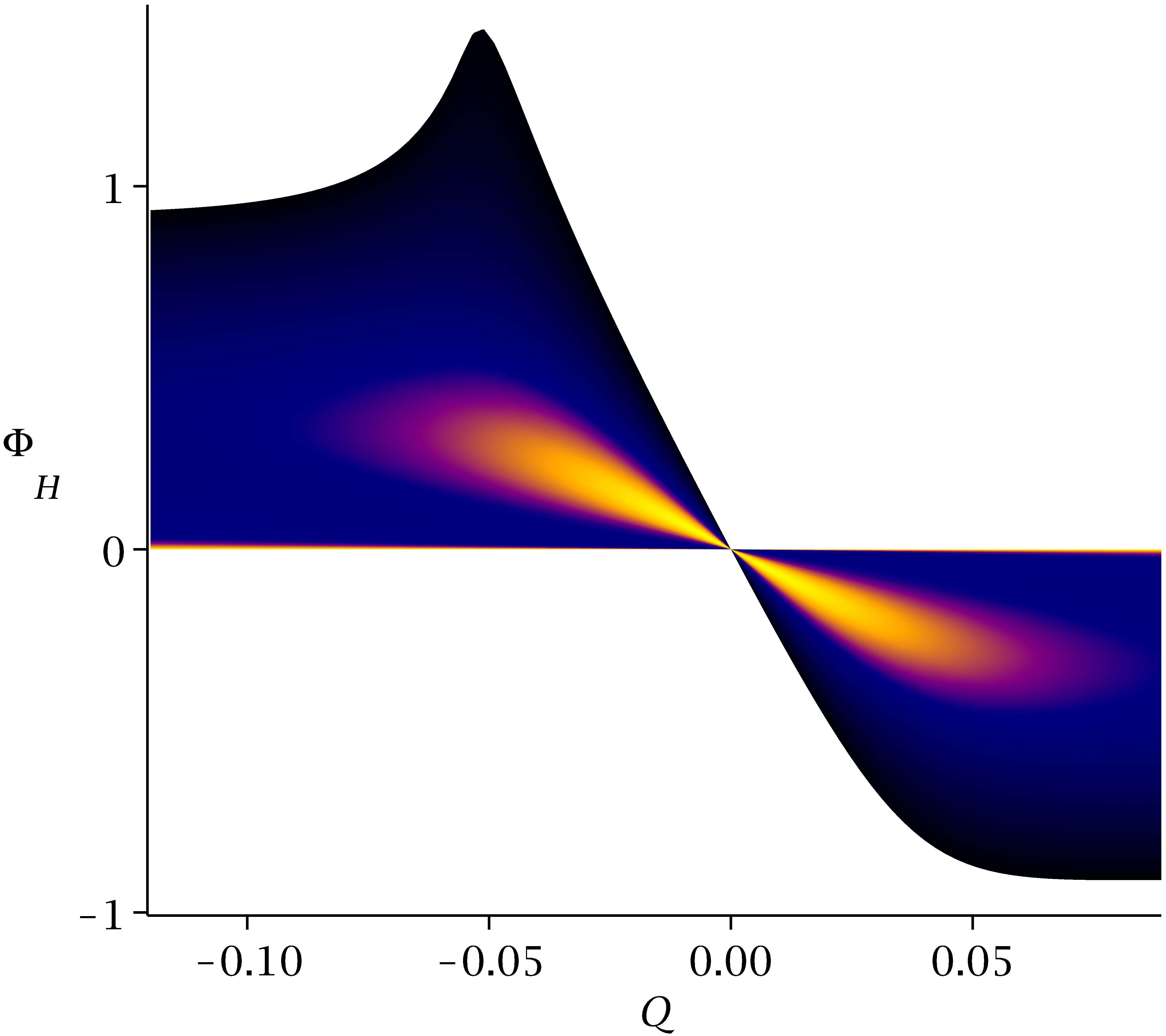}
        \caption{$\lambda = 0.5$}
        \label{fig:Phi_0.5_J0}
    \end{subfigure}
    \begin{subfigure}[b]{0.3\textwidth}
        \includegraphics[width=50mm,scale=0.5]{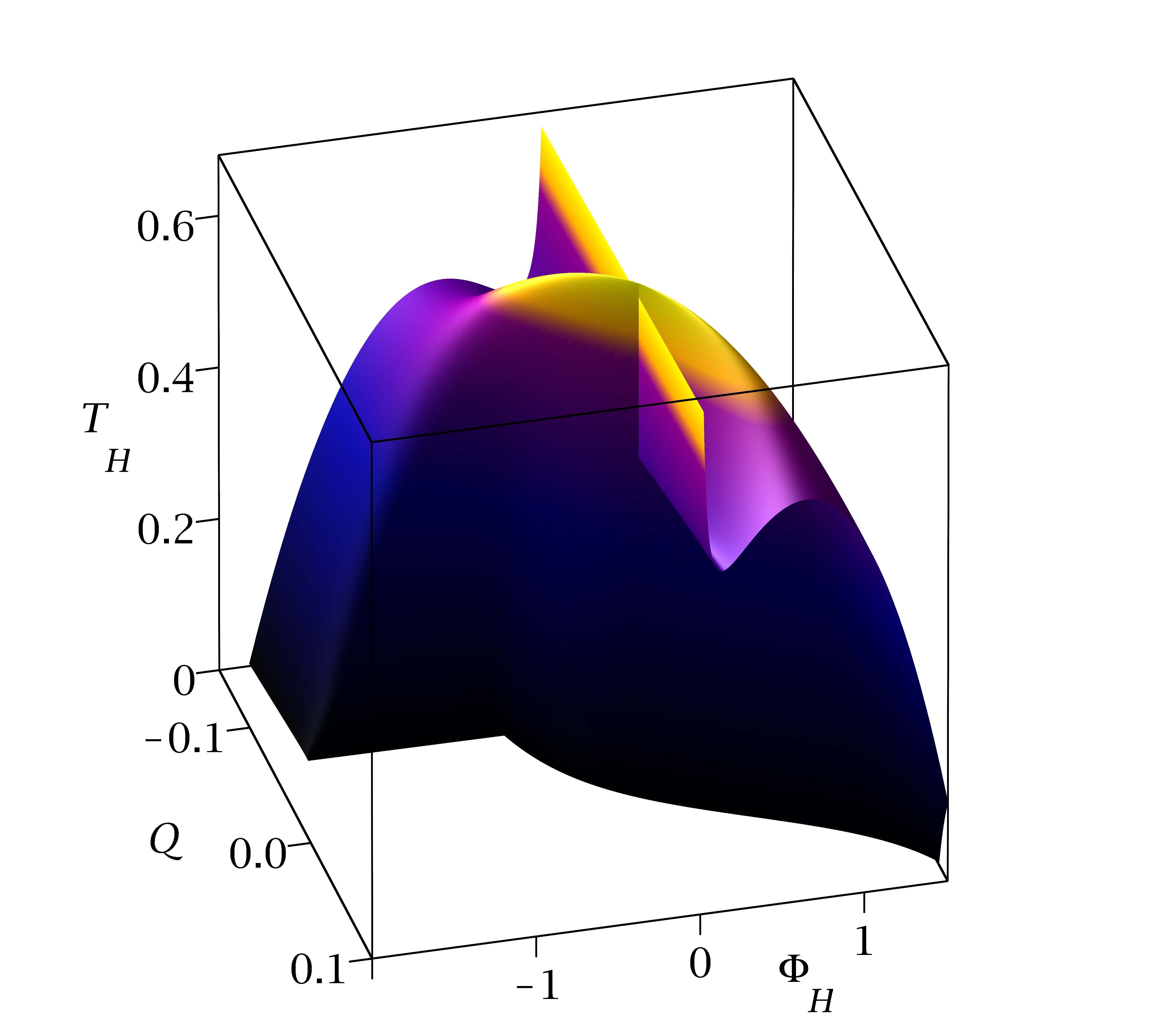}
        \caption{SUGRA}
        \label{fig:Phi_1.0_J0}
    \end{subfigure}
    \begin{subfigure}[b]{0.3\textwidth}
        \includegraphics[width=50mm,scale=0.5]{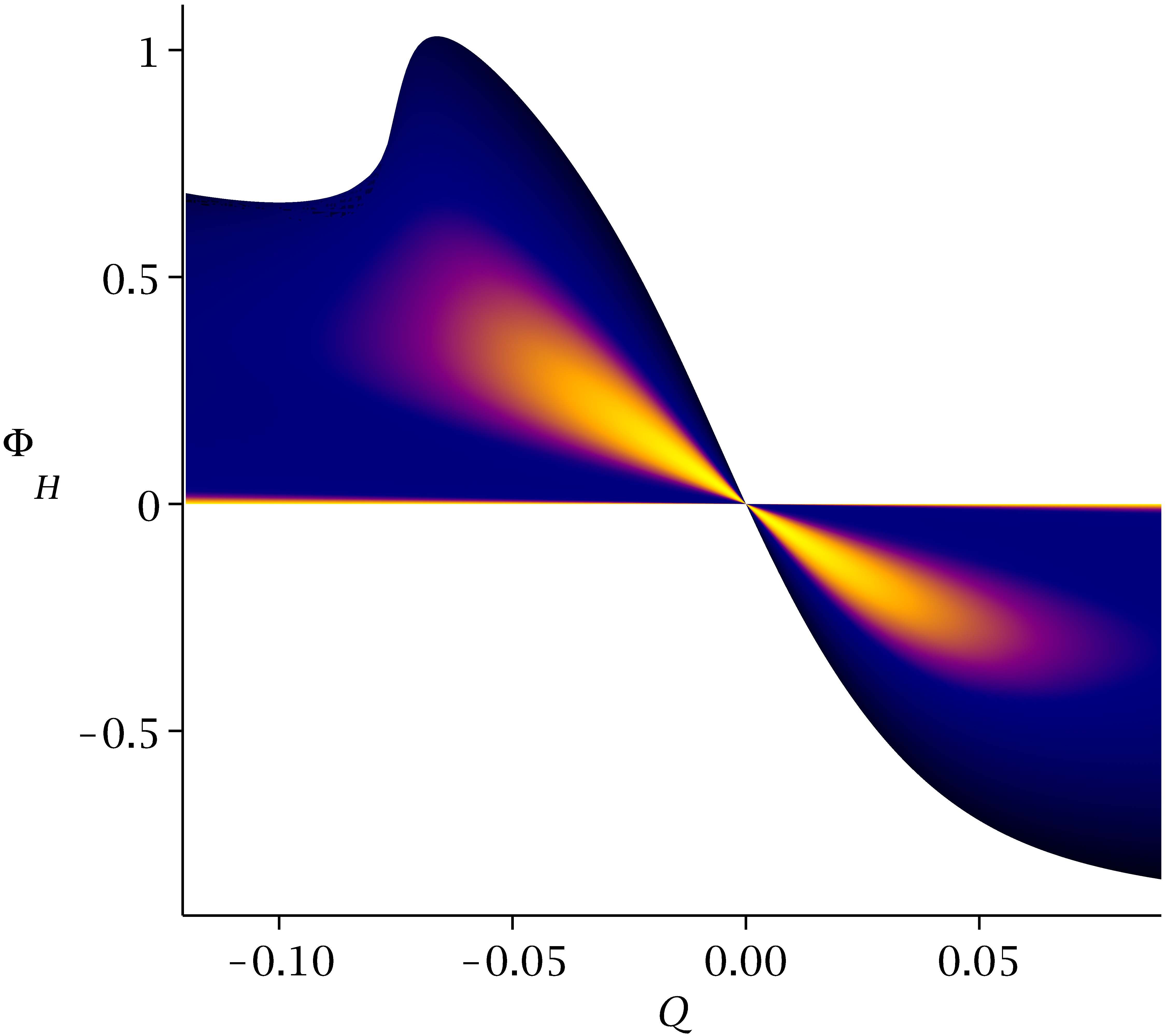}
        \caption{$\lambda = 1.5$}
        \label{fig:Phi_1.5_J0}
    \end{subfigure}
		
    \caption{Electrostatic potential $\Phi_H$ $vs.$ electric charge $Q$ with different temperatures $T_H$ for black holes with fixed angular momentum $J=0.00296$ and $L=1$, for $\lambda = 0.5$ (a), SUGRA $\lambda = 1$ (b), and $\lambda = 1.5$ (c). The behavior of the electrostatic potential is similar to the angular velocity, with the main differences being in the extremal and near extremal situation.}
		
		\label{fig:Phi_J0}
		
\end{figure}

Finally, in Figure \ref{fig:Phi_J0} we show the electrostatic potential $\Phi_H$ 
vs. the electric charge $Q$ and the temperature $T_H$.
In this case, the boundary of the domain of existence is characterized by the set of extremal solutions, 
and the line $\Phi=0$, which is reached as we move away from extremality by increasing the mass and the temperature.

Similarly to what happens in the angular velocity, changing the coupling $\lambda$ has a specially relevant effect on the $Q<Q_0$ solutions. While in the SUGRA case the electrostatic potential is discontinuous and jumps to a larger value when moving from $Q>Q_0$ to $Q<Q_0$, our calculations show that in non-SUGRA the behavior is softened around $Q=Q_0$.

\medskip

Let us now summarize the main features of these configurations with fixed (positive) total angular momentum. 
\begin{itemize}
\item For any $\lambda$, there is an asymmetry between solutions with positive and negative electric charges. 
For positive CS couplings, the $Q<0$ set possesses a \textit{critical} solution with charge $Q_0$, 
which separates two different extremal branches. 
This feature is absent for $Q>0$. 
The \textit{critical} solution has $A_H=M_H=J_H=0$, 
and is approached with a discontinuity in both $\Omega_H$ and  $\Phi_H$.
An analytical understanding of this behavior is given in Appendix A2,  for $\lambda=1$. 
Also, note that the value of $|Q_0|$ increases as $\lambda$ becomes larger.
\item Extremal solutions with $Q <Q_0$ present very different horizon properties depending on the value of the CS coupling. 
For instance, in SUGRA this branch possesses $M_H=\Omega_H=0$, and separates the $\lambda<1$ case, 
with $M_H<0$, $\Omega_H>0$, from the $\lambda>1$ case, where the sign changes to $M_H>0$, $\Omega_H<0$.
\item Non-extremal black holes with $Q<Q_0$ can possess a counter-rotating horizon ($J_H<0$), but the size of this set of solutions contracts when decreasing $\lambda$.
\item Black holes with $\Omega_H<0$ can be found for $\lambda>1$. In particular, and since $\Omega_H$ is discontinuous in the extremal branch, around the \textit{critical} solution one can see that small changes in the electric charge causes large variations of the horizon angular velocity, even changing the direction of rotation. On the contrary, for $\lambda<1$ the angular velocity is always in the direction of the angular momentum.
\end{itemize}

\subsubsection{A fixed electric charge: the generic picture}

\begin{figure}
    \centering
		
   \begin{subfigure}[b]{0.25\textwidth}
				\centering
        \includegraphics[width=30mm,scale=0.5]{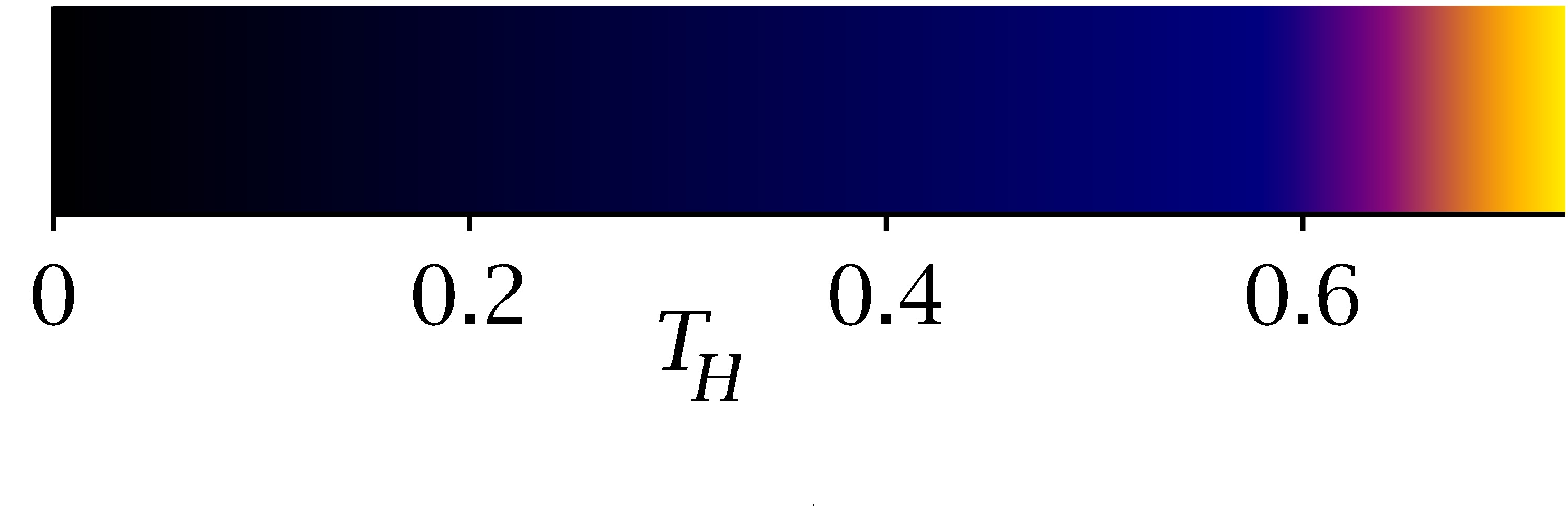}
        \label{fig:T_1.5_Q0}
    \end{subfigure}		
		
    \begin{subfigure}[b]{0.3\textwidth}
        \includegraphics[width=50mm,scale=0.5]{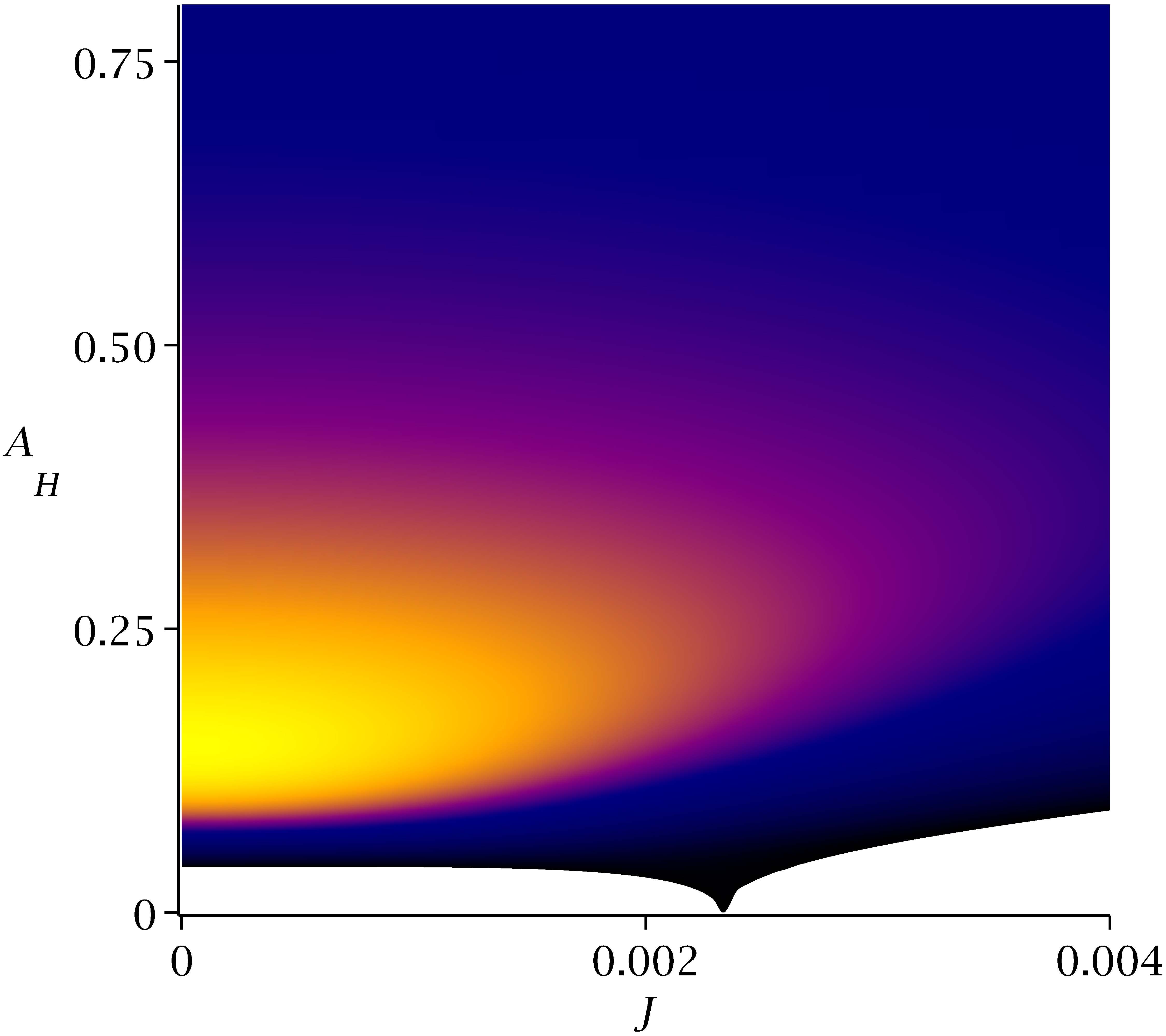}
        \caption{$\lambda = 0.5$}
        \label{fig:Ah_0.5_Q0}
    \end{subfigure}
    \begin{subfigure}[b]{0.3\textwidth}
        \includegraphics[width=50mm,scale=0.5]{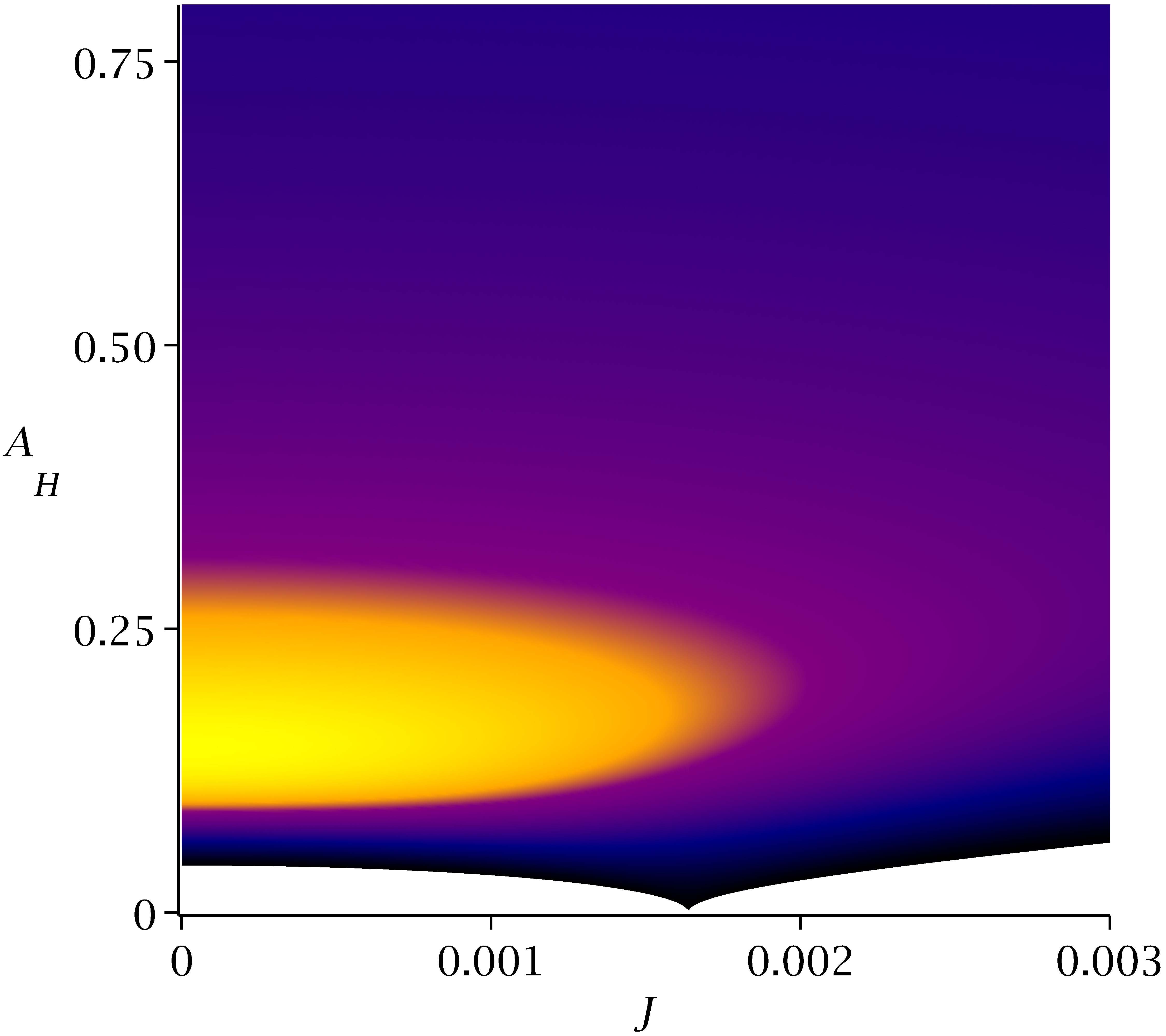}
        \caption{SUGRA}
        \label{fig:Ah_1.0_Q0}
    \end{subfigure}
    \begin{subfigure}[b]{0.3\textwidth}
        \includegraphics[width=50mm,scale=0.5]{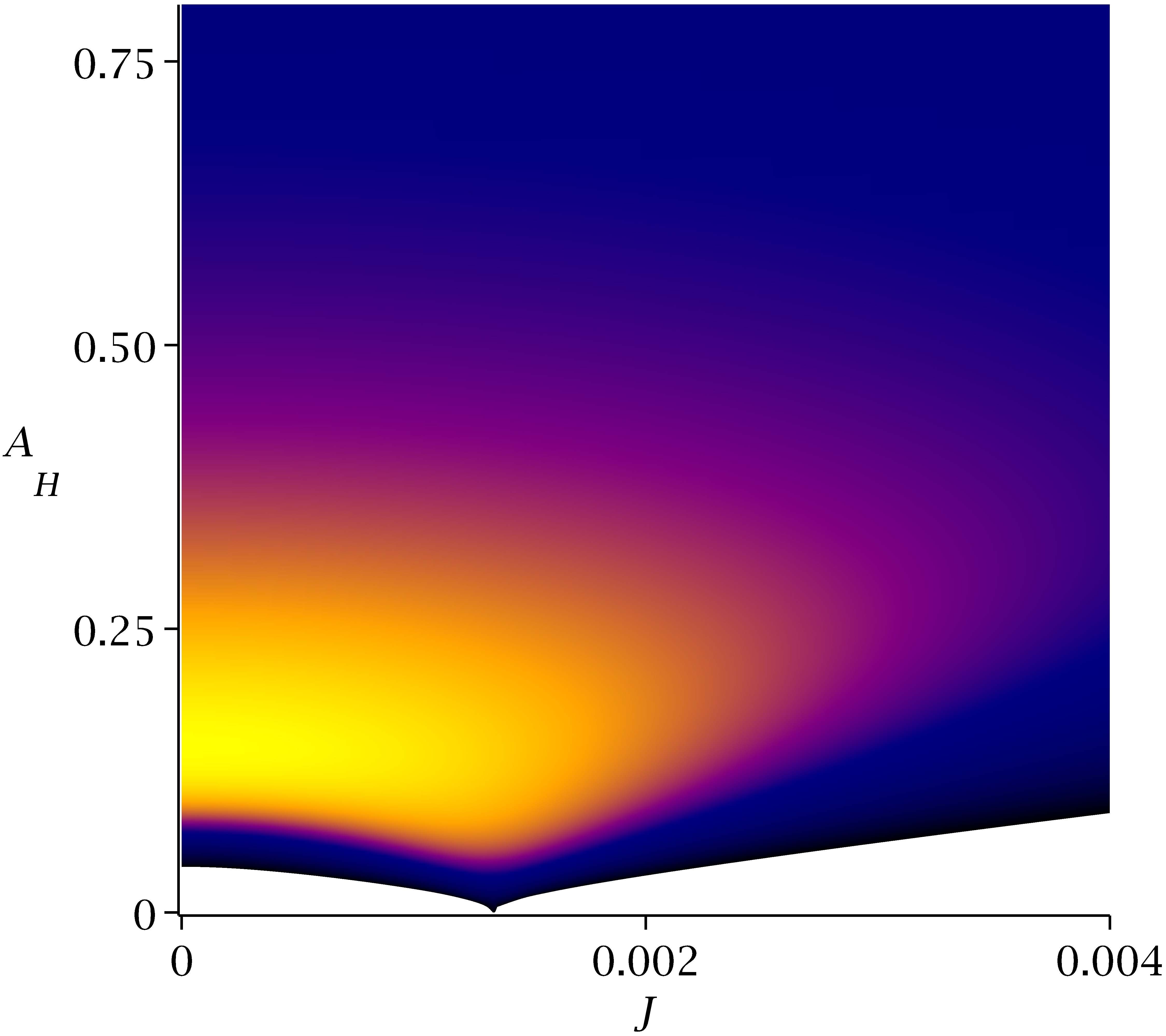}
        \caption{$\lambda = 1.5$}
        \label{fig:Ah_1.5_Q0}
    \end{subfigure}
		
    \caption{Horizon area $A_H$ $vs.$ angular momentum $J$ with different temperatures $T_H$ for black holes with fixed electric charge $Q=-0.044$ and $L=1$, for $\lambda = 0.5$ (a), SUGRA $\lambda = 1$ (b), and $\lambda = 1.5$ (c). The lower bound is given by the set of extremal solutions. The \textit{critical} solution with $A_H=0$ is found at $J=J_0$, where $J_0=0.00233,0.00164,0.00133$ for $\lambda=0.5$, SUGRA and $\lambda=1.5$, respectively. 
}
		
		\label{fig:Ah_Q0}
		
\end{figure}

\begin{figure}
    \centering
		
   \begin{subfigure}[b]{0.25\textwidth}
				\centering
        \includegraphics[width=30mm,scale=0.5]{{T_1.5_Q0}.jpg}
        
    \end{subfigure}		
		
    \begin{subfigure}[b]{0.3\textwidth}
        \includegraphics[width=50mm,scale=0.5]{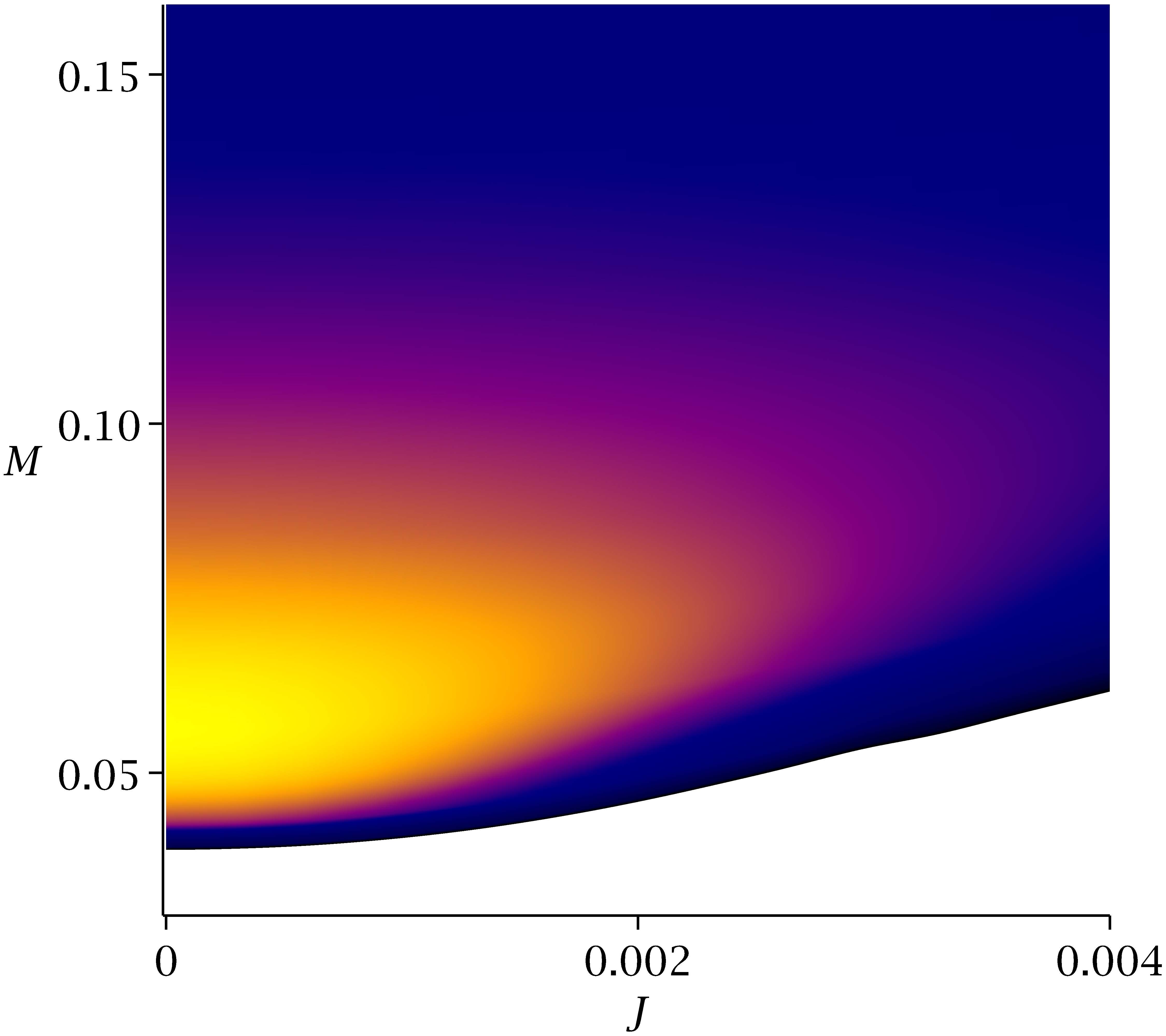}
        \caption{$\lambda = 0.5$}
        \label{fig:M_0.5_Q0}
    \end{subfigure}
    \begin{subfigure}[b]{0.3\textwidth}
        \includegraphics[width=50mm,scale=0.5]{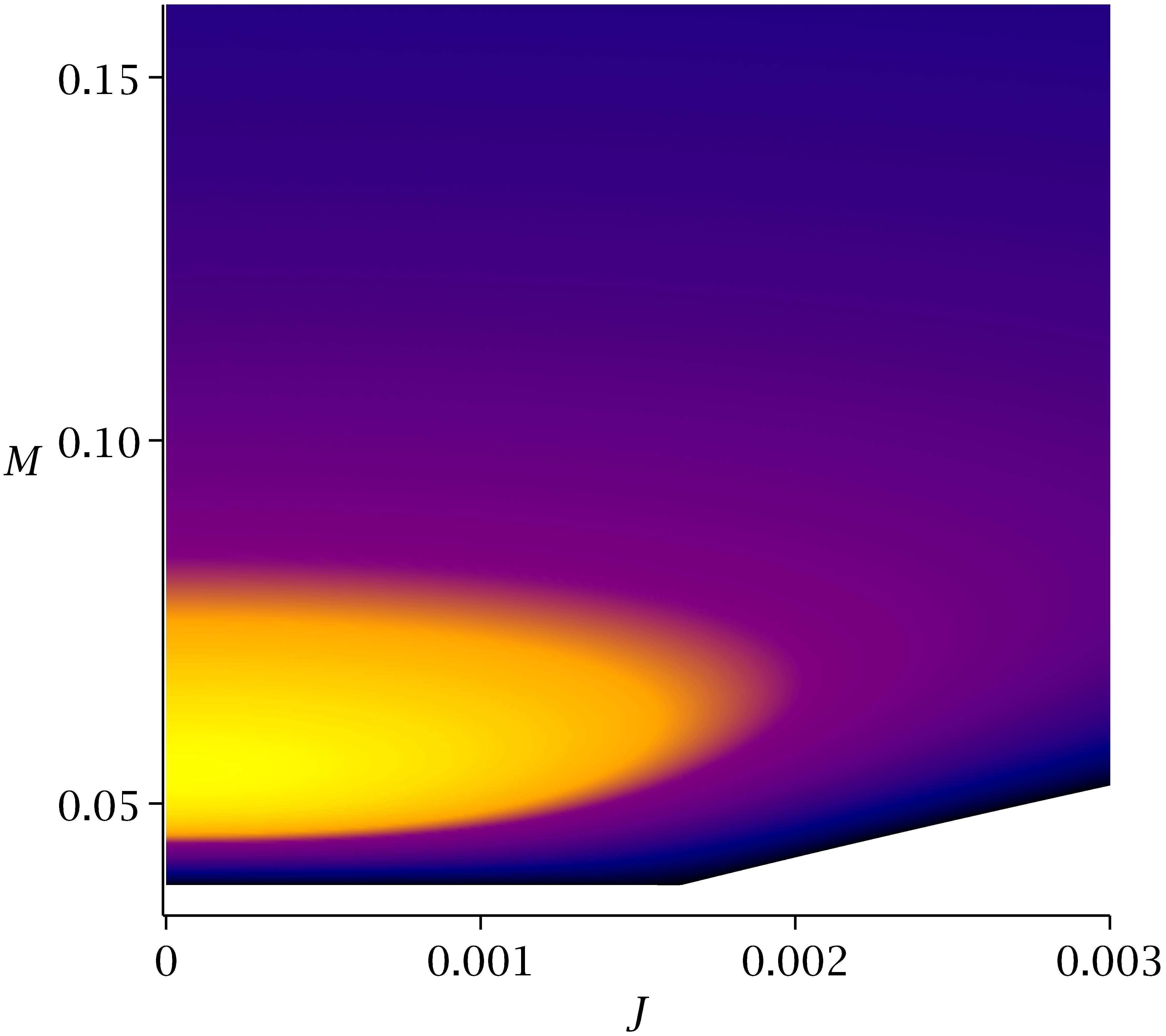}
        \caption{SUGRA}
        \label{fig:M_1.0_Q0}
    \end{subfigure}
    \begin{subfigure}[b]{0.3\textwidth}
        \includegraphics[width=50mm,scale=0.5]{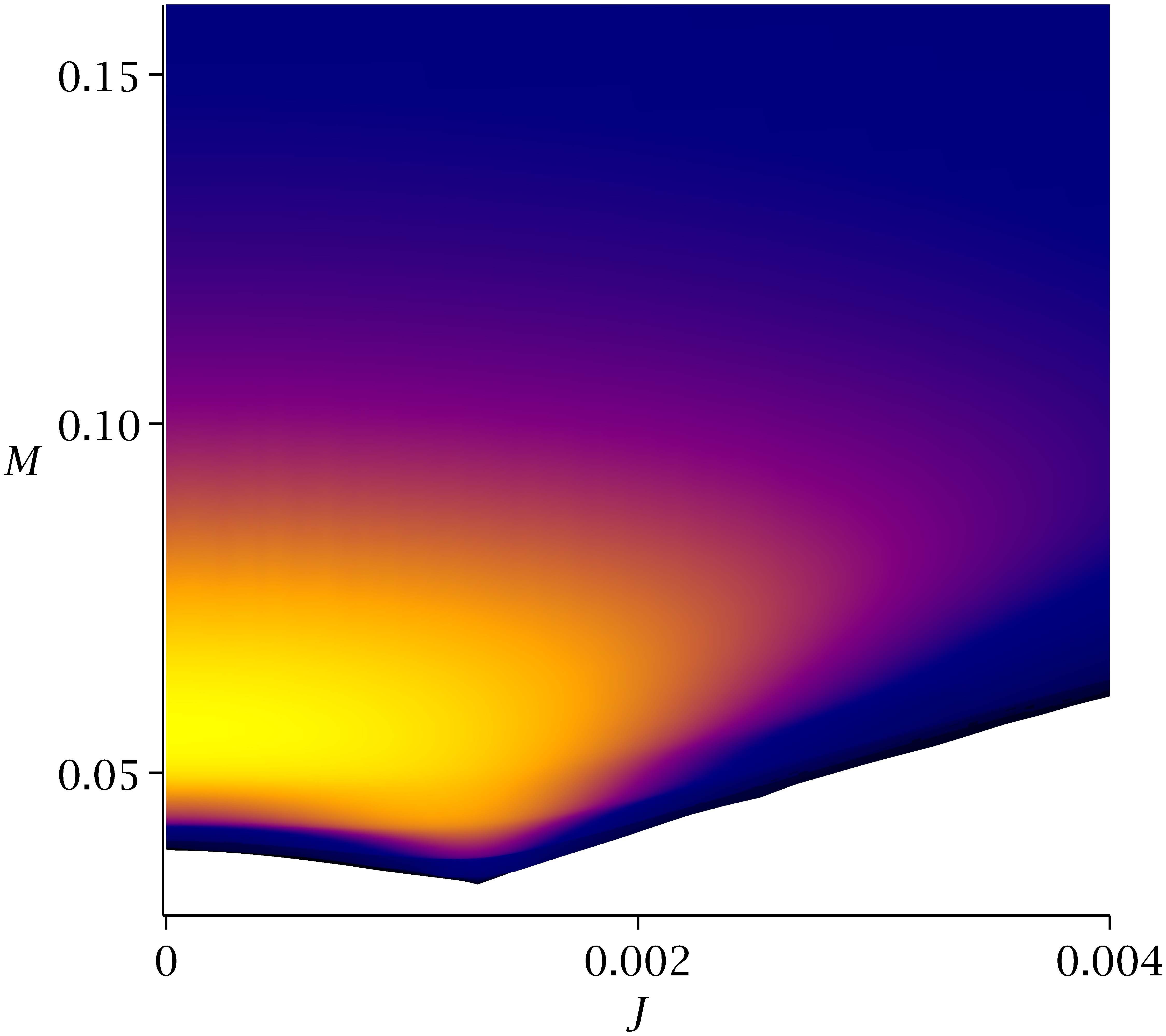}
        \caption{$\lambda = 1.5$}
        \label{fig:M_1.5_Q0}
    \end{subfigure}
		
    \caption{Total mass $M$ $vs.$ angular momentum $J$ with different temperatures $T_H$ for black holes with fixed electric charge $Q=-0.044$ and $L=1$, for $\lambda = 0.5$ (a), SUGRA $\lambda = 1$ (b), and $\lambda = 1.5$ (c). Note that the properties of the extremal set (lower boundary) depend on the CS coupling. When $\lambda \geq 1$, the extremal RN solution is no longer the black hole with minimum mass.}
		
		\label{fig:M_Q0}
		
\end{figure}

\begin{figure}
    \centering
		
   \begin{subfigure}[b]{0.25\textwidth}
				\centering
        \includegraphics[width=30mm,scale=0.5]{{T_1.5_Q0}.jpg}
        
    \end{subfigure}		
		
    \begin{subfigure}[b]{0.3\textwidth}
        \includegraphics[width=50mm,scale=0.5]{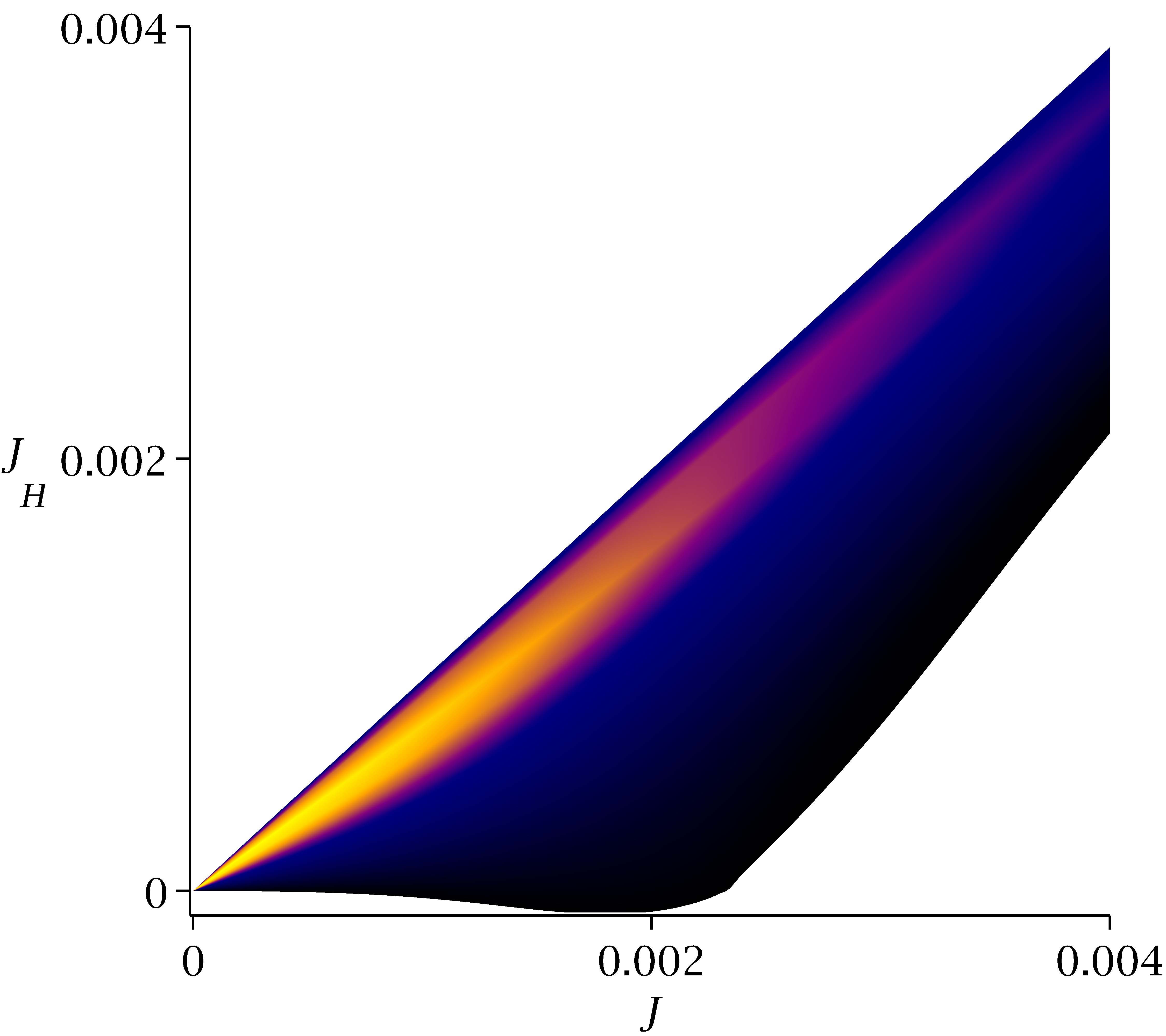}
        \caption{$\lambda = 0.5$}
        \label{fig:Jh_0.5_Q0}
    \end{subfigure}
    \begin{subfigure}[b]{0.3\textwidth}
        \includegraphics[width=50mm,scale=0.5]{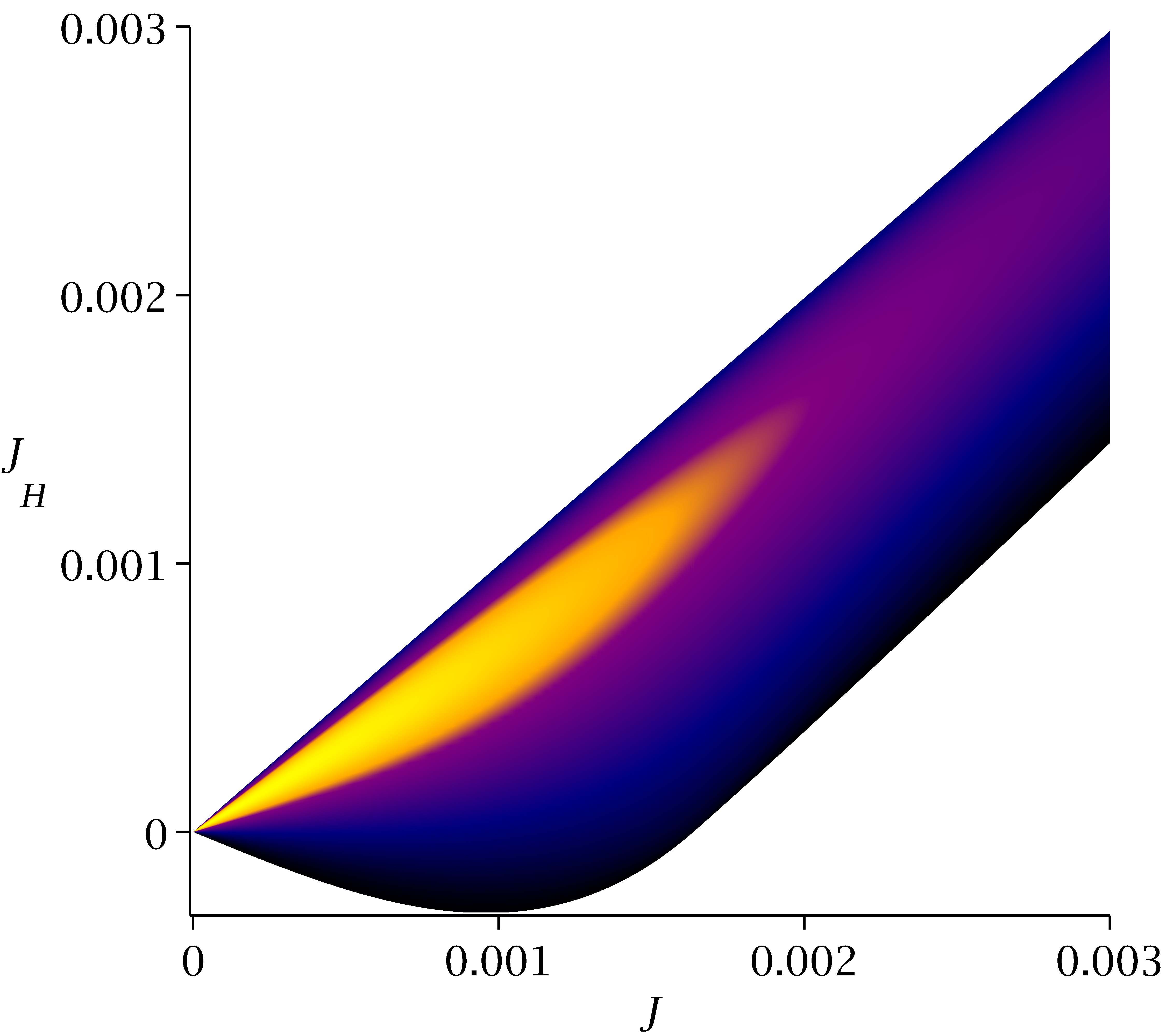}
        \caption{SUGRA}
        \label{fig:Jh_1.0_Q0}
    \end{subfigure}
    \begin{subfigure}[b]{0.3\textwidth}
        \includegraphics[width=50mm,scale=0.5]{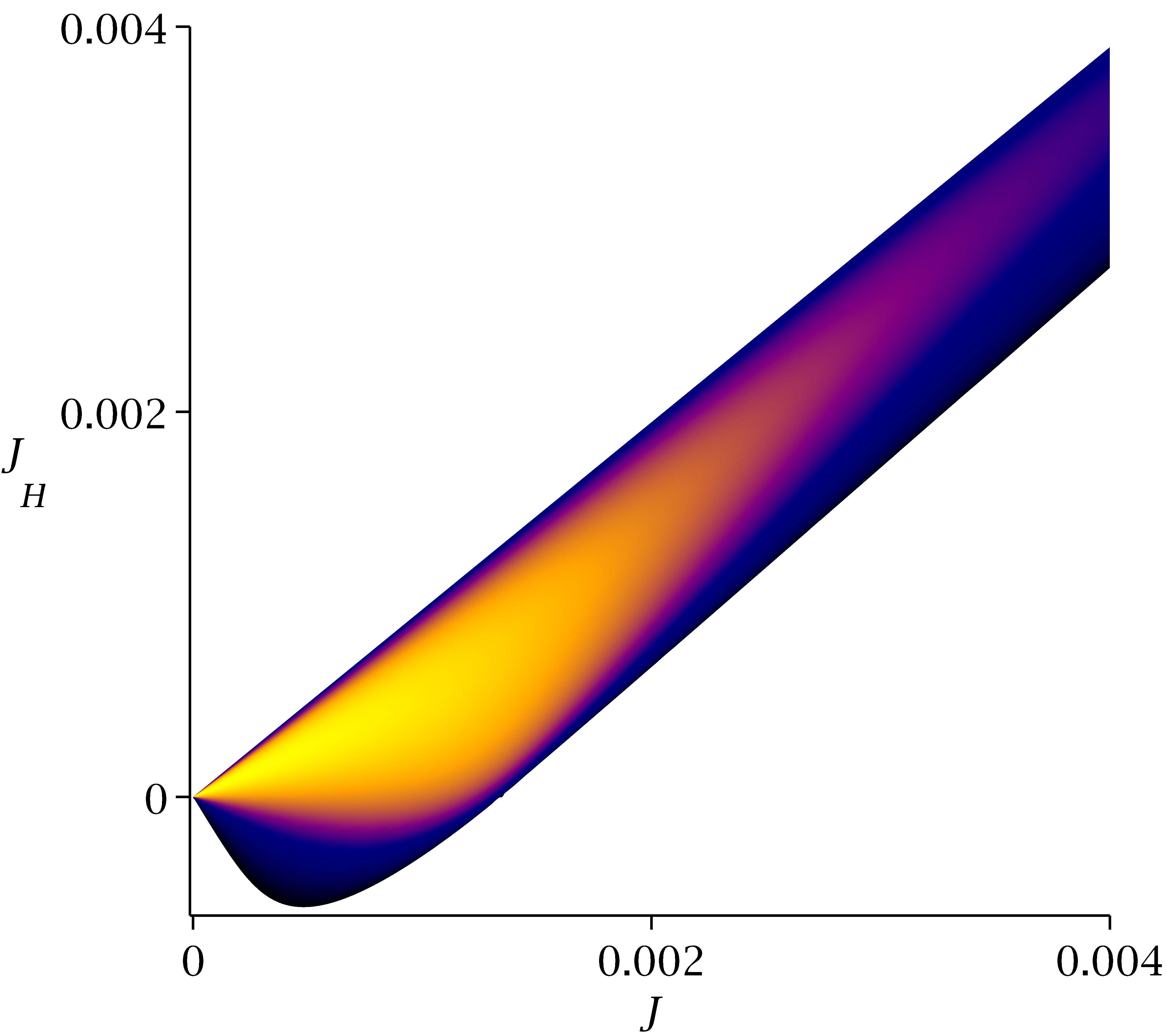}
        \caption{$\lambda = 1.5$}
        \label{fig:Jh_1.5_Q0}
    \end{subfigure}
		
    \caption{Horizon angular momentum $J_H$ $vs.$ angular momentum $J$ with different temperatures $T_H$ for black holes with fixed electric charge $Q=-0.044$ and $L=1$, for $\lambda = 0.5$ (a), SUGRA $\lambda = 1$ (b), and $\lambda = 1.5$ (c). Note that we can find extremal and non-extremal configurations with $J_H<0$. Extremal black holes on the MP branch approximately satisfy ($J_H \approx J - J_0$).}
		
		\label{fig:Jh_Q0}
		
\end{figure}

\begin{figure}
    \centering
		
   \begin{subfigure}[b]{0.25\textwidth}
				\centering
        \includegraphics[width=30mm,scale=0.5]{{T_1.5_Q0}.jpg}
        
    \end{subfigure}		
		
    \begin{subfigure}[b]{0.3\textwidth}
        \includegraphics[width=50mm,scale=0.5]{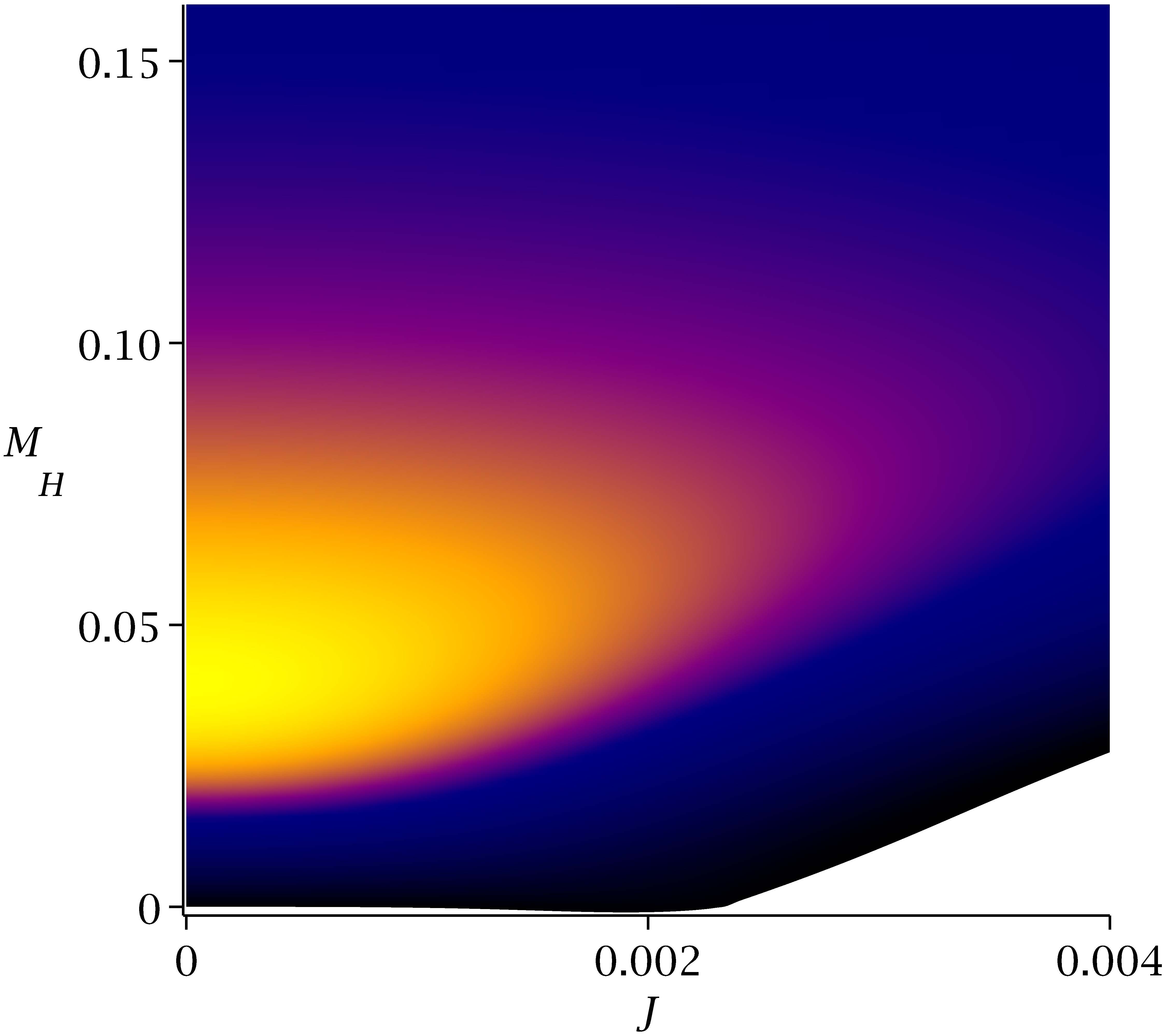}
        \caption{$\lambda = 0.5$}
        \label{fig:Mh_0.5_Q0}
    \end{subfigure}
    \begin{subfigure}[b]{0.3\textwidth}
        \includegraphics[width=50mm,scale=0.5]{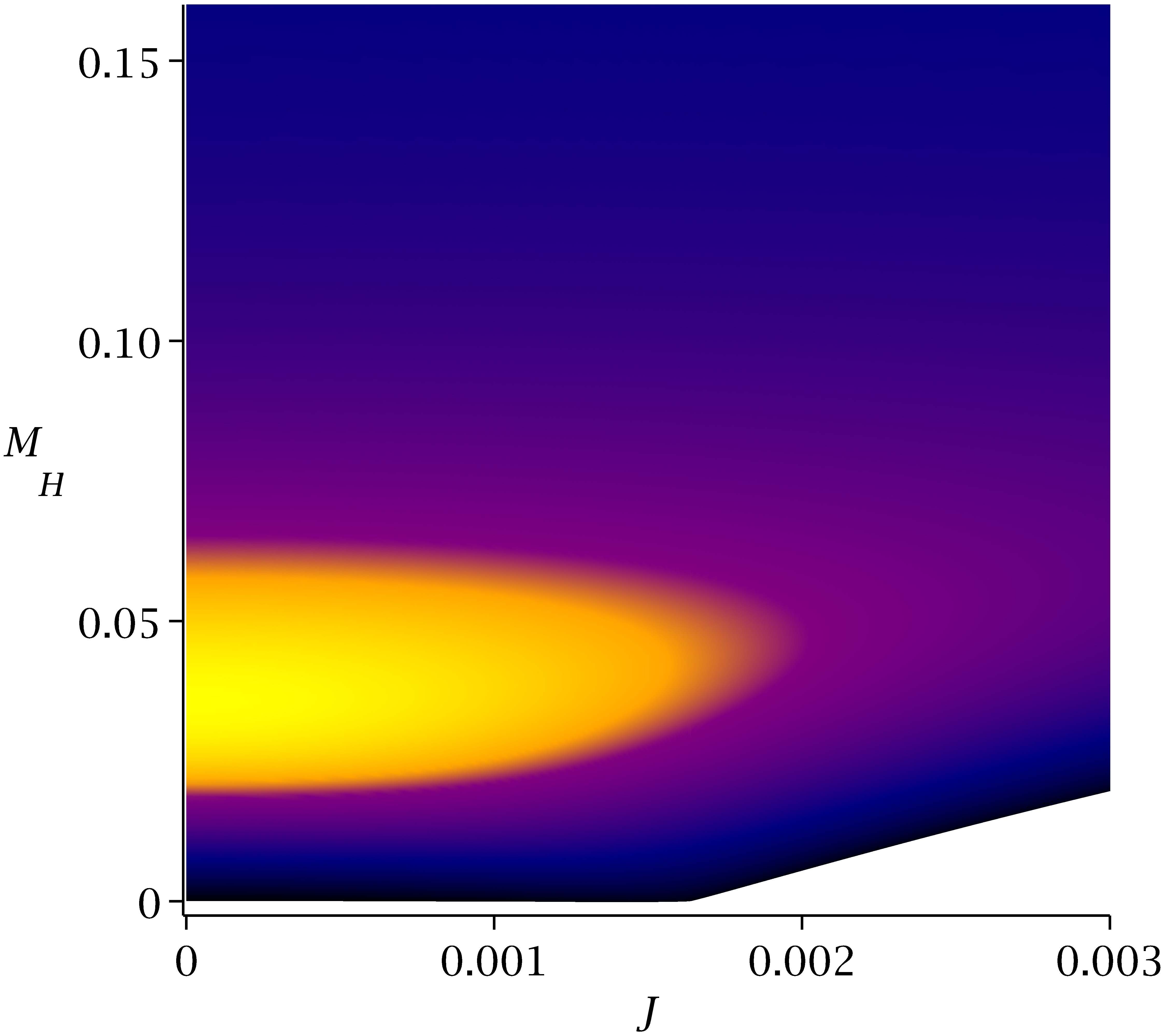}
        \caption{SUGRA}
        \label{fig:Mh_1.0_Q0}
    \end{subfigure}
    \begin{subfigure}[b]{0.3\textwidth}
        \includegraphics[width=50mm,scale=0.5]{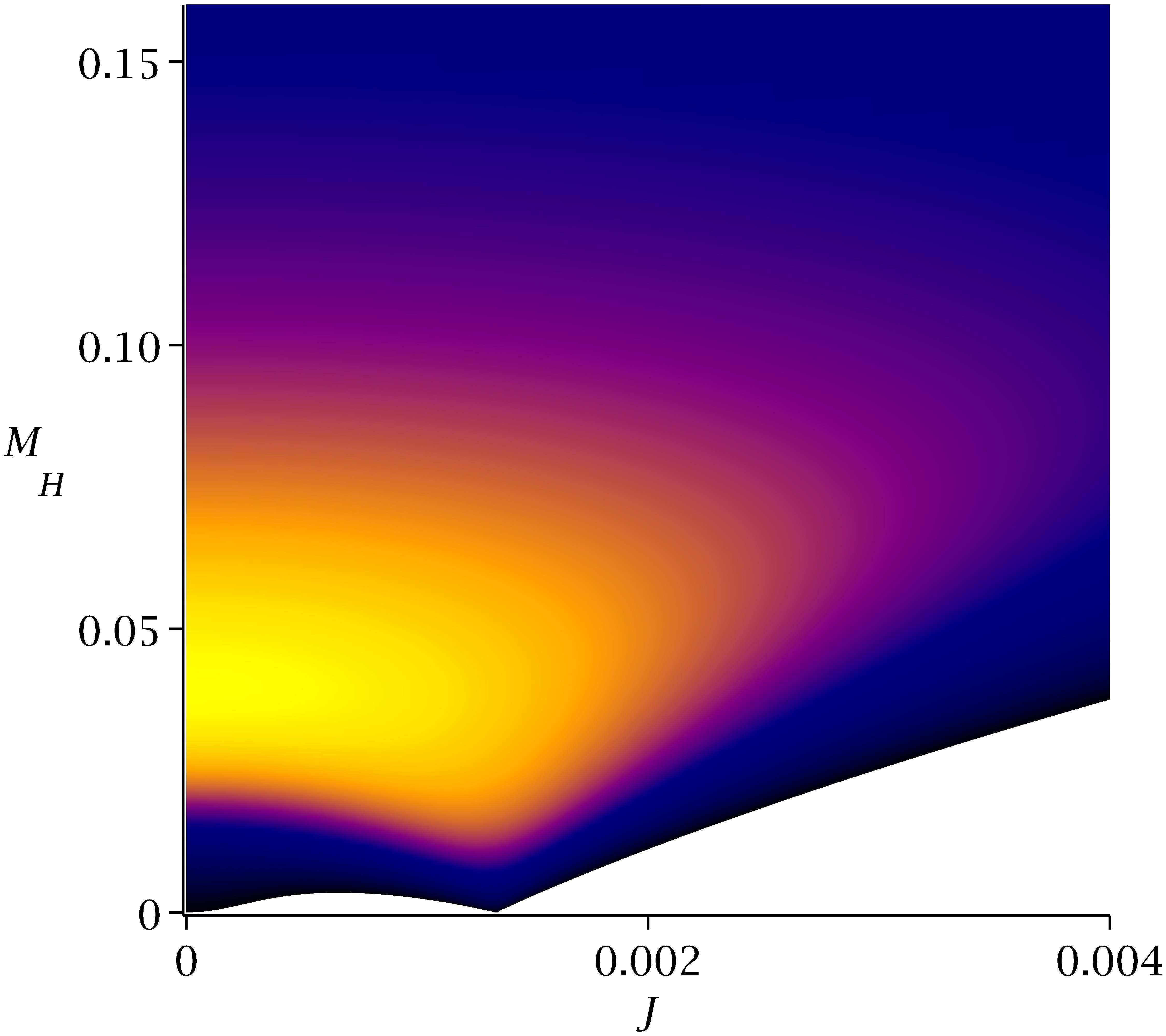}
        \caption{$\lambda = 1.5$}
        \label{fig:Mh_1.5_Q0}
    \end{subfigure}
		
    \caption{Horizon mass $M_H$ $vs.$ angular momentum $J$ with different temperatures $T_H$ for black holes with fixed electric charge $Q=-0.044$ and $L=1$, for $\lambda = 0.5$ (a), SUGRA $\lambda = 1$ (b), and $\lambda = 1.5$ (c). Note SUGRA is a particular case in which the RN-branch has always $M_H=0$. This feature is lost when changing the value of the coupling.}
		
		\label{fig:Mh_Q0}
		
\end{figure}

\begin{figure}
    \centering
		
   \begin{subfigure}[b]{0.25\textwidth}
				\centering
        \includegraphics[width=30mm,scale=0.5]{{T_1.5_Q0}.jpg}
        
    \end{subfigure}		
		
    \begin{subfigure}[b]{0.3\textwidth}
        \includegraphics[width=50mm,scale=0.5]{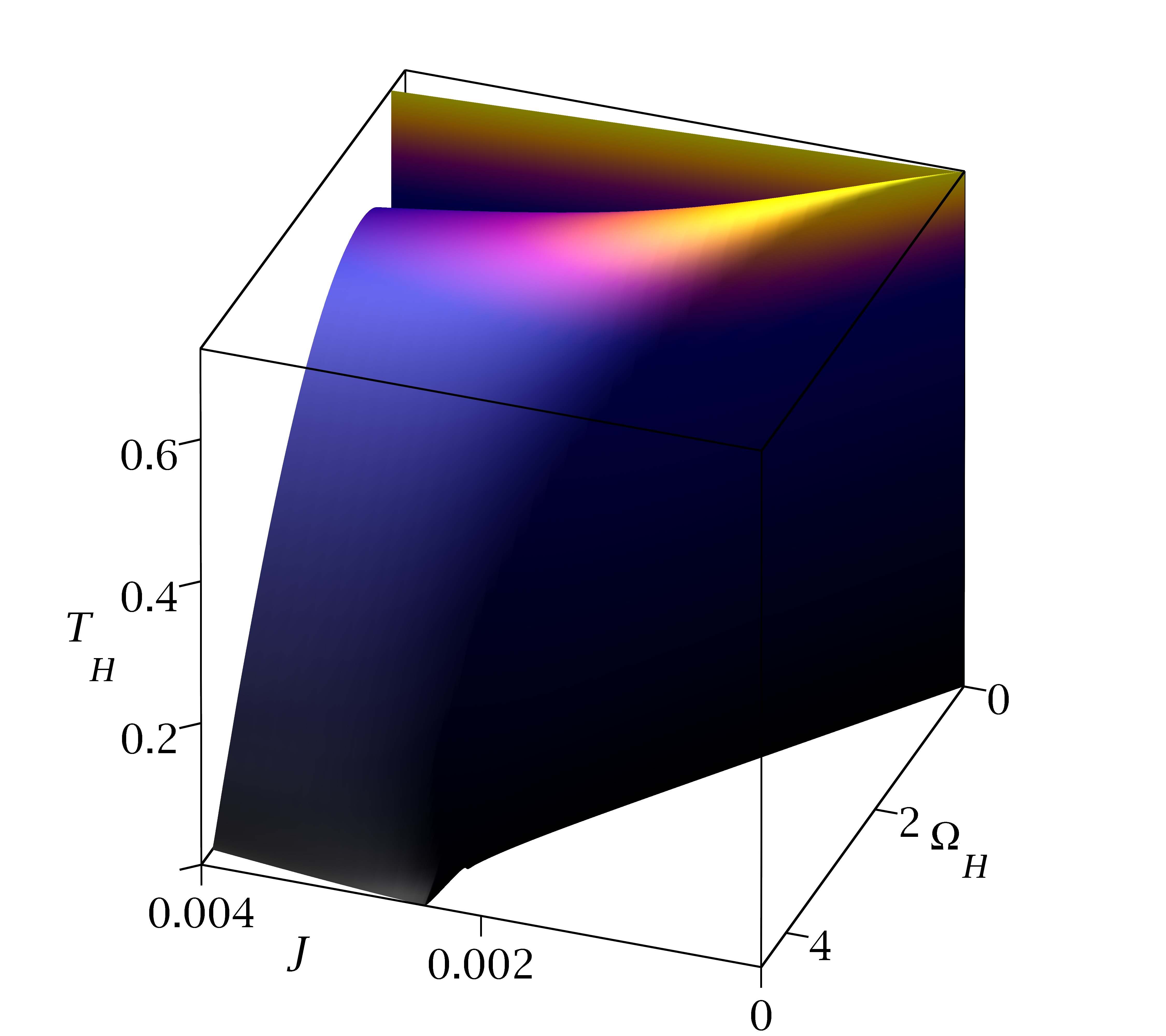}
        \caption{$\lambda = 0.5$}
        \label{fig:Omh_0.5_Q0}
    \end{subfigure}
    \begin{subfigure}[b]{0.3\textwidth}
        \includegraphics[width=50mm,scale=0.5]{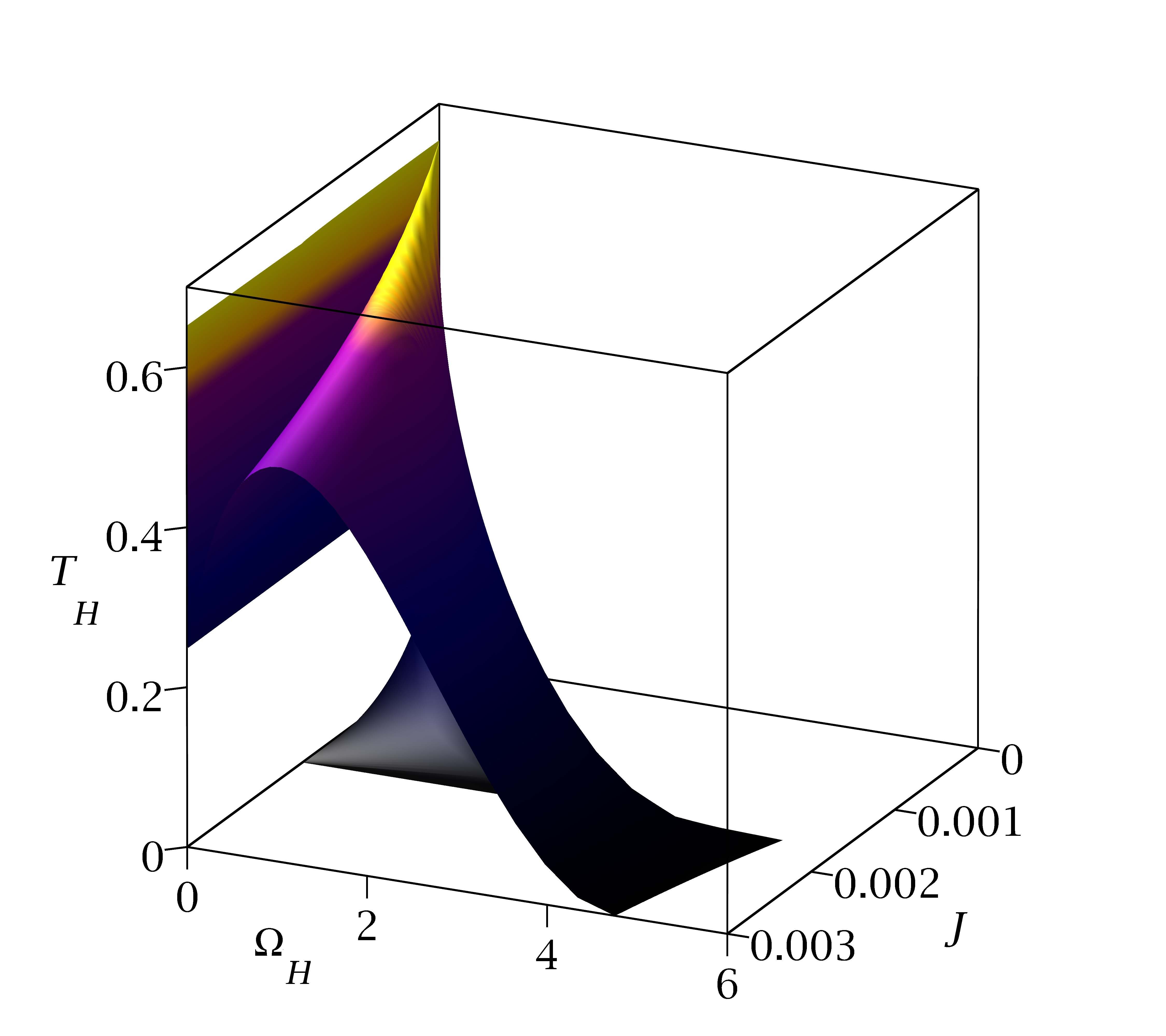}
        \caption{SUGRA}
        \label{fig:Omh_1.0_Q0}
    \end{subfigure}
    \begin{subfigure}[b]{0.3\textwidth}
        \includegraphics[width=50mm,scale=0.5]{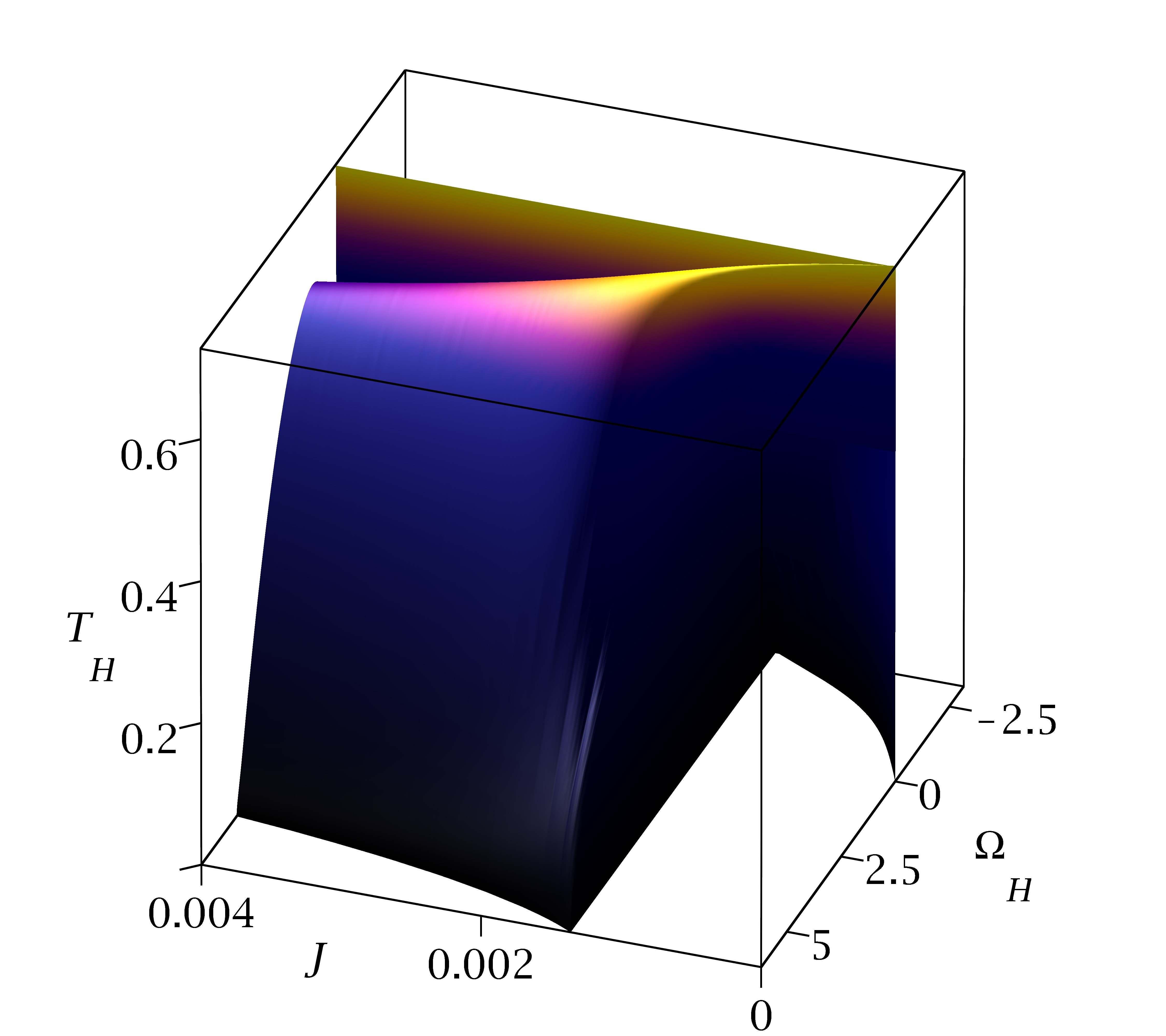}
        \caption{$\lambda = 1.5$}
        \label{fig:Omh_1.5_Q0}
    \end{subfigure}
		
    \caption{Horizon angular velocity $\Omega_H$ $vs.$ angular momentum $J$ with different temperatures $T_H$ for black holes with fixed electric charge $Q=-0.044$ and $L=1$, for $\lambda = 0.5$ (a), SUGRA $\lambda = 1$ (b), and $\lambda = 1.5$ (c). The extremal limit has very different properties in each case. Note that $\Omega_H$ for extremal solutions in $\lambda = 0.5$ (a) is continuous, but discontinuous in the other two cases. For SUGRA $\lambda = 1$ (b), and $\lambda = 1.5$ (c), the RN branch has $\Omega_H=0$ and $\Omega_H<0$, respectively, and the MP branch has always positive $\Omega_H$. This is equivalent to Figure \ref{fig:Omh_J0}.}
		
		\label{fig:Omh_Q0}
		
\end{figure}

\begin{figure}
    \centering
		
   \begin{subfigure}[b]{0.25\textwidth}
				\centering
        \includegraphics[width=30mm,scale=0.5]{{T_1.5_Q0}.jpg}
        
    \end{subfigure}		
		
    \begin{subfigure}[b]{0.3\textwidth}
        \includegraphics[width=50mm,scale=0.5]{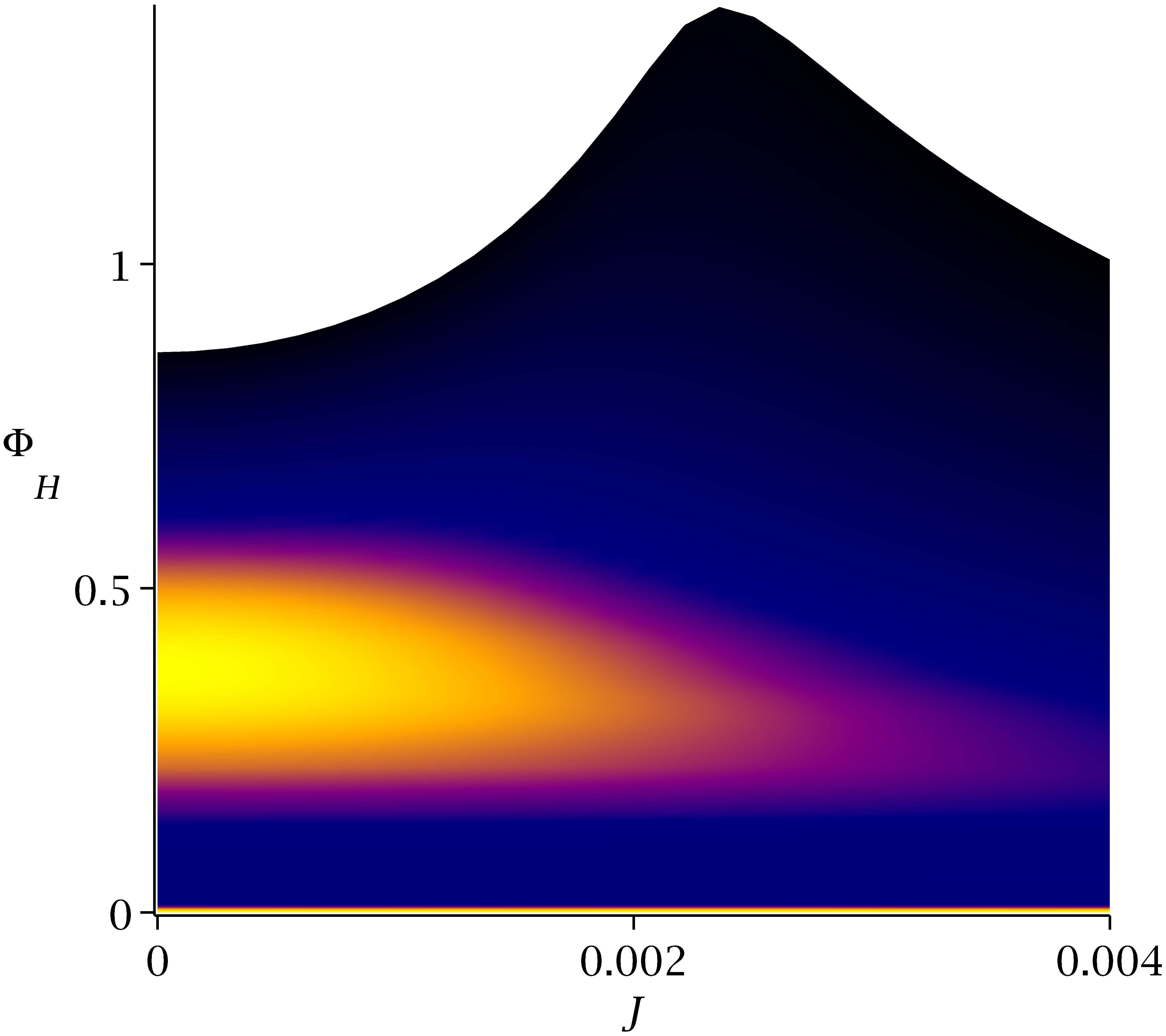}
        \caption{$\lambda = 0.5$}
        \label{fig:Phi_0.5_Q0}
    \end{subfigure}
    \begin{subfigure}[b]{0.3\textwidth}
        \includegraphics[width=50mm,scale=0.5]{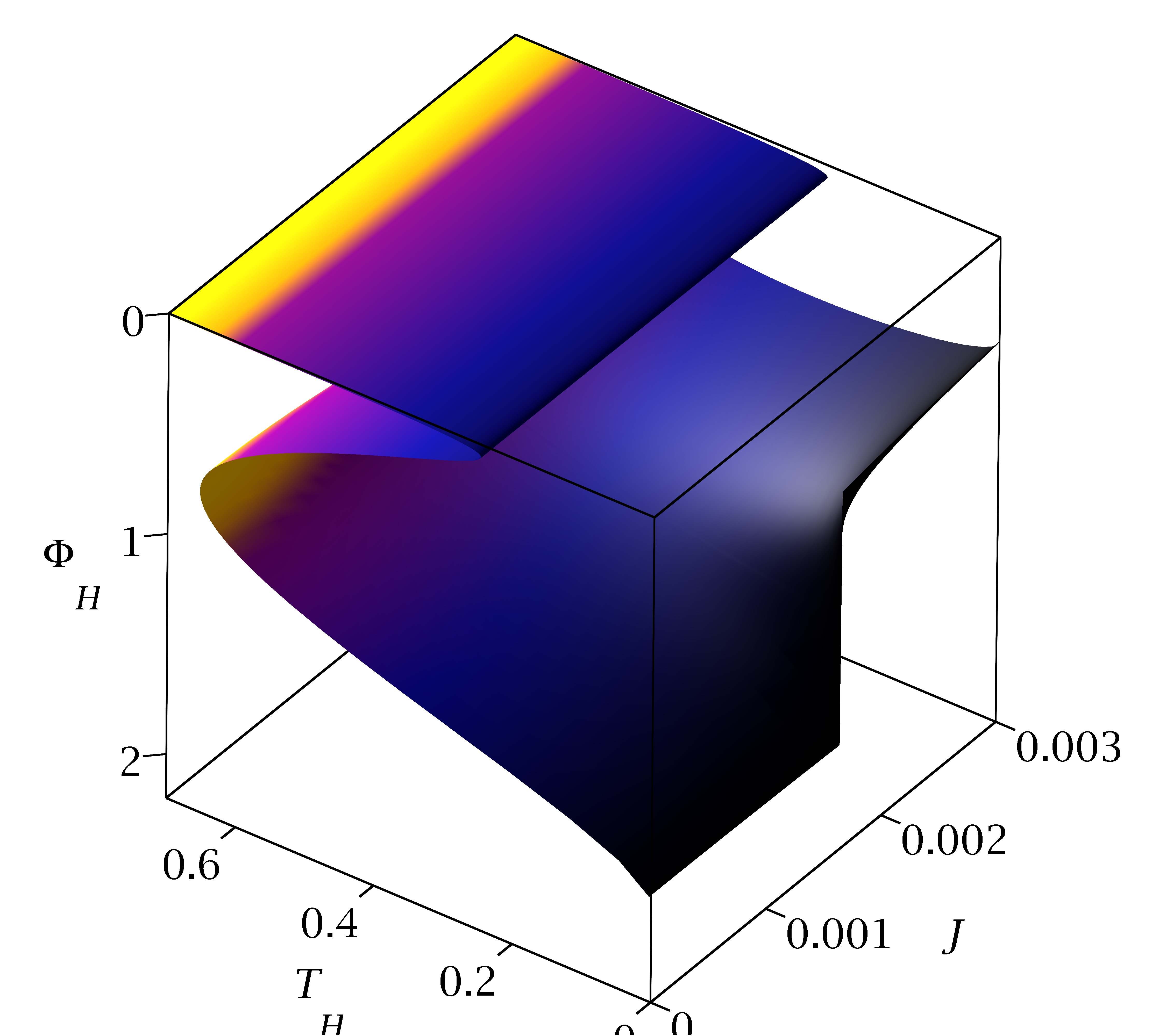}
        \caption{SUGRA}
        \label{fig:Phi_1.0_Q0}
    \end{subfigure}
    \begin{subfigure}[b]{0.3\textwidth}
        \includegraphics[width=50mm,scale=0.5]{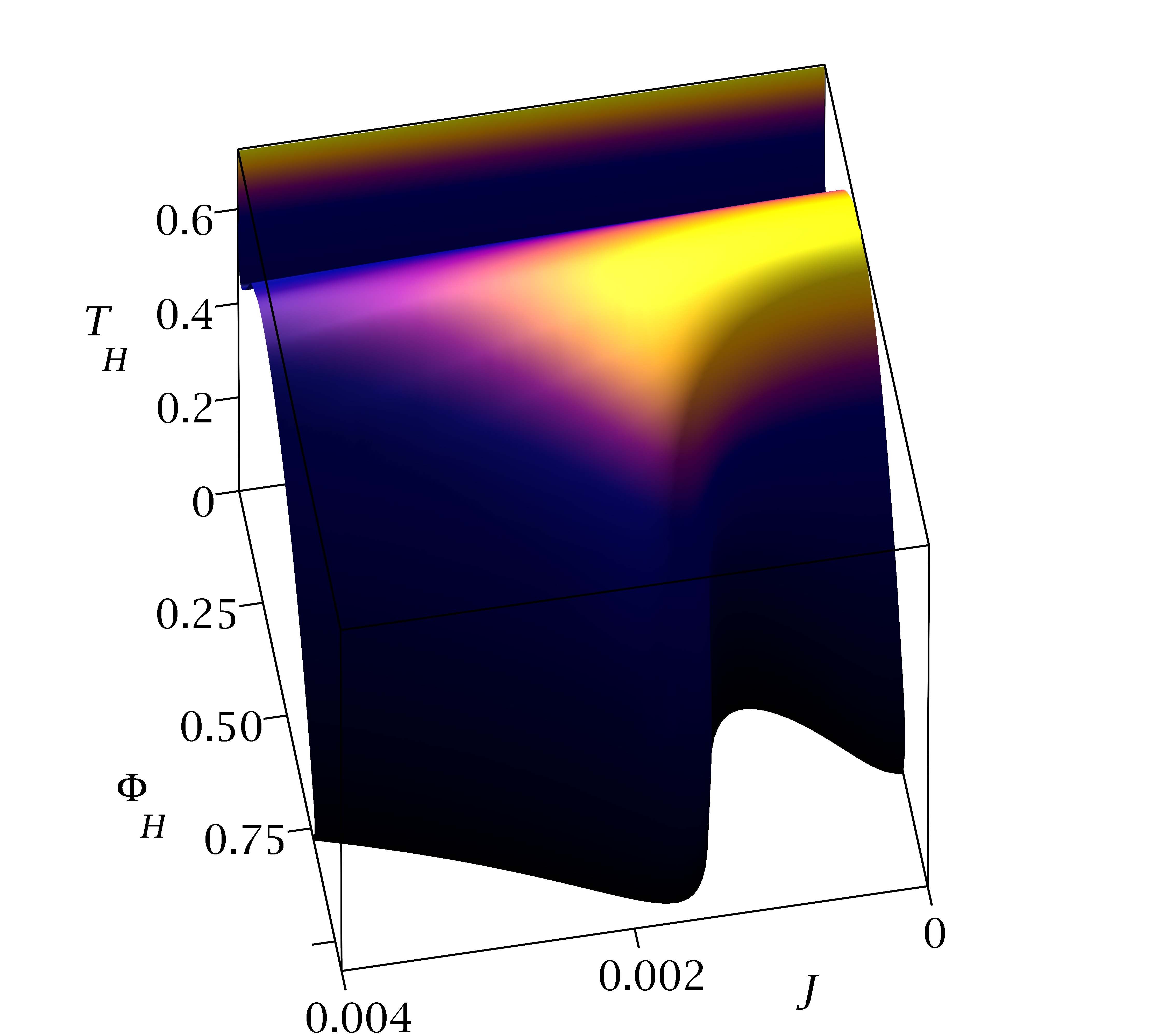}
        \caption{$\lambda = 1.5$}
        \label{fig:Phi_1.5_Q0}
    \end{subfigure}
		
    \caption{Electrostatic potential $\Phi_H$ $vs.$ angular momentum $J$ with different temperatures $T_H$ for black holes with fixed electric charge $Q=-0.044$ and $L=1$, for $\lambda = 0.5$ (a), SUGRA $\lambda = 1$ (b), and $\lambda = 1.0$ (c).}
		
		\label{fig:Phi_Q0}
		
\end{figure}

A complementary picture is found when fixing the electric charge  
and varying both $J$ and $T_H$. 
The counterparts of the plots in the previous Subsection 
are shown in Figures  \ref{fig:Ah_Q0}-\ref{fig:Phi_Q0},
for a fixed electric charge $Q=-0.044$ and  
 the same values of $\lambda$.

Again, the global extremal solutions  
possess two different branches: the MP branch and the RN branch,
which agree 
 with the prediction from the near-horizon formalism.
The RN branch contains the static configuration with $J=0$, and extends for $J\in(-J_0,J_0)$. 
At $J=\pm J_0$ we find the \textit{critical} solution with $A_H=0$.
The MP branch connects with the uncharged and rotating black hole, and is found for $|J|>|J_0|$.
In what follows, we shall present results for positive values of the angular momentum only, 
since all properties are symmetric under a change of sign in $J$.

We start with
the  Figure \ref{fig:Ah_Q0},
where we show the $(A_H,J;T_H)$  diagram. 
The lower boundary of the $A_H$-domain is found for the extremal solutions.
 Hence, one can say that the $T_H=0$ configurations possess the lowest possible horizon area, 
like in the constant $J$ case. 
At $J=0$ we find the set of RN-AdS BHs. 
Note, however, that the extremal RN-AdS solutions do not have the minimum possible entropy. 
Spinning up an extremal RN-AdS BH while keeping $T_H=0$ makes the area to decrease down to zero,
a point where the \textit{critical} solutions are reached. 
These \textit{critical} solutions separate the extremal branch originating in the RN solution ($0\leq J < J_0$) 
from the extremal branch connecting with the MP solution ($J > J_0$). 
Moreover, although in the RN branch, the horizon area decreases with the angular momentum,
in the MP branch $A_H$ increases with $J$. 
An interesting consequence here, 
is that,  for given $Q$, 
it is possible to obtain  EMCS non-extremal spinning and charged BHs 
with a horizon area lower than the area of the extremal RN-AdS BH. 

Concerning the non-extremal solutions, in Figure \ref{fig:Ah_Q0} we can see that the solutions
present a local maximum of the temperature.
 This local maximum is always found for $J=0$, $T_H=0.73$. 
For higher values of the area, the temperature increases again (not displayed in these Figures). 
For instance, in the three cases there is no upper bound for the entropy. 
Also, one can see in Figure \ref{fig:Ah_Q0}  that it is possible 
to define closed sets of isothermal rotating and charged BHs.  

The main effect of changing the CS coupling is on the value of $J_0$: 
increasing $\lambda$ reduces $J_0$ and shrinks the extremal RN branch.

In Figure \ref{fig:M_Q0} we show the $(M,J;T_H)$ plot. 
The local maximum temperature is obtained for a non-extremal RN-AdS configuration  ($J=0$). 
The existence of this local maximum of temperature implies 
the existence of a set of thermally unstable configurations 
with negative thermal capacity. 
Around this set, isothermal sets are closed. 
This means that the mass-angular momentum relation is bounded. 
This is not the case for isothermal sets far away from the local maximum.

The extremal solutions present very different behavior depending on the value of $\lambda$:
For $\lambda=0.5$ (Figure \ref{fig:M_0.5_Q0}), 
we can see that extremal BHs on the RN branch ($J<J_0$) 
have total mass increasing with the angular momentum. 
This means that the minimum mass is obtained for the extremal RN-AdS BH. 
Extremal BHs along the MP branch ($J>J_0$) have total mass increasing with the angular momentum.
For $\lambda=1$ (Figure \ref{fig:M_1.0_Q0}), this is different: 
all extremal BHs on the RN branch have the same (minimal) mass. The mass only starts increasing when $J>J_0$ ($J_0=0.00164$). 
Again, extremal BHs along the MP branch have the total mass increasing with the angular momentum. 
For $\lambda=1.5$ (Figure \ref{fig:M_1.5_Q0}) this changes again, and extremal 
BHs on the RN branch ($J<J_0$) have the total mass decreasing with the angular momentum. 
But extremal BHs along the MP branch have the total mass increasing with the angular momentum. This means the minimum mass is reached at the \textit{critical} solution with $J_0=0.00133$.

This situation contrasts with the one presented in Figure \ref{fig:Ah_J0} 
for the fixed $J$ case, 
where the minimum mass was always found at the extremal uncharged solution. 
Here it clearly depends on the value of the CS coupling being below or beyond the SUGRA case.

In Figure \ref{fig:Jh_Q0} we show the diagram for $(J_H,J;T_H)$.  
One can see that the upper bound there is given by the line $J_H=J$. 
These solutions are reached as the mass of the BHs increases. 
This means that all the angular momentum of these solutions is stored behind the horizon, 
with a vanishing contribution from the gauge field. The lower bound contains the extremal solutions. 
In these three cases, the \textit{critical} solution with fixed $Q$ has $J_H=0$. 
Extremal BHs on the RN branch always present negative horizon angular momentum. 
However, extremal BHs on the MP branch have positive horizon angular momentum, 
and it is approximately given by $J_H \approx J - J_0$.

The RN branch having negative horizon angular momentum means the solutions are counter-rotating,
a feature which, in fact, is also shared by whole sets of  non-extremal BHs. 
 Even more, one can find non-extremal solutions with zero horizon angular momentum, 
but non-zero total angular momentum. 
For such solutions, the angular momentum is stored in the gauge field.

The variation of the coupling $\lambda$ has an important effect on these counter-rotating configurations: reducing the coupling below SUGRA again reduces the size of the space of solutions with counter-rotation.

In Figure \ref{fig:Mh_Q0} we present the $(M_H,J;T_H)$ plot. 
The minimal value of the horizon mass if reached for extremal solutions. 
The static extremal MP-AdS BH at $J=0$ has zero horizon mass. 
In the three cases, the horizon mass of the \textit{critical} solution is also zero. 
But the properties close to extremality of the non-static solutions depend  considerably on the value of $\lambda$.
As seen in Figure \ref{fig:Mh_0.5_Q0}, the
solutions with $J<J_0$ (extremal solutions on the RN branch or near extremal solutions close to it) 
can have negative horizon mass.
 However, one can see in Figure \ref{fig:Mh_1.0_Q0} that this is no longer the case for SUGRA BHs, 
and the extremal RN branch has always zero horizon mass.
In Figure \ref{fig:Mh_1.5_Q0}, we can see that the RN branch has positive horizon mass when $\lambda=1.5$. 
In fact it is interesting to note that the horizon mass of the extremal RN branch increases up to a maximum,
 and then it decreases again to zero.

In Figure \ref{fig:Omh_Q0} we show the $(\Omega_H,J;T_H)$  diagram. 
The differences there occur especially close to extremality.
As seen in Figure \ref{fig:Omh_0.5_Q0},
 the angular velocity of the extremal RN branch is positive, 
and matches with the angular velocity of the extremal MP branch. 
Consider now the CPL solution in Figure \ref{fig:Omh_1.0_Q0}. 
Note that there is a discontinuity in the angular velocity at zero temperature: 
the extremal BHs connecting with the RN solution have zero angular velocity. 
At the \textit{critical} solution, the angular velocity jumps up to a positive value, where the MP branch starts. 
Then the angular velocity decreases with the angular momentum.
The $\lambda=1.5$ case (Figure \ref{fig:Omh_1.5_Q0}), also presents a discontinuity in the angular velocity. 
New features occur here as well.
For example, 
 note that the horizon angular velocity of the extremal RN branch is negative. 
This means one finds counter-rotating solutions.
Such BHs can be non-extremal too. 
Also, note that the angular velocity of non-extremal solutions 
close to the \textit{critical} solution presents very steep changes with 
respect to small changes in the angular momentum.
Hence, if one perturbs the angular velocity of one of these solutions slightly, 
the angular velocity and even the direction of the rotation can change drastically.
 Nevertheless, other quantities ($e.g.$ mass and horizon area) do not change much.

In Figure \ref{fig:Phi_Q0} we show the  ($\Phi,J;T_H)$ diagram. 
Again, the properties of the extremal solutions depend on the particular value of the CS coupling $\lambda$, 
although the  features of BHs far from extremality are rather similar. 
Note that in the CPL solution, Figure \ref{fig:Phi_1.0_Q0}, on the branch connecting with the MP BH,
the electric potential depends on the angular momentum, while on the branch connecting with the RN solution, the electric potential is constant. 
In the other two cases, Figure \ref{fig:Phi_0.5_Q0} and Figure \ref{fig:Phi_1.5_Q0}, 
the particular dependence of $J$ 
on the electrostatic potential on the RN branch depends on the coupling $\lambda$.

\begin{figure}
    \centering
		
        \includegraphics[width=70mm,scale=0.5,angle=-90]{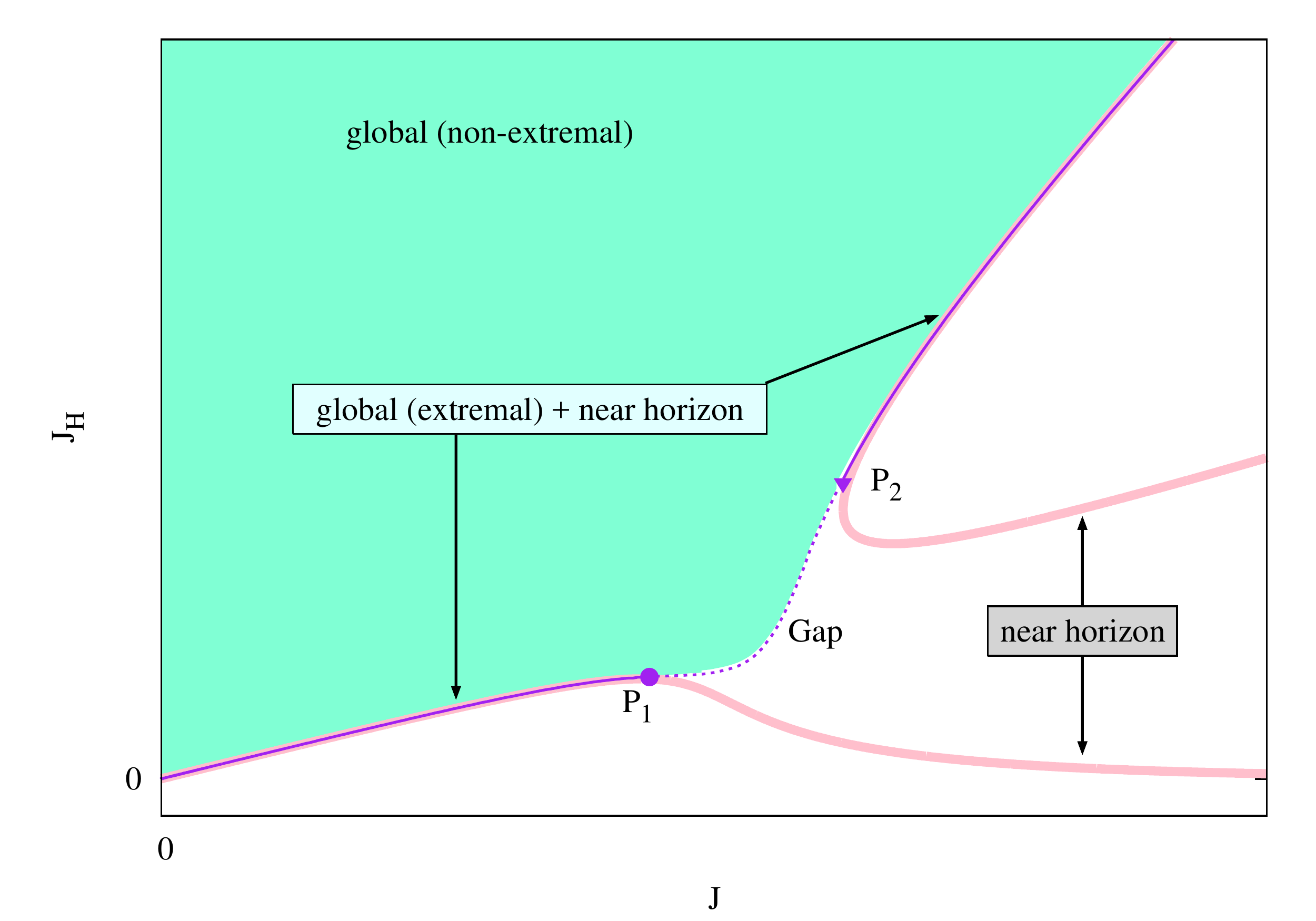}

    \caption{ The generic relation between near-horizon and global solutions
		found for small enough values of $\lambda$
		is shown in a $(J_H,J)$-diagram. 
		}
		
		\label{Em-my}
		
\end{figure}

\begin{figure}
    \centering
		
    \begin{subfigure}[b]{0.4\textwidth}
        \includegraphics[width=50mm,scale=0.5,angle=-90]{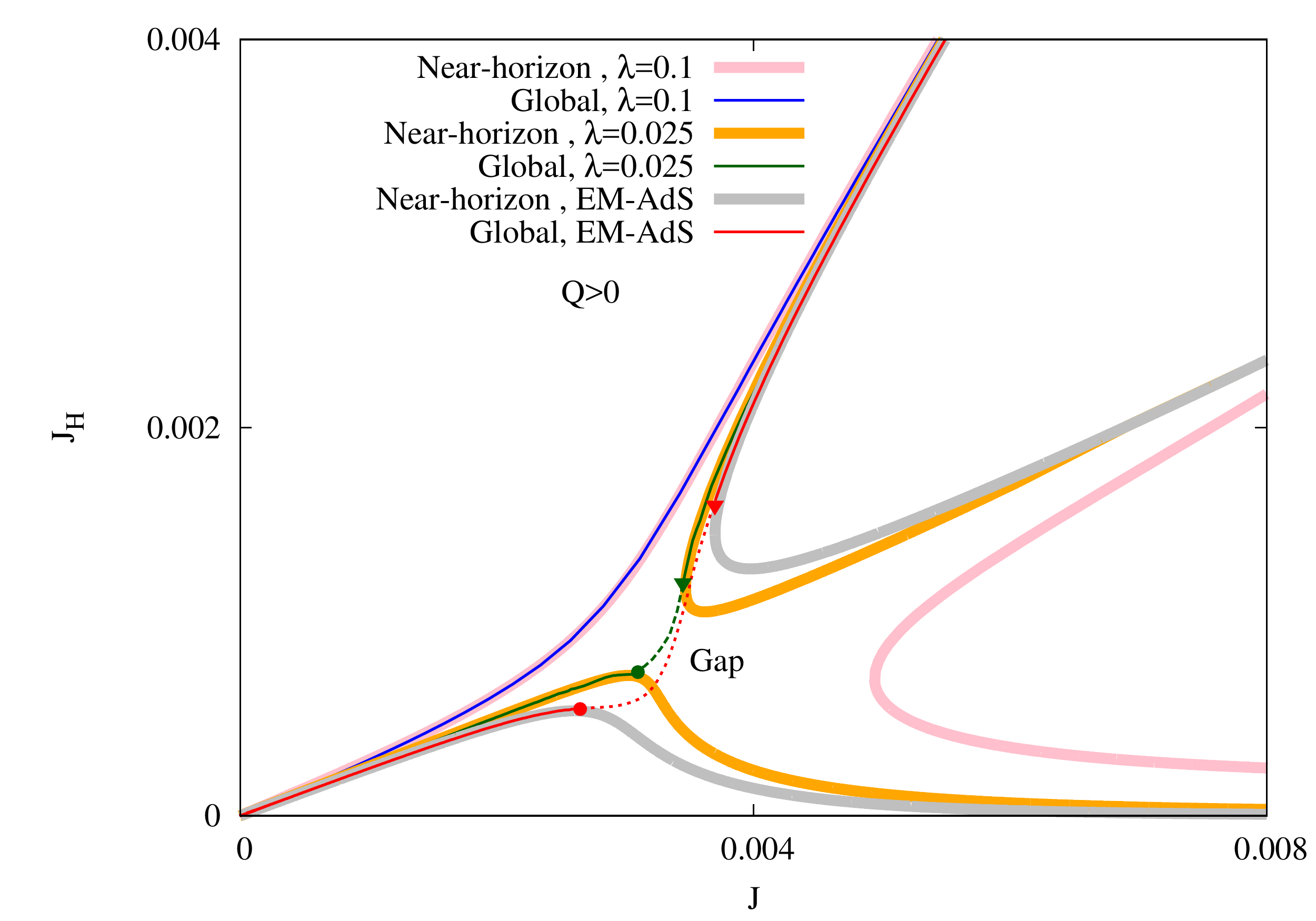}
        \caption{$Q>0$}
        \label{fig:Jh_small_lambda_p}
    \end{subfigure}
    \begin{subfigure}[b]{0.4\textwidth}
        \includegraphics[width=50mm,scale=0.5,angle=-90]{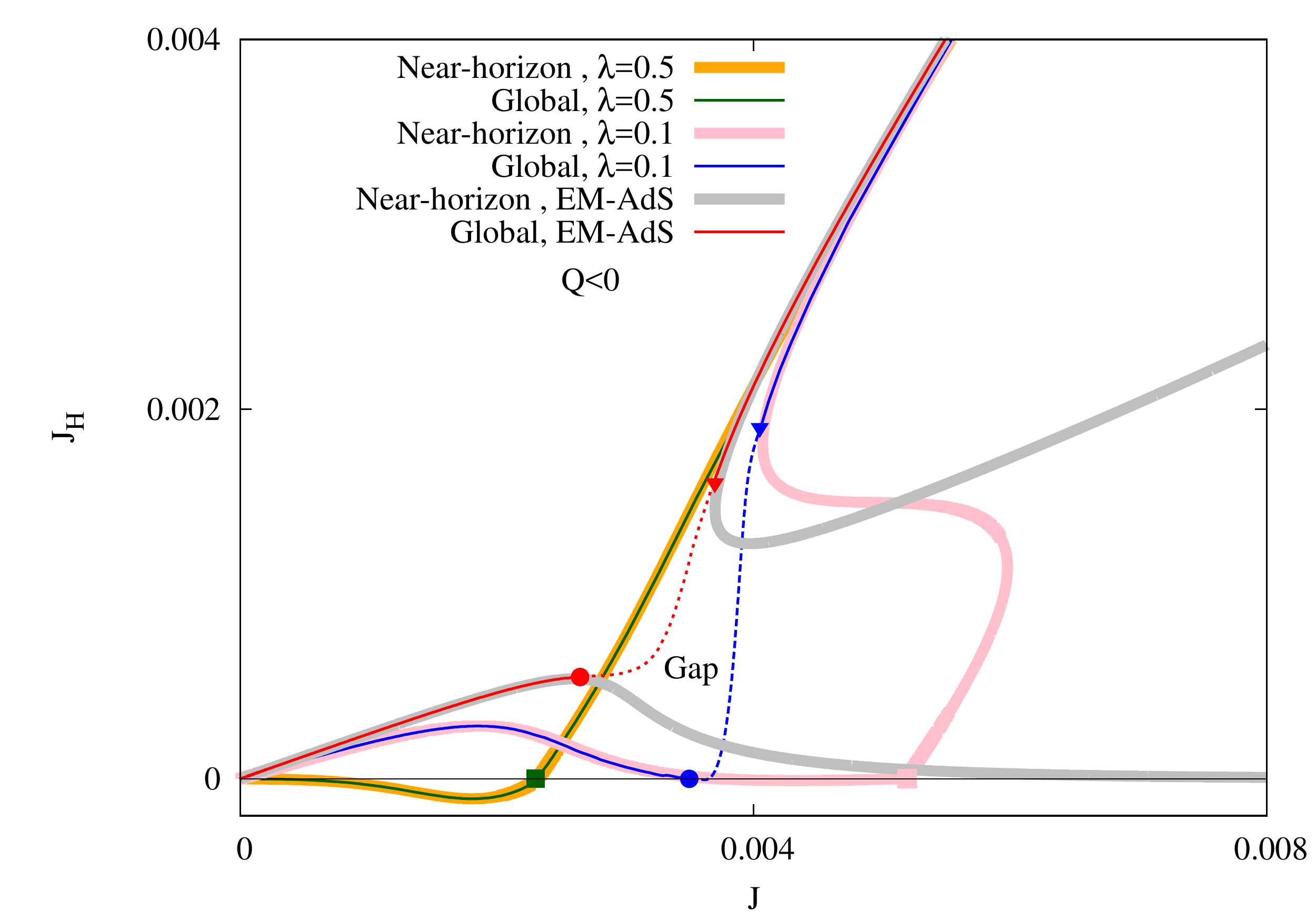}
        \caption{$Q<0$}
        \label{fig:Jh_small_lambda_n}
    \end{subfigure}
		
    \caption{Horizon angular momentum $J_H$ $vs.$ angular momentum $J$ for extremal black holes with fixed electric charge $Q=|0.044|$ and $L=1$, for small values of $\lambda$. Thick curves in grey, orange and pink represent near-horizon solutions. Thin curves in red, blue and green represent global extremal black holes. The corresponding dashed curves represent near-extremal solutions. 
		The dots and triangles represent the \textit{limiting} solutions separating the extremal branches, and the squares the \textit{critical} solutions. 
		One can see that not all near-horizon solutions correspond to a global configuration.
		}
		
		\label{fig:Jh_small_lambda}
		
\end{figure}

\medskip

Finally, let us summarize the main features of the configurations with fixed negative electric charge. 
\begin{itemize}
\item For every $\lambda$, one can find a \textit{critical} solution with angular momentum $J_0$, which separates two different extremal branches: the RN branch with $J<J_0$ and the MP branch with $J>J_0$. The \textit{critical} solution has $A_H=M_H=J_H=0$, 
and a discontinuity in $\Omega_H$ and in $\Phi_H$. The value of $J_0$ decreases with increasing $\lambda$.
\item The RN branch properties depend on the value of the CS coupling. 
For instance, in the SUGRA case the RN branch possesses a constant mass,
but for $\lambda<1$ the mass increases monotonically with $J$, while for $\lambda>1$ it decreases.  
For SUGRA, the RN branch satisfies $M_H=\Omega_H=0$, while for $\lambda<1$ one finds $M_H<0$, $\Omega_H>0$, and for $\lambda>1$ the behavior changes to $M_H>0$, $\Omega_H<0$.
\item Non-extremal black holes with $J<J_0$ can present counter-rotation ($J_H<0$). 
The size of this set reduces when we decrease $\lambda$.
\item Similar to BHs with fixed $J$, $\Omega_H<0$ can be found only for $\lambda>1$, and the angular velocity of non-extremal BHs can change abruptly around the \textit{critical} solution under small changes of the angular momentum, affecting even the direction of rotation.
\end{itemize}

\subsection{Solutions with a small Chern-Simons coupling} 

We have studied also families of solutions with values 
of CS coupling constant
$\lambda<0.5$, 
which was the minimal considered value in the previous Subsection.
We recall that the attractor solutions in Section 3 predict in this case
the existence of new features of extremal BHs,
with a bifurcating branch structure.
Indeed, 
our numerical results for global solutions
show that this is the case and
the picture discussed in the Section 5.1 
fails to capture some properties of
the BHs with small enough $\lambda$. 
This holds in particular for extremal BH solutions, 
whose study will allow us to better understand 
how generic the predictions of the near-horizon formalism in Section 3 are.

The generic picture found in this case is shown in
 Figure \ref{Em-my},
for a $(J_H,J)$-diagram of solutions with a fixed $Q$.
Both near-horizon  and global solutions are shown there.
One can see that, 
starting at $J=J_H=0$ extremal RN-AdS solutions, one finds a branch of BHs 
that ends at a \textit{limiting} solution  
with some nonzero values of  $J,J_H$ (the point $P_1$). 
A second branch of global extremal BHs is found coming from large values of $J,J_H$, 
and ending at the point $P_2\neq P_1$. 
Note that these two branches are in agreement with parts of the branches predicted by the near-horizon formalism. 
The points $P_1$ and $P_2$
are connected by a particular set of  extremal solutions (blue dashed line), 
which is called in what follows the {\it gap} set.
These solutions emerge as the limit of near-extremal global configurations
and appear to possess some pathological properties.
For example, the Kretschmann scalar takes very large values at the horizon,
which makes the direct construction of the extremal solutions connecting $P_1$ and $P_2$ difficult. 
The global non-extremal solutions exist in a domain bounded below by the global extremal BHs connected by the
gap set.
Another important feature one can see in  Figure \ref{Em-my}
is that a part of the near-horizon solutions 
 do $not$ have global counterparts.  

Numerical results 
supporting the above picture 
are shown in Figure \ref{fig:Jh_small_lambda},
 where we consider  extremal solutions with a fixed electric charge $Q=|0.044|$ and several values of $\lambda$
(qualitatively similar pictures have been found for other values of $Q$).
The case $Q>0$ is shown in 
  Figure \ref{fig:Jh_small_lambda_p}.
The grey, orange and pink curves there correspond to near-horizon solutions with 
$\lambda=0$, 
$\lambda=0.025$
 and 
$\lambda=0.1$  respectively. 
Note that these solutions have been presented already in Section 3,
where we have noticed the existence of two different branches of near-horizon solutions
for small enough values of $\lambda$,
 (in particular for $\lambda=0$ and $\lambda=0.025$), 
which bifurcate at $\lambda=0.0305$.
In the same Figure \ref{fig:Jh_small_lambda_p} we add the corresponding sets of 
$global$ extremal solutions 
(for example, the red thin line represents global extremal BHs in pure EM-AdS theory). 
A similar structure is found for other small enough values of the CS coupling constant,
in particular for $\lambda=0.025$, see Figure \ref{fig:Phi_Q0}.
One can see that when $\lambda>0.0305$ the space of configurations no longer presents a {\it gap} set. 
For instance, consider in Figure \ref{fig:Jh_small_lambda_p} the blue line. 
This is the set of global extremal solutions with $\lambda=0.1$, 
and it matches perfectly (in one to one correspondence) with the near-horizon solutions (pink line).

Let us consider now the negative charge case, $Q<0$.
Some results in this case are shown in Figure \ref{fig:Jh_small_lambda_n}. 
One can see that the qualitative picture discussed above for $Q>0$ still holds. 
For example, consider the pure $\lambda=0.1$ case (blue curve). 
Then one finds again the existence of a {\it gap} set (dashed blue line) 
connecting two disconnected branches of BHs that end at two different \textit{limiting} solutions (blue dot and blue triangle). As a consequence, not all near-horizon solutions for this value of $\lambda$ (pink line) correspond to global solutions.
The \textit{critical} solution with $J_H=0$ cannot be reached for this value of $\lambda$.
The situation changes when $\lambda>0.25$, in which case the gap disappears
(see the curve for $\lambda=0.5$). 
The two branches of extremal BHs (green line) are now joined at the \textit{critical} solution with $J_H=0$ (green rectangle).   
Interestingly, in this case with $\lambda=0.5$, all near-horizon solutions (orange line) correspond to a global solution. 
However, in the other cases, this is not true. 
Hence we conclude that also for $Q<0$ there are cases for which near-horizon solutions do not correspond to global solutions.

To summarize, the solutions with a small enough value of $\lambda$
show a number of features which are not captured by the knowledge of the CLP BHs
($e.g.$ the existence of a gap set).

\begin{figure}
     \centering

     \begin{subfigure}[b]{0.4\textwidth}
\includegraphics[width=50mm,scale=0.5,angle=-90]{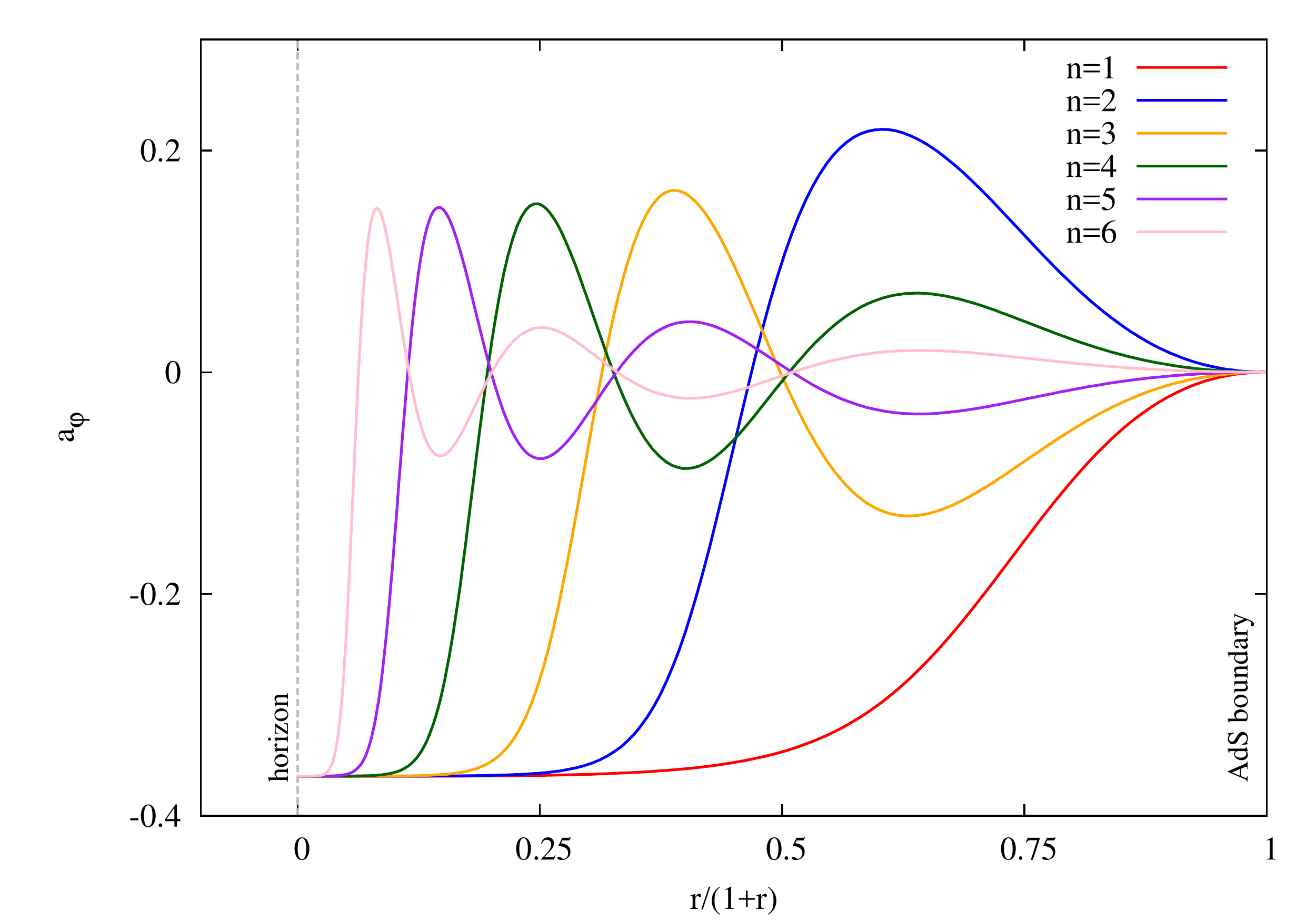}
         \caption{}
         \label{fig:ak_excitations}
     \end{subfigure}
     \begin{subfigure}[b]{0.4\textwidth}
\includegraphics[width=50mm,scale=0.5,angle=-90]{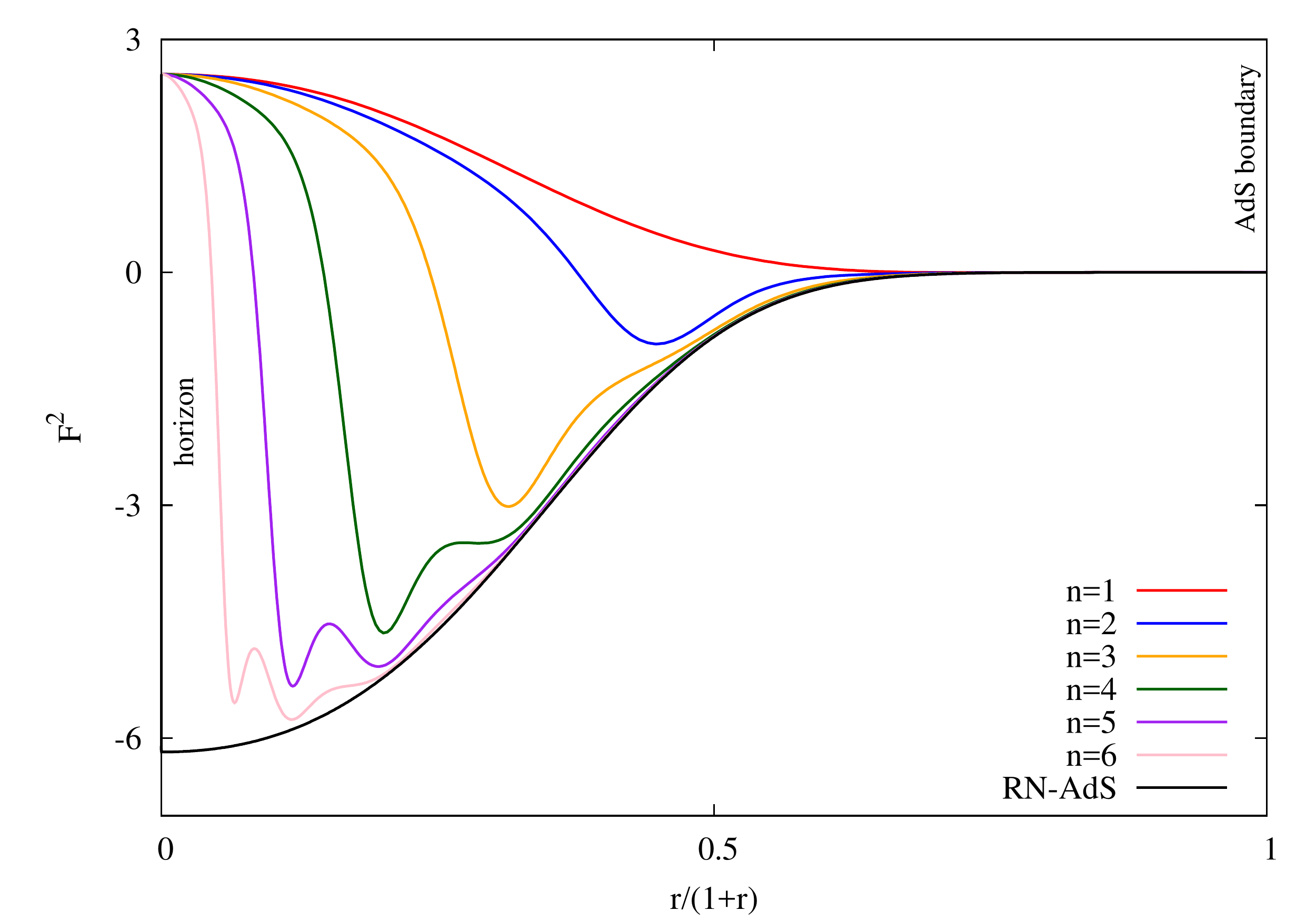}
         \caption{}
         \label{fig:F2_excitations}
     \end{subfigure}

     \caption{(a) The profile of the magnetic gauge potential $a_{\varphi}(r)$ is shown $vs.$ the
compactified radial coordinate for a sequence of  extremal black holes
with a different nodal structure. 
(b) The invariant $F^2=F_{\mu\nu}F^{\mu \nu}$ is shown $vs.$ the
compactified radial coordinate for the same set of solutions;
the corresponding profile for the extremal RNAdS solution is also included for reference. 
These solutions
have 
$J=0$,
$\lambda=5$, 
$L=10$ 
and $Q=-2.72$.}

         \label{fig:excitations}

\end{figure}

\subsection{Large Chern-Simons coupling: discrete sets of radially excited extremal black holes.}
 
New features occur as well for large enough value of $\lambda$.
For example, we have considered solutions with $\lambda=5$
and found an overall picture which  
is qualitatively similar 
to the one obtained in the asymptotically flat case 
\cite{Blazquez-Salcedo:2015kja}.  
The most interesting new feature here is the appearance of a set of non-static 
extremal BHs with vanishing total angular momentum, $J=0$. 
This special set  contains a  large number of $distinct$ solutions,
possibly an infinite one\footnote{So far we have constructed solutions 
with the highest node number $n=40$.
However,  
it is natural to conjecture the existence of solutions with arbitrarily high values of $n$.
}. 
Its members can be labeled 
by an integer   $n\geq 1$, 
which can be identified with the 
 the number of nodes found in the
profiles of the metric function
$\omega(r)$, and of the magnetic gauge potential $a_{\varphi}(r)$.
The solutions reported in the previous Section are the $fundamental$ ones,
with $n=1$ 
(since they still possess a zero of $\omega(r)$, $a_{\varphi}(r)$,
reached as $r\to \infty$). 
The mass of the solutions increases with $n$,
the numerics indicating that the extremal RN-AdS BH mass would be approached 
 as $n\to \infty$. 
Other quantities, like the horizon angular momentum and horizon angular velocity 
decrease with the $n$ number.

The profiles of the magnetic gauge potential $a_{\varphi}$ 
are shown in Figure \ref{fig:ak_excitations} 
for some typical set of $J=0$ extremal BHs with $n=1,..,6$. 
The number of nodes leaves an imprint also
in the invariant quantities. 
For example, in Figure \ref{fig:F2_excitations}  
the
square of the gauge field tensor, $F^2=F_{\mu}F^{\mu\nu}$, 
is shown for the same  configurations.
The 'oscillations' there are found also for the components of the energy-momentum tensor.
As such, 
the excited solutions ($n>1$)
possess a "layer structure", with $n$ distinct radii where the energy density concentrates. 
The more nodes, the more layers the solution develops in the bulk.

However, although
a number $n>1$ modifies the properties of the BHs in the bulk, 
the nodal structure is not seen in the near-horizon behavior.
That is, in that limit,
they are still described by the same squashed $AdS_2\times S^3$ solutions 
discussed in Section 3.
Then, we conclude that similar to the asymptotically flat
case \cite{Blazquez-Salcedo:2015kja},
a given near-horizon 
configuration can correspond to more than one  global solution 
(likely an infinite set). 
 
The  non-extremal BHs possess also excitations, in which case, 
however, we noticed
the existence of a maximal value of  $n$.
Moreover, as expected, excited solutions exist also for $T_H=0$  and
a nonzero $J$.  

A  detailed analysis of the excited configurations, 
with a full study of the branch structure and thermodynamic properties, 
will be presented elsewhere \cite{blazquez:excited_EMCSAdS}.

Finally, we mention that, unfortunately,
no such excited solutions could be found in the special 
$\lambda=1$ case.
In principle, in the absence of a uniqueness proof of the CLP solution,
their existence cannot be excluded.
However, for all input parameters we have considered so far,
the nodal structure disappears for values of $\lambda$
smaller than the SUGRA value.

\section{Conclusions}
This paper has presented a 
discussion of the basic properties of charged rotating
BHs in a $D=5$ EMCS-AdS theory 
with an arbitrary value of the CS coupling constant $\lambda$.
The considered solutions have two equal-magnitude angular momenta, 
possess no pathologies on and outside of an event horizon
of spherical topology,
and approach at infinity a globally AdS background.
So far, the only known solutions of the EMCS-AdS model 
compatible with these assumptions
are the BHs found  
in \cite{Cvetic:2004hs} 
by  
Cveti\v c, L\"u and Pope (CLP)
for
$\lambda=1$, $i.e.$ a minimal gauged supergravity model.
The main questions  we have tried to answer were: 
{\it  "How general are the properties of the CLP solution?"}
and 
{\it  "Are there new features for other values of $\lambda$?"}.
 
\medskip

The main conclusion of our study is that
the intuition based on the CLP BHs $cannot$ 
be safely extrapolated
to solutions of a generic EMCS model. 
New unexpected features occur 
for sufficiently small values of $\lambda$
(in particular for an EM model) and  
for also for large  $\lambda$.

The comparison with the SUGRA exact solutions  
is most easily done in the extremal case.
This limit reduces the parameter space of solutions
and also allows for a partial analytical understanding
based on results within the attractor mechanism.

For sufficiently small values of $\lambda$ 
the most interesting 
 new feature (which is absent in the SUGRA case),
is the existence
of two disconnected branches of extremal BHs.
The  bridge between these BHs is provided by a set of 
extremal solutions which appear to possess pathological properties.
Also, this $gap$ set cannot be described within the attractor mechanism.

New properties  occur as well for large $\lambda$.
The most striking one is the existence
of excited solutions, which are labeled 
by the number of nodes $n$ of the metric function $\omega(r)$ 
(or of the magnetic gauge potential $a_\varphi(r)$).
This nodal structure does not exist in the SUGRA case,
where we could not (numerically) find other solutions 
which in principle could exist apart from the 
$n=1$ in \cite{Cvetic:2004hs}.
Also, it cannot be captured by the near-horizon configurations. 
In fact, the relation between the solutions found within 
 the attractor formalism
and the global ones is quite intricate. 
For example,
 a given near-horizon solution
can correspond to  one global solution,
more than one global solution (possibly even an infinite set),
or, more striking,
no global solution at all.

Other, less spectacular differences, which occur when varying $\lambda$,
are discussed in the Section 5.2.

However,
there are also a number of features which seem to be generic for any $\lambda$.
For example,
the presence of a CS term always implies the occurrence of a 
\textit{critical} set of extremal solutions with a vanishing event horizon area,  $A_H=0$.
Also, the overall thermodynamical 
behavior of the solutions is well captured by
the exact CLP solution, 
the BHs possessing a positive heat capacity 
for large enough values of $J,Q$.
Moreover, the solutions with large temperatures
are less sensitive to changes in $\lambda$. 
 
\medskip
The solutions obtained in this paper
may provide a fertile ground for the further study of
charged rotating configurations in EMCS theory with a negative cosmological constant.
For example, their generalization to include
more scalars is straightforward.
Also, in principle, by using the same techniques,
there should be no difficulty to construct similar AdS solutions
in $D=2N+1$ dimensions,
with $N>2$ equal-magnitude angular momenta.
Also, it would be interesting to find applications of the  solutions 
in this work in an AdS/CFT context.
However, the fact that all $\lambda\neq 1$ solutions
 do not solve a supergravity model
makes it more difficult to obtain a CFT description.

\medskip
\medskip
{\bf Acknowledgements}
\\
J.L.B.S. would like to thank Robert Mann and Carlos Herdeiro for
helpful comments and discussions.
We gratefully acknowledge support by
the DFG Research Training Group 1620 ``Models of Gravity''.
 E. R. acknowledges funding from the FCT-IF programme.
This work was also partially supported 
by  the  H2020-MSCA-RISE-2015 Grant No.  StronGrHEP-690904, 
and by the CIDMA project UID/MAT/04106/2013.  
J.L.B.S. and J.K. gratefully acknowledge support by the grant FP7, Marie Curie Actions, People, International Research Staff Exchange Scheme (IRSES-606096). F. N.-L. acknowledges funding from Complutense University under Project No. PR26/16-20312.

\medskip
\medskip
\textbf{\textit{Note added:}}
\\
During the preparation of this paper for publication, we received communication
from M. Mir and R. B. Mann concerning their research in \cite{Mann:2016}, which overlaps with some of the results presented here.
However, the  approach in \cite{Mann:2016} is complementary to that in our work,
Mir and  Mann presenting closed form solutions obtained via a perturbative approach around the
MPAdS BHs.
 
\appendix
\setcounter{equation}{0}
\renewcommand{\theequation}{A.\arabic{equation}}

\section{The $\lambda=1$ Cveti\v c-L\"u-Pope black holes}

\subsection{Solution}

The most general charged rotating BH solution with two equal angular momenta 
of the EMCS-AdS equations,
 which is known in closed form,
has been reported by Cveti\v c, L\"u and Pope in Ref. \cite{Cvetic:2004hs}
(see also
\cite{Kunduri:2005zg},
\cite{Madden:2004ym},
\cite{Davis:2005ys}
for further investigations of it).

The expression of this solution 
within the Ansatz (\ref{metrici}) and (\ref{gauge1}) reads
\begin{eqnarray}
\label{CLP}
&&
\frac{1}{F_1(r)}=1-\frac{2m(1-\frac{a^2}{L^2})-2q}{r^2}+\frac{2a^2m+(1-\frac{a^2}{L^2})q^2}{r^4}+\frac{r^2}{L^2},
~~
 F_2(r)=r^2,~~
\\
&&
\nonumber
 F_3(r)=r^2({1+\frac{2a^2m}{r^4}}-\frac{a^2q^2}{r^6}),~~
 F_0(r)=\frac{1-\frac{2m(1-\frac{a^2}{L^2})-2q}{r^2}+\frac{2a^2m+(1-\frac{a^2}{L^2})q^2}{r^4}+\frac{r^2}{L^2}}
{1+\frac{2a^2m}{r^4} -\frac{a^2q^2}{r^6}},
\\
\nonumber
&&
W(r)=\frac{a(2m-q-\frac{q^2}{r^2})}{r^4 (1+\frac{2a^2m}{r^4}-\frac{a^2q^2}{r^6})},~~
a_\varphi(r)=-\frac{\sqrt{3}aq}{2r^2},~~a_0(r)=\frac{\sqrt{3}q}{2r^2},
\end{eqnarray}
where $a,q$ and $m$ are three constants.
To make contact with the approach in this work, 
we express $m$ as a function of the 
event horizon radius $r_H$ (with $1/F_1(r_H)=F_0(r_H)=0$):
\begin{eqnarray}
m= \frac{\frac{r_H^4}{2L^2}+(\frac{q+r_H^2}{2r_H^2})^2-\frac{a^2q^2}{2L^2r_H^2}}
{1-a^2(\frac{1}{L^2} +\frac{1}{r_H^2})}~.
\end{eqnarray}
Working again in a non-rotating frame at infinity, 
the quantities which enter a thermodynamic description of the solutions, as expressed in terms of $r_H,a$ and $q$,
 read:
\begin{eqnarray}
\nonumber
&&
M=\frac{\pi}{8}\frac{r_H^2}
{( 1-a^2(\frac{1}{L^2} +\frac{1}{r_H^2}))}
\bigg[
(3+\frac{a^2 }{L^2})(1+\frac{r_H^2}{L^2})
+\frac{q}{r_H^4}
\left(q(3-\frac{a^4}{L^4}-\frac{2a^2}{L^2})
+2a^2(3+\frac{4r_H^2}{L^2})
\right)
\bigg],
\\
\nonumber
&&
Q=\frac{\sqrt{3}\pi}{2}q,~~~~\Phi_H=\frac{\sqrt{3}q}{2r_H^2}\frac{1-\frac{a^2}{r_H^2}(1+\frac{r_H^2}{L^2})}{1+\frac{a^2q}{r_H^4}},
~~~~
\Omega_H=\frac{a}{L^2}\frac{1+\frac{L^2(q+r_H^2)}{r_H^4}}{1+\frac{a^2q}{r_H^4}},
\\ 
&&
J=\frac{\pi a}{4} 
\bigg[
\frac{\frac{r_H^4}{L^2}+\frac{(q+r_H^2)^2}{r_H^2}-\frac{a^2q^2}{L^2r_H^2}}
{  1-a^2(\frac{1}{L^2} +\frac{1}{r_H^2}) }-q\bigg],~~~~
A_H=\frac{2\pi^2 r_H^3(1+\frac{a^2q}{r_H^4})}{\sqrt{ 1-a^2(\frac{1}{L^2} +\frac{1}{r_H^2})}},~~
\\
\nonumber
&&
\label{TH-CLP}
T_H=\frac{1}{2\pi r_H}\frac{(1+\frac{a^2 q}{r_H^4})^{-1}}{\sqrt{  1-a^2(\frac{1}{L^2} +\frac{1}{r_H^2}) }}
\bigg[
(1-\frac{2a^2+q}{r_H^2})(1+\frac{q}{r_H^2})
-\frac{a^2(a^2q^2+2r_H^4)}{L^4r_H^4}
+\frac{2(r_H^6+a^2(q^2-2r_H^4)}{L^2r_H^4}
\bigg]~.
\end{eqnarray}
The corresponding expressions for mass and angular momentum of the horizon are
\begin{eqnarray}
&&
M_H=\frac{\pi}{4}\frac{a^2 q +r_H^4}{L^2}
\frac{2+\frac{a^2 q}{r_H^4}\left(1+(1+\frac{L^2}{r_H^2})(1+\frac{q}{r_H^2}) \right)+\frac{L^2}{r_H^2}(1-\frac{q^2}{r_H^4})}
{  1-a^2(\frac{1}{L^2} +\frac{1}{r_H^2}) },
\\
\nonumber
&&
J_H=- \frac{\pi}{8}\frac{a (a^2 q +r_H^4)}{L^2}
\frac{(1+\frac{L^2}{r_H^2})(2+\frac{2a^2q^2}{r_H^6}+\frac{L^2q}{r_H^4}(2-\frac{q}{r_H^2}))}
{ 1-a^2(\frac{1}{L^2} +\frac{1}{r_H^2}) }.
\end{eqnarray}
The parameters $(r_H,a,q)$ are subject to the condition
\begin{eqnarray}
(1-\frac{2a^2+q}{r_H^2})(1+\frac{q}{r_H^2})
-\frac{a^2(a^2q^2+2r_H^4)}{L^4r_H^4}
+\frac{2(r_H^6+a^2(q^2-2r_H^4)}{L^2r_H^4}
 \geq 0.
\end{eqnarray}
If the inequality in the equation above is saturated, the horizon is degenerate and we get an extremal BH.
With $q=0$, the CLP solutions reduce to MP-AdS spinning BHs with equal angular momenta \cite{Hawking:1998kw}.
Another limit of interest corresponds to $a=0$, in which case one recovers the
RN-AdS BHs.

\subsection{Extremal limit and the critical solutions}
In discussing the $T_H=0$ limit of these solutions,
it is convenient to reparametrize the constants $a,r_H$ as
\begin{eqnarray} 
\label{param} 
a=L x,~~ r_H=\frac{L x y}{\sqrt{1-x^2}},
\end{eqnarray}  
with 
$0\leq x<1$,
$ 1\leq y <\infty$.
	Then the conditions $T_H=0$ is written as
\begin{eqnarray}  
q=q_{\pm}=\frac{L^2 x^2}{(1-x^2)^2}\left(-1 \pm (y^2-1)\sqrt{1+2y^2 x^2} \right), 
\end{eqnarray} 	
which reveals the existence of two branches of extremal solutions,
in terms of the parameters $(x,y)$. 
In particular,
 BHs with $T_H=0$ can be found for any value of $(Q,J)$.

The extremal BHs possess an interesting limit with a zero event horizon area,
corresponding to the $\lambda=1$
 \textit{critical} solution discussed above.
This limit is approached for
\begin{eqnarray}  
\label{limit}
y=\frac{r_H}{a}\sqrt{1-\frac{a^2}{L^2}} \to 1,
\end{eqnarray} 
on both branches of solutions. 
Interesting enough, the $\pm$ global charges are the same as $y  \to 1$:
\begin{eqnarray}  
M_{\pm}\to \frac{L^2\pi}{8}\frac{x^2(x^4-3x^2+6)}{(1-x^2)^3},~~
J_{\pm}\to \frac{L^3\pi}{4}\frac{x^3}{(1-x^2)^3},~~
Q_\pm \to -\frac{L^2\pi}{2}\frac{x^2}{(1-x^2)^2}<0,
\end{eqnarray} 
while the $\pm$ expressions of 
electrostatic potential
and
horizon angular velocity are different
\begin{eqnarray}  
\Phi_{H\pm}\to-\frac{\sqrt{3} }{2(2\pm \sqrt{1+2x^2}) },~~
\Omega_{H\pm} \to \frac{1}{L}\frac{1+x^2\pm\sqrt{1+2x^2}}{x(2\pm \sqrt{1+2x^2})} .
\end{eqnarray} 
This shows the existence of a discontinuity 
in both $\Phi_H$ and $\Omega_H$
as the limit $y\to 1$ is approached, 
with different
limiting values for these quantities on each branch 
(although the global charges are equal).

The solution with $y=1$
has an interesting closed-form expression.
After replacing 
(\ref{param}),
(\ref{limit}) in 
(\ref{CLP})
one finds
\begin{eqnarray}  
&&
\nonumber
F_0(r)=\frac{r^2(1-x^2)}{L^2}\frac{\left(r^2(1-x^2)+L^2(1+x^2) \right)
 \left(r^2(1-x^2)-L^2 x^2 \right) }{r^4(1-x^2)^3+r^2L^2x^2(1-x^2)^2+L^4x^4},
\\
&&
F_1(r)=\frac{L^2r^4 (1-x^2)^3}{(r^2(1-x^2)+L^2(1+x^2))(r^2(1-x^2)-L^2x^2)^2},~~F_2(r)=r^2,
\\
&&
\nonumber
F_3(r)=\frac{r^6(1-x^2)^4+r^2 L^4 x^6 (1-x^2)-L^6x^6}{r^4 (1-x^2)^4},~~
~~
W(r)=\frac{L^3 x^3}{r^4(1-x^2)^3+r^2 L^2 x^2(1-x^2)^2+L^4 x^4},
\\
\nonumber
&&
a_\varphi(r)=\frac{\sqrt{3}L^3}{2r^2}\frac{x^3}{(1-x^2)^2},~~
a_0(r)=-\frac{\sqrt{3}L^2}{2r^2}\frac{x^2}{(1-x^2)^2}.
\end{eqnarray} 
A direct inspection shows that this describes a BH spacetime,
with standard AdS asymptotics.
The event horizon is  located at
\begin{eqnarray}  
r=r_H=\frac{Lx}{\sqrt{1-x^2}}\geq 0.
\end{eqnarray} 
Despite possessing a zero horizon area,
this configuration shows
no (obvious) signs of a pathological behavior.
For example, 
both the Ricci and Kretschmann scalar are finite on and outside the horizon. 

Its near horizon expansion, $r\to r_H$
reads
 \begin{eqnarray}  
&&
F_0(r)=\frac{2}{L}\frac{\sqrt{1-x^2}(1+2x^2)}{x(3-2x^2)}(r-r_H)+\dots,
~~
\frac{1}{F_1(r)}=\frac{4}{L^2} \frac{1+2x^2}{x^2 } (r-r_H)^2+\dots,~~
\\
\nonumber
&&
F_2(r)=r_H^2+\dots,~~
F_3(r)=2L\frac{x(3-2x^2)}{(1-x^2)^(3/2)}(r-r_H),~~
\\
\nonumber
&&
a_\varphi(r)=\frac{\sqrt{3}L}{2}\frac{x}{1-x^2}+\dots,
~~
a_0(r)=-\frac{\sqrt{3} }{2}\frac{1}{1-x^2}+\dots,
~~
w(r)=-\frac{1}{L}\frac{1}{x(3-2x^2)}+\dots,
\end{eqnarray}
which, to leading order, describes an AdS$_3\times S^2$ geometry.  
Other properties of this special solution have been already 
reported in the main text.

 \begin{small}

 \end{small}


\begin{thebibliography}{99}

\bibitem{Hawking:1973uf} 
  S.~W.~Hawking and G.~F.~R.~Ellis,
  ``The Large scale structure of space-time,''
  Cambridge University Press, Cambridge, 1973 
\bibitem{Witten:1998qj}
E.~Witten,
Adv.\ Theor.\ Math.\ Phys.\  {\bf 2} (1998) 253
[arXiv:hep-th/9802150].
\bibitem{Maldacena:1997re}
J.~M.~Maldacena,
Adv.\ Theor.\ Math.\ Phys.\  {\bf 2} (1998) 231
[Int.\ J.\ Theor.\ Phys.\  {\bf 38} (1999) 1113]
[arXiv:hep-th/9711200].

\bibitem{Myers:1986un}
  R.~C.~Myers and M.~J.~Perry,
  Annals Phys.\  {\bf 172} (1986) 304.
\bibitem{Hawking:1998kw}
  S.~W.~Hawking, C.~J.~Hunter and M.~M.~Taylor-Robinson,
  Phys.\ Rev.\ D {\bf 59} (1999) 064005
  [arXiv:hep-th/9811056].
\bibitem{Kunz:2007jq}
  J.~Kunz, F.~Navarro-L\'erida and E.~Radu,
  Phys.\ Lett.\ B {\bf 649} (2007) 463
  [gr-qc/0702086].
 \bibitem{Cvetic:2004hs}
  M.~Cveti\v c, H.~L\"u and C.~N.~Pope,
  Phys.\ Lett.\ B {\bf 598} (2004) 273
  [arXiv:hep-th/0406196].
\bibitem{Gutowski:2004ez}
  J.~B.~Gutowski and H.~S.~Reall,
  JHEP {\bf 0402} (2004) 006
  [hep-th/0401042].
\bibitem{Chong:2005hr}
  Z.~W.~Chong, M.~Cveti\v c, H.~L\"u and C.~N.~Pope,
  Phys.\ Rev.\ Lett.\  {\bf 95} (2005) 161301
  [arXiv:hep-th/0506029];
\\ 
  Z.~W.~Chong, M.~Cveti\v c, H.~L\"u and C.~N.~Pope,
  Phys.\ Lett.\ B {\bf 644} (2007) 192
  [arXiv:hep-th/0606213];
\\
  Z.~W.~Chong, M.~Cveti\v c, H.~L\"u and C.~N.~Pope,
  Phys.\ Rev.\ D {\bf 72} (2005) 041901
  [arXiv:hep-th/0505112];
\\
  M.~Cveti\v c, H.~L\"u and C.~N.~Pope,
  Phys.\ Rev.\ D {\bf 70} (2004) 081502
  [arXiv:hep-th/0407058];
	
\bibitem{Blazquez-Salcedo:2013muz}
  J.~L.~Bl\'azquez-Salcedo, J.~Kunz, F.~Navarro-L\'erida and E.~Radu,
  Phys.\ Rev.\ Lett.\  {\bf 112} (2014) 011101
  [arXiv:1308.0548 [gr-qc]].
\bibitem{Blazquez-Salcedo:2015kja}
  J.~L.~Bl\'azquez-Salcedo, J.~Kunz, F.~Navarro-L\'erida and E.~Radu,
  Phys.\ Rev.\ D {\bf 92} (2015),  044025
  [arXiv:1506.07802 [gr-qc]].

\bibitem{Blazquez-Salcedo:2016hae} 
  J.~L.~Bl\'azquez-Salcedo, J.~Kunz, F.~Navarro-L\'erida and E.~Radu,
  arXiv:1602.00822 [gr-qc].
\bibitem{Figueras:2014dta}
  P.~Figueras and S.~Tunyasuvunakool,
  JHEP {\bf 1503} (2015) 149
   [JHEP {\bf 1503} (2015) 149]
  [arXiv:1412.5680 [hep-th]].
	
	
	
\bibitem{Brihaye:2010wx}
  Y.~Brihaye, B.~Kleihaus, J.~Kunz and E.~Radu,
  JHEP {\bf 1011} (2010) 098
  [arXiv:1010.0860 [hep-th]].
\bibitem{Dias:2011at}
  O.~J.~C.~Dias, G.~T.~Horowitz and J.~E.~Santos,
  JHEP {\bf 1107} (2011) 115
  [arXiv:1105.4167 [hep-th]].


\bibitem{Stotyn:2011ns} 
  S.~Stotyn, M.~Park, P.~McGrath and R.~B.~Mann,
  Phys.\ Rev.\ D {\bf 85}, 044036 (2012)
  [arXiv:1110.2223 [hep-th]].
  
  
\bibitem{Stotyn:2013yka}
  S.~Stotyn, C.~D.~Leonard, M.~Oltean, L.~J.~Henderson and R.~B.~Mann,
  Phys.\ Rev.\ D {\bf 89} (2014) 044017
  [arXiv:1307.8159 [hep-th]].
\bibitem{Kunz:2006eh}
  J.~Kunz, F.~Navarro-L\'erida and J.~Viebahn,
  Phys.\ Lett.\ B {\bf 639} (2006) 362
  [arXiv:hep-th/0605075].
\bibitem{Kunz:2006yp} 
  J.~Kunz and F.~Navarro-L\'erida,
  Phys.\ Lett.\ B {\bf 643} (2006) 55
  [hep-th/0610036].
  
\bibitem{Brihaye:2014nba} 
  Y.~Brihaye, C.~Herdeiro and E.~Radu,
  Phys.\ Lett.\ B {\bf 739}, 1 (2014)
  [arXiv:1408.5581 [gr-qc]].
  
\bibitem{Brihaye:2016vkv} 
  Y.~Brihaye, C.~Herdeiro and E.~Radu,
  Phys.\ Lett.\ B {\bf 760}, 279 (2016)
  [arXiv:1605.08901 [gr-qc]].
  

\bibitem{Aliev:2006yk} 
  A.~N.~Aliev,
  Phys.\ Rev.\ D {\bf 74} (2006) 024011

\bibitem{Aliev:2008bh} 
  A.~N.~Aliev and D.~K.~Ciftci,
  Phys.\ Rev.\ D {\bf 79} (2009) 044004


\bibitem{Allahverdizadeh:2010xx}
  M.~Allahverdizadeh, J.~Kunz and F.~Navarro-L\'erida,
  Phys.\ Rev.\ D {\bf 82} (2010) 024030
  [arXiv:1004.5050 [gr-qc]].
\bibitem{Allahverdizadeh:2010fn}
  M.~Allahverdizadeh, J.~Kunz and F.~Navarro-L\'erida,
  Phys.\ Rev.\ D {\bf 82} (2010) 064034
  [arXiv:1007.4250 [gr-qc]]. 
\bibitem{Kunduri:2007qy}
  H.~K.~Kunduri and J.~Lucietti,
  JHEP {\bf 0712} (2007) 015
  [arXiv:0708.3695 [hep-th]].	

\bibitem{Sen:2005wa} 
  A.~Sen,
  JHEP {\bf 0509}, 038 (2005)
  [hep-th/0506177].
\bibitem{Astefanesei:2006dd} 
  D.~Astefanesei, K.~Goldstein, R.~P.~Jena, A.~Sen and S.~P.~Trivedi,
  JHEP {\bf 0610}, 058 (2006)
  [hep-th/0606244].
\bibitem{Goldstein:2007km}
  K.~Goldstein and R.~P.~Jena,
  JHEP {\bf 0711}, 049 (2007)
  [arXiv:hep-th/0701221].
\bibitem{Suryanarayana:2007rk} 
  N.~V.~Suryanarayana and M.~C.~Wapler,
  Class.\ Quant.\ Grav.\  {\bf 24}, 5047 (2007)
  [arXiv:0704.0955 [hep-th]].	

\bibitem{Wald:1993nt} 
  R.~M.~Wald,
  Phys.\ Rev.\ D {\bf 48}, 3427 (1993)
  [gr-qc/9307038].
\bibitem{Lee:1990nz} 
  J.~Lee and R.~M.~Wald,
  J.\ Math.\ Phys.\  {\bf 31}, 725 (1990).
\bibitem{Rogatko:2007pv} 
  M.~Rogatko,
  Phys.\ Rev.\ D {\bf 75}, 024008 (2007)
  [hep-th/0611260].
\bibitem{Ashtekar:1984zz}
  A.~Ashtekar and A.~Magnon,
  Class.\ Quant.\ Grav.\  {\bf 1} (1984) L39;
	\\
  A.~Ashtekar and S.~Das,
  Class.\ Quant.\ Grav.\  {\bf 17}, L17 (2000)
  [hep-th/9911230].
\bibitem{Balasubramanian:1999re}
V.~Balasubramanian and P.~Kraus,
Commun.\ Math.\ Phys.\ \textbf{208}
(1999) 413.  


  
\bibitem{Kastor:2009wy} 
  D.~Kastor, S.~Ray and J.~Traschen,
  Class.\ Quant.\ Grav.\  {\bf 26}, 195011 (2009)
  [arXiv:0904.2765 [hep-th]].

 
  
\bibitem{Cvetic:2010jb} 
  M.~Cveti\v c, G.~W.~Gibbons, D.~Kubiznak and C.~N.~Pope,
  Phys.\ Rev.\ D {\bf 84}, 024037 (2011)
  [arXiv:1012.2888 [hep-th]].


\bibitem{COLSYS}
 U. Ascher, J. Christiansen, R.~D. Russell,
 Mathematics of Computation {\bf 33} (1979) 659;
 ACM Transactions {\bf 7} (1981) 209.	

	
\bibitem{Cvetic:2005zi}
  M.~Cveti\v c, G.~W.~Gibbons, H.~L\"u and C.~N.~Pope,
{\it "Rotating black holes in gauged supergravities: Thermodynamics, supersymmetric limits, topological solitons and time machines"'},
  hep-th/0504080.

\bibitem{Hawking:1982dh}
  S.~W.~Hawking and D.~N.~Page,
  Commun.\ Math.\ Phys.\  {\bf 87} (1983) 577.


%
\bibitem{Aliev:2004ec}
  A.~N.~Aliev and V.~P.~Frolov,
  Phys.\ Rev.\ D {\bf 69} (2004) 084022
  [hep-th/0401095].

\bibitem{Aliev:2006tt}
  A.~N.~Aliev,
  Class.\ Quant.\ Grav.\  {\bf 24} (2007) 4669
  [hep-th/0611205];
\\	
  A.~N.~Aliev,
  Phys.\ Rev.\ D {\bf 75} (2007) 084041
  [hep-th/0702129].


\bibitem{blazquez:excited_EMCSAdS} 
  J. L. Bl\'azquez-Salcedo,
{\it 
"Radially excited AdS$_5$ black holes in Einstein-Maxwell-Chern-Simons theory"}, to appear.


\bibitem{Mann:2016} 
  M.~Mir and R.~B.~Mann,
  Phys.\ Rev.\ D {\bf 95} (2017) 024005




\bibitem{Kunduri:2005zg} 
  H.~K.~Kunduri and J.~Lucietti,
  Nucl.\ Phys.\ B {\bf 724}, 343 (2005)
  [hep-th/0504158].
\bibitem{Madden:2004ym}
  O.~Madden and S.~F.~Ross,
  Class.\ Quant.\ Grav.\  {\bf 22} (2005) 515
  [hep-th/0409188].


\bibitem{Davis:2005ys}
  P.~Davis, H.~K.~Kunduri and J.~Lucietti,
  Phys.\ Lett.\ B {\bf 628} (2005) 275
  [hep-th/0508169].





 \end{thebibliography}
\end{document}